\documentstyle[12pt,cite,psfig]{article}

\renewcommand{\today}{January 30, 1998}

\newcommand{\nc}{\newcommand}
\nc{\mb}[1]{\makebox[#1]{}}

\begin{document}
\thispagestyle{empty}
\begin{center}
{\large{\bf The Validity of Charge Symmetry for Parton Distributions}}
\\
\vspace{1.8cm}
J. T. Londergan \\
\vspace{1.0cm}
{\it 
Department of Physics and Nuclear Theory Center, \\
Indiana University, Bloomington, IN 47404, USA  } \\
\vspace{1.2cm}
A.W.~Thomas \\
\vspace{1.0cm}
{\it
Department of Physics and Mathematical Physics and \\
Special Research Centre for the Subatomic Structure of Matter \\
University of Adelaide, Adelaide 5005, Australia } \\
\vspace{1.0cm}
\today
\vspace{1.0cm}
\begin{abstract}
Recent measurements of the Gottfried Sum Rule have focused attention
on the possibility of substantial breaking of flavor symmetry
in sea quark distributions of the proton.  This has been confirmed
by by pp and pD Drell-Yan processes measured at FNAL.  The theoretical 
models used to infer flavor symmetry breaking rely on the
assumption that parton distributions are charge symmetric; 
it is conceivable that current tests of flavor symmetry could
be affected by substantial charge symmetry violation.    
Since all phenomenological parton distributions assume the validity
of charge symmetry, in this paper we examine the possibility that
charge symmetry is violated [CSV].  We first list definitions for 
structure functions which do not make the usual assumption
that parton distributions obey charge symmetry.  We then give some 
simple model estimates of CSV for both valence and sea quark 
distributions.  Next, we list a set of relations which must hold if
charge symmetry is valid, and we review 
the current experimental limits on charge symmetry
violation in parton distributions. We then propose a series of possible
experimental tests of charge symmetry.  The proposed 
experiments could either detect charge symmetry violation in 
parton distributions, or they could provide more stringent
upper limits on CSV.  We discuss CSV contributions to sum
rules, and we propose new sum rules which could differentiate
between flavor symmetry, and charge symmetry, violation in
nuclear systems.   
\end{abstract}

\end{center}
\begin{flushleft}

\end{flushleft}

\newpage
\tableofcontents
\newpage

\newpage

\section{Introduction}
\mb{.5cm}

It has long been recognized that the strong interaction
respects charge symmetry to a high degree.  We review
the definition of charge symmetry, which is sometimes  
confused with isospin symmetry.  For details we refer
the reader to comprehensive reviews on charge symmetry by Miller, 
Nefkens and Slaus \cite{Mil90}, and Henley and Miller 
\cite{Hen79}.  The assumption of isospin independence 
of hadronic forces requires that the Hamiltonian of
the system commutes with the isospin operator {\bf T}, i.e.\ 
\begin{equation}
\left[ H, {\bf T}\right] = \left[ H,  T^2 \right] = 0
\end{equation}
Whereas isospin symmetry requires invariance of the Hamiltonian
with respect to all rotations in isospin space, charge
symmetry requires invariance only with respect to rotations
of $180^\circ$ about the $T_2$ axis, where the charge corresponds 
to the
third axis in isospin space.  Consequently, isospin symmetry
necessarily implies the validity of charge symmetry; however,
the converse is not necessarily true.  As charge symmetry
is a more restricted symmetry than isospin symmetry, it is
generally conserved in strong interactions to a greater 
degree than isospin symmetry.  Thus, while in many nuclear
reactions isospin symmetry is violated at the few percent
level, in most cases charge symmetry is obeyed to better
than one percent.  

For a system of $A$ particles, the charge symmetry operator
can be written as 
\begin{equation}
P_{cs} = \exp(i\pi T_2) = \Pi_{i=1}^A \exp[i\pi T_2(i)] ~~.
\end{equation}
The operation of charge symmetry maps up quarks to down, 
and protons to neutrons.  Specifically, under charge symmetry 
\begin{eqnarray}
e^{i\pi T_2} &:& u \rightarrow d, p \rightarrow n \nonumber \\
  &:& d \rightarrow -u, n \rightarrow -p 
\end{eqnarray}

At the quark level, charge symmetry
implies the invariance of a system under the interchange
of up and down quarks.  The proton and neutron each contain
three valence quarks, plus a ``sea'' of quark-antiquark pairs.  
Coulomb effects aside, the ``proton'' is converted to a ``neutron'' by
interchanging up and down quarks in the two nucleons.  
At the level of parton distributions, charge symmetry 
implies the relations
\begin{eqnarray}
u^p(x,Q^2) &=& d^n(x,Q^2) \nonumber \\ 
d^p(x,Q^2) &=& u^n(x,Q^2) 
\end{eqnarray} 
Charge symmetry is broken by electromagnetic interactions, 
but these should play a minor role at high energies.  We 
can get an estimate for the magnitude of charge symmetry 
violation [CSV] at the parton level; we would naively expect parton
CSV to be of the order of the up--down
current quark mass difference divided by some average mass 
expectation value of the strong Hamiltonian, or $(m_d - m_u)/<M>$, 
where $<M>$ has a value of roughly 0.5--1.0 GeV.  This would naturally
put CSV effects at a level of 1\% or smaller.  Note that we expect 
charge symmetry to be valid at this level, despite the fact that the 
current quark masses themselves, 
i.e.\ $m_u \sim 4$ MeV, $m_d \sim 7$ MeV, differ by 50\%!  However, 
our understanding is that dynamical chiral symmetry breaking and/or 
confinement masks this very large ``primordial'' 
violation of charge symmetry, and observable quantities are expected 
to respect charge symmetry to roughly one percent.

In nuclear physics, charge symmetry involves the interchange 
of protons and neutrons in a system.  At low energies, charge
symmetry appears to be generally valid at the level of 1\% or
better in nuclear systems, although there are some notable
exceptions to this rule of thumb \cite{Mil90}.  The proton
and neutron masses are equal to about 0.1\%; the binding
energies of tritium and $^3$He are equal to 1\%, after 
Coulomb corrections.  We can compare energy levels in
``mirror'' nuclei (nuclei related to one another by $P_{cs}$), 
and generally find agreement to better than 1\%, after correcting
for electromagnetic interactions.  

From our experience with charge symmetry in nuclear systems, 
and because of the order of magnitude estimates of CSV in parton
systems, charge symmetry has been 
universally assumed in quark/parton phenomenology.  With this
assumption, one reduces the number of independent quark distribution 
functions by a factor of two.  One simply defines all quark 
distribution functions in the neutron to be equal to the 
corresponding functions in the proton, while 
interchanging up and down quarks in the process.  
The assumption of charge symmetry is sufficiently ingrained in 
quark/parton phenomenology that its validity is a necessary condition 
for many relations between structure functions.   Thus, it is not 
apparent to many physicists that several sum rules or  
structure function equalities may be valid only to the extent that 
charge symmetry is exact.  

Recently, much attention has been focused on the apparent violation
of what is called SU(2) flavor symmetry in the nucleon\footnote{We 
adopt this terminology, which is widespread, despite the fact that 
there is no underlying SU(2) symmetry which is broken.  The term 
refers to the fact that proton sea quark distributions are not
equal, i.e.\ $\bar{d}^p$ and $\bar{u}^p$ are not equal at all Bjorken
$x$.}.  The 
measurements of the Gottfried sum rule \cite{Got67}
reported by the New Muon Collaboration (NMC) \cite{Ama91} 
sparked a great deal of interest in the sea-quark flavor
distributions of nucleons 
\cite{Pre91,Hen90,Lon94,Lon95,Lev91,Fra89,Kum91}. 
The ``natural'' explanation of the NMC results 
is that ``SU(2) flavor symmetry'' is
broken in the proton sea quark distributions 
(i.e. $\bar{u}^p(x) \not = \bar{d}^p(x)$).  This has been
widely cited, and several theoretical investigations have been 
carried out to investigate the possible origin of this flavor
symmetry violation.  At the time of the NMC measurements, another 
possible explanation for this data was that the Gottfried
sum rule was obeyed, and that the apparent experimental violation
resulted from significant contributions to the sum rule at
extremely small values of $x$ \cite{Mar90}.  

Since the NMC experiment suggested large flavor symmetry violating 
antiquark distributions, this led people towards experiments
which had the possibility 
of ``direct'' observation of flavor symmetry violation [FSV] 
in the proton sea.  Ellis and Stirling \cite{Ell91} pointed 
out that this information could be obtained by studying 
proton-induced Drell-Yan cross sections with 
both proton and neutron ({\it i.e.}, deuteron) targets.  
Two subsequent experiments have been carried out, by
the NA51 group at CERN \cite{Bal94}, and the E866 experiment 
at FNAL \cite{Haw97}.  Experimental results from these 
collaborations also seem to
show a large flavor symmetry violation [FSV] in the proton sea
quark distributions.  We will discuss these results in detail in Sect.\ 
4 of this review.   

However, the conclusion that these three experiments demonstrate 
large FSV effects, as well as the magnitude of the FSV effects 
extracted, relies on the implicit assumption of charge
symmetry.  It has been pointed out \cite{Ma92,Ma93} that
all three experimental results (NMC, NA51 and E866) could 
in principle be explained even if flavor symmetry were conserved, 
if we assume 
charge symmetry violation [CSV] in the nucleon sea.  As we will show, 
the CSV terms necessary to account for the results of these 
experiments would be surprisingly large -- much larger than 
theoretical models would predict.  If charge symmetry were violated to this
degree in parton distributions, it would be amazing that at low
energies it would be nearly an exact symmetry.  

However, the history of our understanding
of nucleon structure has involved a series of similar ``surprises''
from experiment: 
e.g., the significant contribution of antiquarks to nucleon structure
functions at small $x$; the large fraction of the nucleon's momentum  
carried by glue; the persistence of substantial spin effects at high 
energies despite perturbative QCD [pQCD] predictions that single-spin
asymmetries should vanish; the ``spin crisis,'' which suggests that a 
surprisingly small fraction of the proton's spin may be carried by valence 
quarks; and the behavior of quark distributions at very small $x$.  
So, despite the strong indirect evidence from low-energy physics, 
and straightforward pQCD arguments  
which suggest that charge symmetry should be valid to about the 
1\% level in parton
distributions, we urge the reader to keep an open mind on this 
question.  

In this paper, we will provide a comprehensive review of the following
question: how valid is the assumption of charge symmetry for parton 
distributions?  First, we will redefine the nucleon 
structure functions in terms of quark/parton
distributions, without assuming charge symmetry.  Next, we
will show how relations between structure functions become
modified when we allow CSV terms.  We then calculate the
CSV contributions to various observables.  We examine
the current experimental evidence for charge symmetry.    
As we will show, all experiments to date are consistent with
parton charge symmetry.  However, in some regions present experimental 
upper limits on parton charge symmetry violation are rather weak.  
On the other hand, new experimental neutrino deep inelastic 
scattering data, 
when taken together with high energy muon scattering, can provide
rather strong constraints on parton CSV, at least for a certain
range of Bjorken $x$.  

We will also present some simple model estimates of charge
symmetry violation in both valence quark and sea quark
distributions.  For the ``majority'' valence quark distributions,
({\it i.e.}, $u^p_{\rm v}(x) - d^n_{\rm v}(x)$), we predict very small 
CSV amplitudes, no larger than 1\%.  However, our model calculation of 
charge symmetry violation in the ``minority'' nucleon valence quark 
distributions ($d^p_{\rm v}(x) - u^n_{\rm v}(x)$) \cite{Sat92,Rod94} 
suggests surprisingly large CSV terms.  We discuss several 
experiments which could detect CSV in parton distributions, or
which could improve the current upper limits on quark CSV.    

The structure of our paper is as follows.  In Sect.\ 2 we 
review the general expressions between cross sections and
structure functions for deep inelastic scattering processes.  
We write down the most general form of the structure functions,
without assuming charge symmetry.  In Sect.\ 3, we give 
derivations, from simple models, of charge symmetry breaking
for both valence quarks and sea quarks.  We show the 
magnitude and sign of the expected CSV terms in these
models. We also review relations between structure functions which
hold if charge symmetry is valid.  

In Sect.\ 4  we review those experiments which currently
place the best upper limits on CSV in parton distributions.  
Because of the current interest in flavor symmetry violation 
in the proton sea, and because current ``tests'' of FSV in fact
are testing a combination of FSV and CSV, we review at length 
the recent Drell-Yan measurements which are presented as
evidence for FSV.  We review the constraints which recent
experiments place on CSV and FSV in antiquark parton 
distributions.  Preliminary results from the E866 Drell-Yan experiment 
suggests that they can measure the relative magnitude of $\bar{d}$ 
and $\bar{u}$ in the proton, over a fairly wide kinematic
region.  In this same general 
region, we also have data from the NMC measurement of $F_2$ 
structure functions in protons and neutrons, using high energy
muon beams.  In addition, we have the structure function $F_2^{W^\pm}$
measured by the CCFR group, from charge changing weak 
interactions induced by neutrinos and antineutrinos on iron.  
All three experiments obtain measurements at similar values of
$Q^2$ and $x$.    
 
In Sect.\ 5, we propose experiments
which could in principle reveal charge symmetry violation in
the valence quark distributions (these would also differentiate
between FSV and CSV effects).   
In Sect.\ 6, we review QCD sum rules.  We show how these are
modified if we include sea quark CSV contributions.  We review
the best known unpolarized sum rules, the Gottfried, 
Adler and Gross-Llewellyn Smith sum rule.  In Sect.\ 7, 
we show that
by defining two new sum rules, it would be possible to
measure separately CSV, and FSV, contributions to sea
quark distributions.  We call these the ``charge symmetry''
and ``flavor symmetry'' sum rules, respectively.  We also
review the status of existing experiments to determine current 
upper limits on sea quark CSV via the charge symmetry sum
rule.  In Sect.\ 8 we present our conclusions.  

\section{Relations Between High Energy Cross Sections and 
Parton Distributions}
\mb{.5cm}

\subsection{General form of high energy cross sections}

\begin{figure}
\centering{\hbox{ \hspace{0in} \psfig{figure=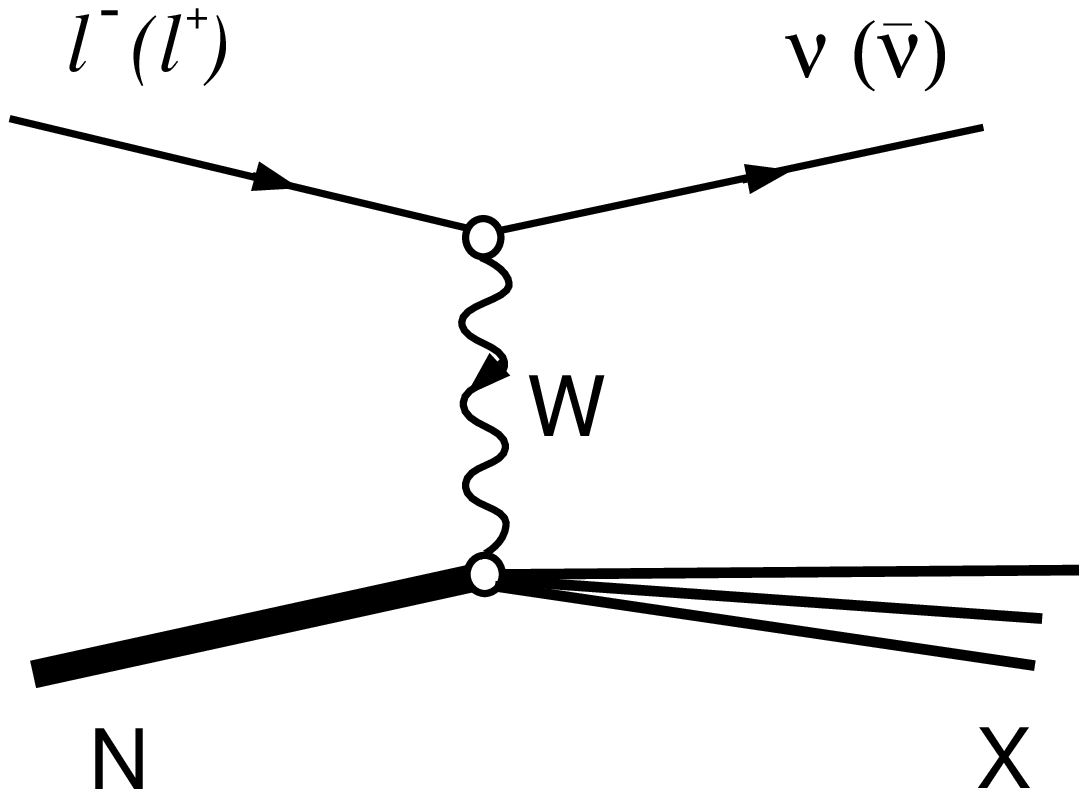,height=5.7cm}
\hspace{0.1in} \psfig{figure=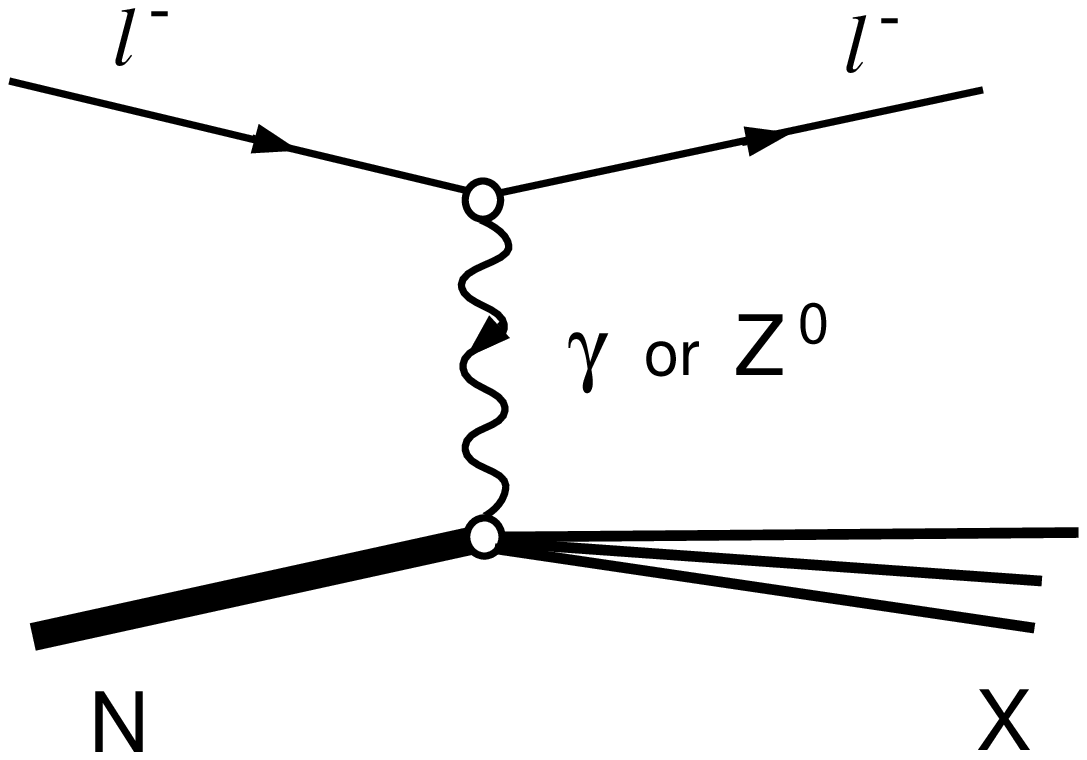,height=5.7cm}}}
\vspace{0.15truein}
\caption{Schematic picture of deep inelastic scattering of charged 
leptons from a nucleon. a) Charged-current weak interactions.  An 
intermediate $W$ is absorbed on the nucleon. b) Neutral-current 
electroweak interactions.}
\vspace{0.1truein}
\label{fig21}
\end{figure}

We can write the cross sections for deep inelastic scattering
in terms of a set of structure functions, which depend on
the relativistic kinematics of the reaction.  Through the
quark/parton model, these structure
functions can in turn be written in terms of quark/parton
distributions \cite{Lea96}.  For example, the most general form of
the cross section for charged 
current interactions initiated by charged leptons on nucleons
has the form 
\begin{eqnarray}
{d^2\sigma^{l^+ (l^-)}_{CC} \over dx\,dy} &=& {\pi s \over
  2} \left( {\alpha\over 2 \sin^2\theta_W M_W^2} \right)^2
  \left({M_W^2\over M_W^2 + Q^2}\right)^2
  \, \left[ xy^2 F_1^{W^\pm}(x, Q^2) \right.
  \nonumber \\ &+& \left. 
  \left( 1-y - {xym_N^2\over s}\right)\, F_2^{W^\pm}(x, Q^2) 
  \mp (y-y^2/2) xF_3^{W^\pm}(x, Q^2) \right] 
\label{siglCC}
\end{eqnarray}   
This process is shown schematically in Fig.\ \ref{fig21}a.  It
involves a charged virtual $W^\pm$ of momentum $q$ being interchanged 
between the lepton/neutrino vertex, and the hadronic vertex.  
The relativistic invariants in Eq.\ \ref{siglCC} 
are $Q^2 =-q^2$, the square of the four 
momentum transfer for the reaction, $x$ and $y$.  For four
momentum $k$ ($p$) for the initial state lepton (nucleon), 
we have the relations 
\begin{eqnarray} 
  x &=& {Q^2 \over 2p\cdot q}; \qquad y = {p\cdot q \over p\cdot k}  
  \nonumber \\
  s &=& (k+p)^2  
\label{kinem}  
\end{eqnarray}
In Eq.\ \ref{siglCC}, $M_W$ is the mass of the charged weak
vector boson, and $\theta_W$ is the Weinberg angle.  

Similarly, the cross section for charged 
current interactions initiated by neutrinos or antineutrinos 
on nucleons has the form 
\begin{eqnarray}
{d^2\sigma^{\nu (\bar\nu)}_{CC} \over dx\,dy} &=& \pi s 
   \left( {\alpha\over 2 \sin^2\theta_W M_W^2} \right)^2
  \left({M_W^2\over M_W^2 + Q^2}\right)^2
  \, \left[ xy^2 F_1^{W^\pm}(x, Q^2) \right.
  \nonumber \\ &+& \left. 
  \left( 1-y - {xym_N^2\over s}\right)\, F_2^{W^\pm}(x, Q^2) 
  \pm (y-y^2/2) xF_3^{W^\pm}(x, Q^2) \right] 
\label{signuCC}
\end{eqnarray}   
This process is obtained by interchanging the initial and final
state leptons in Fig.\ \ref{fig21}a.  

Neutral current (NC) reactions initiated by neutrinos or
antineutrinos have the form 
\begin{eqnarray}
{d^2\sigma^{\nu (\bar\nu)}_{NC} \over dx\,dy} &=& \pi s 
   \left( {\alpha\over 2 \sin^2\theta_W\cos^2\theta_W M_Z^2} 
   \right)^2
  \left({M_Z^2\over M_Z^2 + Q^2}\right)^2
  \, \left[ xy^2 F_1^Z(x, Q^2) \right.
  \nonumber \\ &+& \left. 
  \left( 1-y - {xym_N^2\over s}\right)\, F_2^Z(x, Q^2) 
  \pm (y-y^2/2) xF_3^Z(x, Q^2) \right] 
\label{signuNC}
\end{eqnarray}   

Finally, the cross section for scattering of a left (L) or
right (R) handed charged lepton in NC reactions has the
form 
\begin{eqnarray}
{d^2\sigma^{L,R}_{NC} \over dx\,dy} &=& {4\pi\alpha^2 s \over
   Q^4}\,\left( \left[ xy^2 F_1^\gamma(x, Q^2) + 
  (1-y)F_2^\gamma(x, Q^2) \right]  \right. \nonumber \\ 
  &-& {Q^2\over (Q^2 + M_Z^2)}
  {v_\ell \pm a_\ell \over 2\sin \theta_W \cos \theta_W} 
  \left[ xy^2 F_1^{\gamma Z}(x, Q^2) + 
  (1-y) F_2^{\gamma Z}(x, Q^2) \right. \nonumber \\ 
   &\pm&\left. (y-y^2/2) xF_3^{\gamma Z}(x, Q^2) \right] 
  + \left({Q^2\over (Q^2 + M_Z^2)}\right)^2 {v_\ell \pm a_\ell \over 
  2\sin \theta_W \cos \theta_W} \left[ xy^2 F_1^Z(x, Q^2) 
  \right. \nonumber \\ &+& \left.\left. (1-y) F_2^Z(x, Q^2)  \pm 
  (y-y^2/2) xF_3^Z(x, Q^2) \right] \right)
\label{sigLRNC}
\end{eqnarray}   
This process is shown schematically in Fig.\ \ref{fig21}b.  Either a
photon or $Z^0$ boson can be exchanged in this process.  

In Eq.\ \ref{sigLRNC}, we have 
\begin{eqnarray}
v_e &=& {-1 + 4 \sin^2\theta_W \over 4 \sin\theta_W\cos\theta_W}
  \nonumber \\ v_{\nu_e} &=& a_{\nu_e} = -a_e = {1 
  \over 4 \sin\theta_W\cos\theta_W} 
\end{eqnarray}
Eq.\ \ref{sigLRNC} describes the deep inelastic scattering
for an $L$ ($R$) handed charged lepton from a nucleon.  
For momentum transfers which are sufficiently small (relative
to $M_Z^2$), we can neglect the contribution from $Z$ bosons, 
in which case the scattering is a function only of the
two electromagnetic structure functions, $F_1^\gamma$ and 
$F_2^\gamma$, respectively.  
 
\subsection{Structure functions in terms of quark/parton
distributions}

\begin{figure}
\centering{\ \psfig{figure=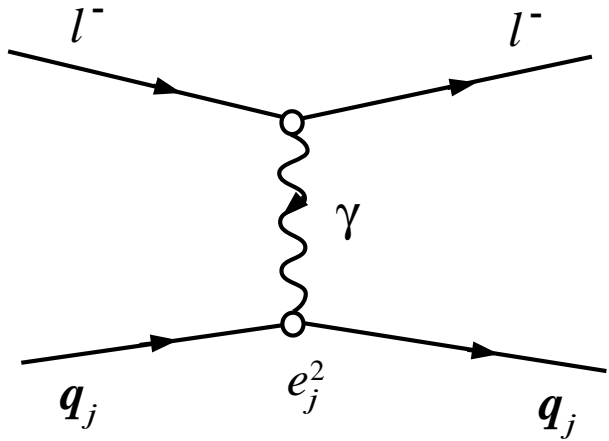,width=8cm}}
\vspace{0.15truein}
\caption{Coupling of a charged lepton to quarks through exchange
of a virtual photon.}
\vspace{0.1truein}
\label{fig25}
\end{figure}

The form of the proton structure functions, obtained from 
deep inelastic scattering of an electron or muon, can be written
in terms of interaction of the charged leptons and quarks 
with the virtual photon $\gamma$ \cite{Lea96}.  Here we assume 
we are at sufficiently low energies 
that we can neglect contributions from $Z^0$ in electroweak 
processes.  From Eq.\ \ref{sigLRNC} we see that the resulting 
cross section can be written in terms of two structure functions, 
$F_1^\gamma$ and $F_2^\gamma$.  Furthermore, we work in an energy
regime where both $Q^2$ and the energy transfer are very large, while $x$ 
remains finite, so that scaling is valid, i.e.\ the structure
functions (to first approximation) depend only on $x$ and not on 
$Q^2$.  The resulting structure function 
$F_1^{\gamma p}(x)$ 
can be written in terms of the parton distributions as  
\begin{eqnarray}
F_1^{\gamma p}(x,Q^2)  &\equiv& {1\over 2} \left( 
  {4\over 9}\,\left[ u^p(x) + \bar{u}^p(x) + 
  c^p(x) + \bar{c}^p(x) \right] \right. \nonumber \\
  &+& \left. {1\over 9}\,\left[ d^p(x) + 
  \bar{d}^p(x) + s^p(x) + \bar{s}^p(x)\right] \right) 
%
\label{F1gamm}
\end{eqnarray}
This process is shown schematically in Fig.\ \ref{fig25}.
In Eq.\ \ref{F1gamm}, we assume we can neglect any contribution 
from bottom or top quarks 
in the proton.  The virtual photon couples to the 
squared charge of the struck quarks.  To obtain the
corresponding $F_1$ structure function for the neutron, we
simply change the superscript $p \rightarrow n$
everywhere in Eq.\ \ref{F1gamm}.  

In Eq.\ \ref{F1gamm} (and in most subsequent equations), we 
have neglected the dependence of the parton distributions on
the scale at which they are evaluated.  As is well known 
\cite{Ste95,Lea96}, there is an uncertainty in the parton
distributions with respect to the scale $\mu^2$ at which
they are evaluated.  Once we calculate the parton distributions 
$q_k(x,\mu^2)$ and gluon distributions at some starting scale, we 
can evolve the parton distributions
to some higher $Q^2$ through the QCD evolution equations 
of Dokshitzer, Gribov, Lipatov, Altarelli and Parisi 
\cite{Gri72}.  For convenience, we will generally omit 
this scale in equations involving parton distributions.  

In the lowest order quark/parton model, the structure function
$F_2^\gamma$ is related to the structure function $F_1^\gamma$ 
by the Callan-Gross relation \cite{Cal69}
\begin{equation}
F_2^\gamma (x,Q^2) = 2x F_1^\gamma (x,Q^2) \nonumber
\end{equation}
As is well known, the Callan-Gross relation is valid if the
virtual photon which initiates this process is completely 
transverse.  The more general relation between the two
structure functions is 
\begin{equation}
F_2^\gamma(x,Q^2) = {1+R(x,Q^2)\over 1+4M^2x^2/Q^2}\,2x\,
  F_1^\gamma(x,Q^2)~~.
\label{Rdenn}
\end{equation}
In Eq.\ \ref{Rdenn}, $R = \sigma_L/\sigma_T$ is the ratio of
the cross section for longitudinally to transversely polarized 
photons.  An analogous relation will hold for the weak
structure functions $F_i^{W^\pm}$.  An empirical relation fit to the
world's available data on $R$ has been made by Whitlow {\it et al.} 
\cite{Whi90}.  The formula is 
\begin{eqnarray}
R(x, Q^2) &=& {b_1\theta\over \ln (Q^2/0.04)} + {b_2\over Q^2} + 
  {b_3\over Q^4 + 0.09} ~~, \nonumber \\ 
  \theta &=& 1 + {12 Q^2\over Q^2 + 1}\left({c^2\over c^2 + x^2}\right)  .
\label{Rfit}
\end{eqnarray}
The coefficients in Eq.\ \ref{Rfit} can be found in Ref.\ \cite{Whi90}. 
This fit covers the region accessible at that time, i.e.\ $x > 0.1$ and 
$Q^2 < 125$ GeV$^2$.  

\begin{figure}
\centering{\hbox{ \hspace{-.1in} \psfig{figure=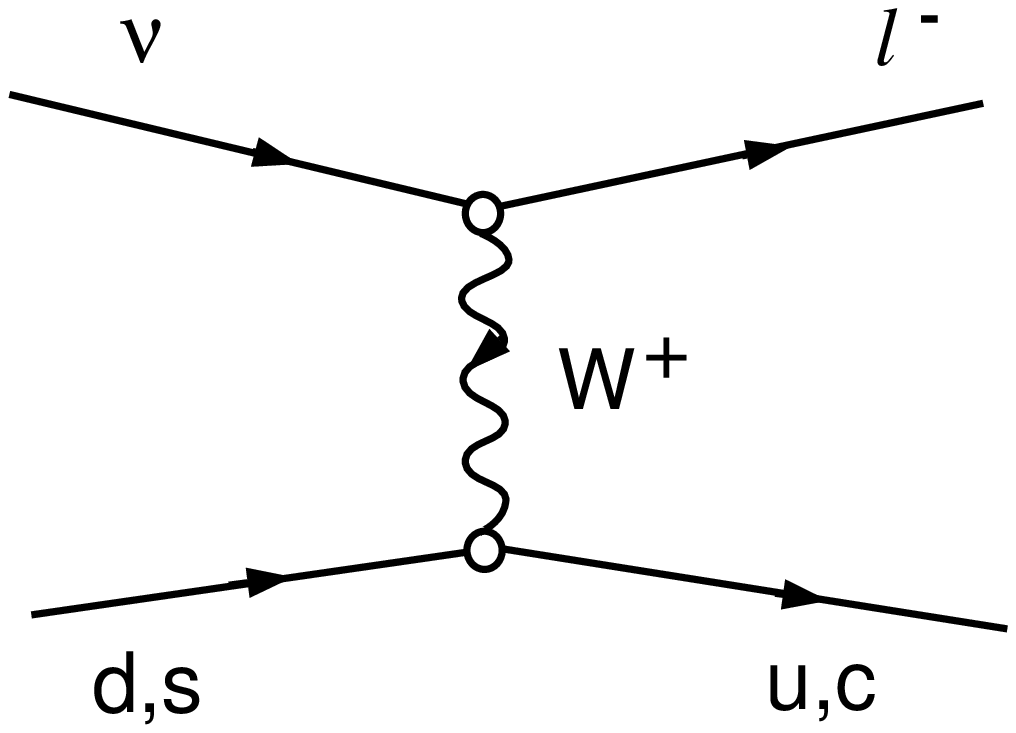,height=5.6cm}
\hspace{.2in} \psfig{figure=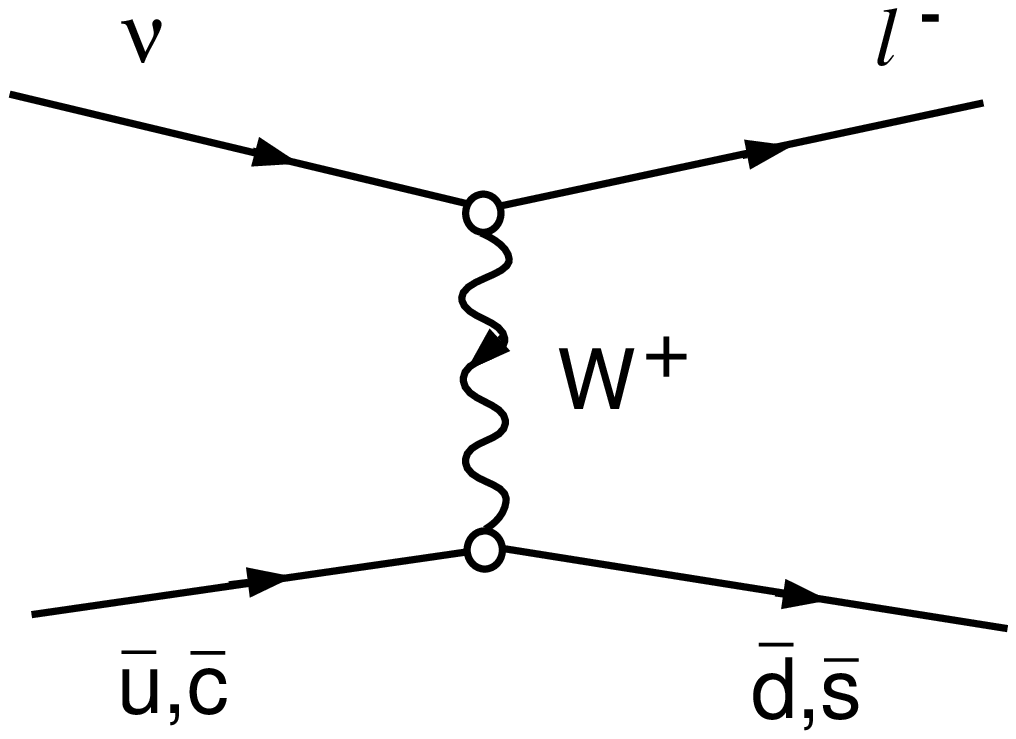,height=5.6cm}}}
\vspace{0.15truein}
\caption{a) Quark contributions to the charged-current reactions 
induced by neutrinos.  The virtual $W^+$ is absorbed by negatively 
charged quarks. b) Antiquark contributions to neutrino induced 
charged-current reactions.}
\vspace{0.1truein}
\label{fig26}
\end{figure}

Charged current neutrino scattering on hadrons is mediated by   
emission of the weak vector boson $W^\pm$ by the leptons 
and subsequent absorption
of the $W^\pm$ on the proton or neutron.  
Thus the structure function $F_1$ corresponding to charge-changing
interactions of neutrinos on protons can be written in terms of the
quark distribution functions as  
\begin{eqnarray}
F_1^{W^+ p}(x)  &\equiv& d^p(x)|V_{ud}|^2 + 
  d^p(\xi_c)|V_{cd}|^2 \theta (x_c-x) + \bar{u}^p(x) [ 
  |V_{ud}|^2  + |V_{us}|^2 ]  \nonumber \\ 
  &+& \bar{u}^p(\xi_b)|V_{ub}|^2 \theta(x_b-x) + s^p(x)|V_{us}|^2 
  + s^p(\xi_c)|V_{cs}|^2 \theta(x_c-x) \nonumber \\ 
  &+& \bar{c}^p(\xi_b)|V_{cb}|^2 \theta(x_b-x) + \bar{c}^p(x) [ 
  |V_{cd}|^2  + |V_{cs}|^2 ] ,  
\label{Wplus}
\end{eqnarray}
In Fig.\ \ref{fig26}a we show the coupling of the virtual $W^+$ to quarks; 
the coupling is to quarks with negative charge.  In Fig.\ \ref{fig26}b we 
show the coupling to antiquarks.  

Similarly, the structure function $F_1$ corresponding to charge-changing 
reactions for antineutrinos on protons can be written as  
\begin{eqnarray}
F_1^{W^- p}(x)  &\equiv& u^p(x)[ |V_{ud}|^2 + |V_{us}|^2 ]
  + u^p(\xi_b)|V_{ub}|^2 \theta(x_b-x) + \bar{d}^p(x)|V_{ud}|^2  
  \nonumber \\ &+& \bar{d}^p(\xi_c)|V_{cd}|^2 \theta (x_c-x)  
  + \bar{s}^p(x)|V_{us}|^2 
  + \bar{s}^p(\xi_c)|V_{cs}|^2 \theta(x_c-x) \nonumber \\ 
  &+& c^p(\xi_b)|V_{cb}|^2 \theta(x_b-x) + c^p(x) [ 
  |V_{cd}|^2  + |V_{cs}|^2 ] .  
\label{Wminus}
\end{eqnarray}
For antineutrinos the virtual $W^-$ is absorbed by positively charged
quarks and antiquarks.   
In Eqs.\ \ref{Wplus} and \ref{Wminus}, quantities like $V_{ud}$ are 
elements of the Cabibbo-Kobayashi-Maskawa (CKM) quark mixing
matrix \cite{Lea96}.  We have also introduced the so-called 
``slow rescaling'' formalism \cite{Geo76,critic} 
to account for threshold corrections in heavy
quark production.  For production of a heavy quark with
current quark mass $m_k$, we define the quantities
\begin{eqnarray}
  \xi_k(Q^2) &=& x\left( 1 + {m_k^2\over Q^2} \right) \nonumber \\
  x_k(Q^2) &=& {Q^2\over Q^2 + \Delta_k^2} \nonumber \\
  \Delta_k^2 &\equiv& (M_X^{min})^2 - m_N^2
\label{rescal}
\end{eqnarray}
In Eq.\ \ref{rescal}, the quantity $M_X^{min}$ is the minimum
mass of the final state for the light quark to heavy
quark transition.  
For the various quark flavors, we have $M_X^{min} \approx$ 
2.8 GeV, 3 GeV, and 6.2 GeV for transitions $d \rightarrow c$, 
$s \rightarrow c$, and $u \rightarrow b$, respectively (in this
review we neglect all contributions from top quarks).  
  
Note that with this rescaling model, the Callan-Gross relation
fails to apply in the region of heavy quark thresholds, even in
the event that the interactions are purely transverse.  The structure 
function $F_2^{W^+ p}$ can 
be obtained from $F_1^{W^+ p}$, Eq.\ \ref{Wplus}, by the replacement
\begin{equation}
q_i(x) \rightarrow 2x q_i(x) \quad\quad q(\xi_k) \rightarrow
2\xi_k q(\xi_k) ~~, 
\label{f1f2}
\end{equation}
and identical replacements for the antiquark distributions.  
With the same replacement, Eq.\ \ref{f1f2}, one can obtain the 
structure function $F_2^{W^-p}$ from $F_1^{W^- p}$.   
Similarly, the structure function ${1\over 2} F_3^{W^\pm p}$ can be 
obtained from $F_1^{W^\pm p}$ by the replacement 
\begin{equation}
\bar{q}_k(x) \rightarrow -\bar{q}_k(x) 
\label{f1f3}
\end{equation}

\subsection{High energy limiting form for weak structure functions}

At sufficiently high momentum transfers, e.g.\ well above charm 
threshold, we have $\xi_c \rightarrow x$.  
In the limit of very high momentum transfer, $Q^2 \rightarrow
\infty$, we have $\xi_k(Q^2) \rightarrow x$ and $x_k 
\rightarrow 1$ for all heavy quark flavors.  In this limit
we once again recover the Callan-Gross relation (if 
$\sigma_L/\sigma_T \rightarrow 0$).  
At very high momentum transfers and at very high
energy, if we neglect
correction terms of magnitude $|V_{ub}|^2 = |V_{td}|^2 \approx 
\sin(\theta_C)^6 \approx 1\times 10^{-4}$, then the structure
functions $F_i^{W^\pm}(x,Q^2)$ reduce to 
\begin{eqnarray}
 \lim_{Q^2 \rightarrow \infty} 
 F_1^{W^+ p}(x, Q^2) &\rightarrow& d^p(x) + \bar{u}^p(x) +s^p(x) + 
 \bar{c}^p(x)\nonumber \\
 F_1^{W^- p}(x, Q^2) &\rightarrow& u^p(x) + \bar{d}^p(x) + 
 \bar{s}^p(x) + c^p(x) \nonumber \\
 {1\over 2}F_3^{W^+ p}(x, Q^2) &\rightarrow& d^p(x) - \bar{u}^p(x) 
  + s^p(x) - \bar{c}^p(x)\nonumber \\
 {1\over 2}F_3^{W^- p}(x, Q^2) &\rightarrow& u^p(x) - \bar{d}^p(x) 
 - \bar{s}^p(x) + c^p(x) , 
 \label{FhighE}
\end{eqnarray}
with the corresponding structure functions for neutrons obtained
by replacing superscripts $p \rightarrow n$ everywhere in Eqs.\ 
\ref{FhighE}.  Because of their simplicity we will generally
use Eq.\ \ref{FhighE} in deriving relations between structure
functions, although we should revert to Eqs.\ \ref{Wplus}
and \ref{Wminus} when comparing with data.  This is particularly
relevant for experiments at relatively low $Q^2$, where 
threshold effects can be rather important.   

The assumption of charge symmetry for parton distributions is
that 
\begin{eqnarray}
d^n(x) &=& u^p(x) \nonumber \\
u^n(x) &=& d^p(x) \nonumber \\
s^n(x) &=& s^p(x) = s(x) \nonumber \\
c^n(x) &=& c^p(x) = c(x) .
\label{csudsc}  
\end{eqnarray}
We have identical relations for antiquark distributions.  
With this assumption, all neutron parton distributions can be
replaced by the corresponding distributions in the proton.  To
retain the charge symmetry violating parton distributions, we
introduce the CSV parton distributions for up and down quarks via
\begin{eqnarray}
d^n(x) &\equiv& u^p(x) - \delta u(x) \nonumber \\
u^n(x) &\equiv& d^p(x) - \delta d(x) .
\label{csvdef}
\end{eqnarray}
If the quantities $\delta u(x)$ and $\delta d(x)$ vanish, then
charge symmetry is exact.  We have analogous relations
for CSV in antiquark distributions.  We assume that the strange 
quark (and antiquark) distributions are the same in both the proton
and neutron, as is given in Eq.\ \ref{csudsc}.  We make the same
assumption for charm quarks.  There is no theoretical or 
experimental reason to expect strange and charm distributions to 
vary from proton to neutron.   

It is useful to divide parton distributions into valence quark and 
sea quark parts.  The
valence up quark distribution in the proton is defined by 
$u^p_{\rm v}(x) \equiv u^p(x) - \bar{u}^p(x)$.  The valence
quark distributions obey the following quark normalization 
conditions  
\begin{eqnarray}
\int_0^1\,dx\,u^p_{\rm v}(x)  &=& \int_0^1\,dx\, \left( 
  u^p(x) - \bar{u}^p(x) \right) = \nonumber \\ 
 \int_0^1\,dx\,d^n_{\rm v}(x)  &=& \int_0^1\,dx\, \left( 
  d^n(x) - \bar{d}^n(x) \right) = 2; \nonumber \\ 
  \int_0^1\,dx\,d^p_{\rm v}(x) &=& \int_0^1\,dx\, \left( 
  d^p(x) - \bar{d}^p(x) \right) = \nonumber \\ 
  \int_0^1\,dx\,u^n_{\rm v}(x) &=& \int_0^1\,dx\, \left( 
  u^n(x) - \bar{u}^n(x) \right) = 1;  \nonumber \\
  \int_0^1\,dx\, \left( s(x) - \bar{s}(x) \right) &=&    
  \int_0^1\,dx\, \left( c(x) - \bar{c}(x) \right) = 0.   
\label{GSRnrm}
\end{eqnarray}
The CSV quantities defined in Eq.\ \ref{csvdef} 
can have both valence and sea pieces.  The 
valence quark charge symmetry violating distributions are 
defined as
\begin{eqnarray}
\delta u_{\rm v}(x) &=& \delta u(x) - \delta\bar{u}(x)
  \nonumber \\ 
\delta d_{\rm v}(x) &=& \delta d(x) - \delta\bar{d}(x)
\label{deluv}
\end{eqnarray}
From these definitions of valence quark CSV, it is straightforward to 
show that the first moment of the valence quark CSV distributions
(i.e., the integral over $x$) must vanish.  
We see that 
\begin{eqnarray}
\int_0^1 dx \delta u_{\rm v}(x)  &=& \int_0^1 dx \left(u^p(x)- d^n(x) - 
  \bar{u}^p(x) + \bar{d}^n(x) \right) \nonumber \\ 
  &=&  \int_0^1 dx \left( u^p_{\rm v}(x) - d^n_{\rm v}(x)\right) = 0
\label{deluint}
\end{eqnarray}
In Eq.\ \ref{deluint}, the integral over the valence quark
distributions is fixed by the normalization condition to the 
number of valence up quarks in the proton (down quarks
in the neutron).  Since both of these are equal to 2, the 
integral of $\delta u_{\rm v}(x)$ must give zero.   

From Eq.\ \ref{GSRnrm} we see that the first moment of the
heavy quark and antiquark distributions are identical.  Until 
recently, it was customary to assume that the strange and charmed 
quark and antiquark distributions were equal for all values
of $x$.  That is, one assumed that 
\begin{equation}
 s(x) = \bar{s}(x) \equiv s(x) ~~,  
\label{seadef}
\end{equation}
with an identical relation for the charmed quark distributions. 
However, recently there has been both theoretical and
experimental interest in whether the strange and antistrange
distributions are in fact equal.  We will review how strange
quark distributions are extracted, and the experimental
situation regarding strange and antistrange quark 
distributions, in Sect.\ 2.6.    

\subsection{CSV Contributions to Structure Functions}

In the high energy limit, well above heavy quark thresholds, the 
structure functions for charged current weak interactions on neutrons 
take the form 
\begin{eqnarray}
 \lim_{Q^2 \rightarrow \infty} 
 F_1^{W^+ n}(x, Q^2) &=& d^n(x) + \bar{u}^n(x) +s(x) +
 \bar{c}(x) \nonumber \\
 F_1^{W^- n}(x, Q^2) &=& u^n(x) + \bar{d}^n(x) + 
 \bar{s}(x) + c(x) \nonumber \\
 {1\over 2}F_3^{W^+ n}(x, Q^2) &=& d^n(x) - \bar{u}^n(x) + s(x) 
 - \bar{c}(x) \nonumber \\
 {1\over 2}F_3^{W^- n}(x, Q^2) &=& u^n(x) - \bar{d}^n(x) 
 - \bar{s}(x) + c(x) . 
 \label{FnhighE}
\end{eqnarray}
Introducing the CSV parton distributions from Eq.\ \ref{csvdef}, 
Eq.\ \ref{FnhighE} becomes 
\begin{eqnarray}
 \lim_{Q^2 \rightarrow \infty} 
 F_1^{W^+ n}(x, Q^2) &=& u^p(x) + \bar{d}^p(x) +s(x) + 
 \bar{c}(x)- \delta u(x) - \delta\bar{d}(x) \nonumber \\
 F_1^{W^- n}(x, Q^2) &=& d^p(x) + \bar{u}^p(x) + 
 \bar{s}(x) + c(x) - \delta d(x)  - \delta\bar{u}(x) \nonumber \\
 {1\over 2}F_3^{W^+ n}(x, Q^2) &=& u^p(x) - \bar{d}^p(x)  + s(x)
 - \bar{c}(x) - \delta u(x)  + \delta\bar{d}(x) \nonumber \\
 {1\over 2}F_3^{W^- n}(x, Q^2) &=& d^p(x) - \bar{u}^p(x) 
 - \bar{s}(x) + c(x) - \delta d(x)  + \delta\bar{u}(x) . 
 \label{FnCSV}
\end{eqnarray}

For completeness, we include the electromagnetic structure
function for neutron targets.  Eq.\ \ref{F1gamm} becomes 
\begin{eqnarray}
F_1^{\gamma n}(x,Q^2)  &=& {1\over 2} \left( 
  {4\over 9}\,\left[ u^n(x) + \bar{u}^n(x) + 
  c(x) + \bar{c}(x) \right] + {1\over 9}\,\left[ d^n(x) + 
  \bar{d}^n(x) + s(x) + \bar{s}(x)\right] \right) 
  \nonumber \\
 &=& {1\over 2} \left( 
  {4\over 9}\,\left[ d^p(x) + \bar{d}^p(x) + c(x)+ \bar{c}(x) \right] +   
  {1\over 9}\,\left[ u^p(x) + \bar{u}^p(x) + s(x)+ \bar{s}(x) \right] 
  \right. \nonumber \\ &-&  \left. {4\over 9}\,\left[ \delta d(x) + 
   \delta \bar{d}(x) \right]- {1\over 9}\,\left[ \delta u(x) + 
   \delta \bar{u}(x) \right]  \right)  .
\label{F1gammn}
\end{eqnarray}

Many tests of charge symmetry will involve deep inelastic
scattering on isoscalar targets, which we label as $N_0$.  Such
reactions involve equal contributions from protons and 
neutrons.  Under the assumptions we have listed previously, the
weak and electromagnetic structure functions on isoscalar
targets can be written (in terms of structure functions 
per nucleon) 
\begin{eqnarray}
F_1^{W^+ N_0}(x, Q^2) &=& {1\over 2}\left(u^p(x) + d^p(x) + 
 \bar{u}^p(x) + \bar{d}^p(x) +2s(x) +2\bar{c}(x) - \delta u(x) - 
 \delta\bar{d}(x) \right)  \nonumber \\
 F_1^{W^- N_0}(x, Q^2) &=&  {1\over 2}\left(u^p(x) + d^p(x) + 
 \bar{u}^p(x) + \bar{d}^p(x) +2\bar{s}(x) +2c(x)  - \delta d(x)  
 - \delta\bar{u}(x) \right) \nonumber \\
 {1\over 2}F_3^{W^+ N_0}(x, Q^2) &=& {1\over 2}\left(u^p(x) + d^p(x) 
 - \bar{u}^p(x) - \bar{d}^p(x) +2s(x) -2\bar{c}(x) - \delta u(x)  
 + \delta\bar{d}(x) \right) \nonumber \\
 {1\over 2}F_3^{W^- N_0}(x, Q^2) &=& {1\over 2}\left(u^p(x) + d^p(x) 
 - \bar{u}^p(x) - \bar{d}^p(x) -2\bar{s}(x) +2c(x) - \delta d(x)  
 + \delta\bar{u}(x) \right) ~~,\nonumber \\ 
  F_1^{\gamma N_0}(x,Q^2)  &=& {1\over 4} \left( 
  {5\over 9}\,\left[ u^p(x) + d^p(x) + \bar{u}^p(x) +
   \bar{d}^p(x) \right] + {2\over 9}\,\left[s(x)+ \bar{s}(x)\right]  
   \right. \nonumber \\ &+& \left. {8\over 9}\,\left[c(x)+ 
  \bar{c}(x)\right] - {4\over 9}\,\left[ \delta d(x) + 
   \delta \bar{d}(x) \right]- {1\over 9}\,\left[ \delta u(x) + 
   \delta \bar{u}(x) \right]  \right) 
\label{F1Nzero}
\end{eqnarray}

\subsection{Isolating CSV Effects in Structure Functions}

In order to isolate and measure CSV effects, we need to find relations 
between various structure functions which depend on the validity of
charge symmetry, and which can be tested.  There are two
such relations: the relation between the $F_1$ (or $F_2$) charge-changing
electroweak structure functions from neutrino and antineutrino 
reactions, 
and the relation between the $F_2$ structure functions obtained
from neutrino deep inelastic scattering, and the $F_2$ structure
function from deep inelastic scattering induced by
charged leptons (muons or electrons).  

We first discuss the $F_1$ structure functions for charge-changing
weak interactions.  For deep inelastic scattering on an isoscalar 
target, we can derive the following identity from Eq.\ \ref{F1Nzero}: 
\begin{eqnarray}
\lim_{Q^2 \rightarrow \infty}\left( F_1^{W^+ N_0}(x,Q^2) - 
 F_1^{W^- N_0}(x,Q^2) \right) &=& 
  {1\over 2}\left(\delta d_{\rm v}(x) - \delta u_{\rm v}(x) \right) 
  \nonumber \\ &+& 
  s(x) - \bar{s}(x) - c(x) + \bar{c}(x)  \nonumber \\ 
\,&{CS\atop =}&\, s(x) - \bar{s}(x) - c(x) + \bar{c}(x)
\label{F1rel}
\end{eqnarray}
For Eq.\ \ref{F1rel} to be valid, we must be at sufficiently high
values of $Q^2$ that we are well above both charm and bottom 
thresholds.  Furthermore, we neglect terms of order 
$|V_{ub}|^2 = |V_{td}|^2 \approx 1\times 10^{-4}$.  
Eq.\ \ref{F1rel} should be true at all values of $x$.  The final line 
of this equation holds in the limit that charge symmetry is exact.  
To avoid confusion in this review, we
have introduced the notation ${CS\atop =}$.  This means
that an equation is true provided charge symmetry is
exact.  In this way we hope one can 
distinguish between relations which are generally true, 
and those which require the (generally implicit) assumption
of charge symmetry.  

At sufficiently high energies (well above heavy quark production 
thresholds) threshold effects should become negligible, and then 
Eq.\ \ref{F1rel} should be valid.  If these structure functions are not 
equal at all values
of $x$, this implies either charge symmetry violation in
parton distributions, or inequality of the strange quark and
antiquark distributions (the charm quark contributions should be 
quite small, and we know of no theoretical reason why charm and
anticharm distributions should be unequal).  In Sect.\ 5.1 we present
theoretical estimates of valence quark CSV and strange/antistrange
quark contributions to this relation.    

From Eqs.\ \ref{FhighE} and \ref{FnCSV} we can also derive relations 
between the $F_1$ structure functions for neutrinos on protons, and
antineutrinos on neutrons, 
\begin{eqnarray}
F_1^{W^+ p} (x,Q^2) - F_1^{W^- n} (x,Q^2)  &=& \delta d(x) + 
  \delta \bar{u}(x) + s(x) -\bar{s}(x) -c(x) 
  + \bar{c}(x) \nonumber \\
  F_1^{W^+ n} (x,Q^2) - F_1^{W^- p} (x,Q^2) &=& -\delta u(x) - 
  \delta \bar{d}(x) + s(x) -\bar{s}(x) -c(x) 
  + \bar{c}(x).  
\label{F1csv}
\end{eqnarray}
Eqs.\ \ref{F1csv} are valid under the same conditions as Eq.\ \ref{F1rel}, 
namely that we are well above heavy quark thresholds, and that we neglect
Kobayashi-Maskawa matrix elements of order $10^{-4}$.  If parton charge
symmetry were exact, and strange quark and antiquark parton distributions 
are identical at all $x$, then we would expect  
\begin{eqnarray}
F_1^{W^+ p} (x,Q^2) \,&=&\, F_1^{W^- n} (x,Q^2) \nonumber \\
F_1^{W^- p} (x,Q^2) \,&=&\, F_1^{W^+ n} (x,Q^2) 
\label{F1pn}
\end{eqnarray}

At sufficiently high energies, if charge symmetry is valid
for valence quark distributions, and if the strange quark and antiquark 
distributions are equal 
at all $x$, then the $F_1^{W^\pm}$ structure functions
are identical for isoscalar nuclear targets.  Identical relations
hold for the $F_2$ structure functions, when we include the 
longitudinal/transverse ratio $R$ of Eq.\ \ref{Rfit}. 
These equations have been used to extract the structure functions
$F_3$ in electroweak reactions.  Using Eqs.\ \ref{F1pn} and
\ref{signuCC}, we can derive  
\begin{eqnarray}
{3\pi \over 2 G^2M_NE} \left( d\sigma^{\nu p}/dx - 
  d\sigma^{\bar\nu n}/dx \right) &=& 
  F_2^{W^+ p}(x,Q^2) - F_2^{W^- n}(x,Q^2) \nonumber \\ 
  &+& {1\over 2}\left( xF_3^{W^+ p}(x,Q^2) + xF_3^{W^- n}(x,Q^2) 
  \right) \nonumber \\ &{CS\atop =}&\  
  {1\over 2}\left( xF_3^{W^+ p}(x,Q^2) + xF_3^{W^- n}(x,Q^2) 
  \right) \nonumber \\  {3\pi \over 2 G^2M_NE} \left( 
  d\sigma^{\nu n}/dx - d\sigma^{\bar\nu p}/dx \right) &=& 
  F_2^{W^+ n}(x,Q^2) - F_2^{W^- p}(x,Q^2) \nonumber \\  
  &+& {1\over 2}\left( xF_3^{W^+ n}(x,Q^2) + xF_3^{W^- p}(x,Q^2) 
  \right) \nonumber \\  &{CS\atop =}&\   
  {1\over 2}\left( xF_3^{W^+ n}(x,Q^2) + xF_3^{W^- p}(x,Q^2) 
  \right) \nonumber \\  {3\pi \over 2 G^2M_NE} \left( 
  d\sigma^{\nu N_0}/dx - d\sigma^{\bar\nu N_0}/dx \right) &=& 
  F_2^{W^+ N_0}(x,Q^2) - F_2^{W^- N_0}(x,Q^2) \nonumber \\  
  &+& {1\over 2}\left( xF_3^{W^+ N_0}(x,Q^2) + xF_3^{W^- N_0}(x,Q^2) 
  \right) \nonumber \\ &{CS\atop =}&\   
  {1\over 2}\left( xF_3^{W^+ N_0}(x,Q^2) + xF_3^{W^- N_0}(x,Q^2) 
  \right) , 
\label{F3rel}
\end{eqnarray}
where in Eq.\ \ref{F3rel} we define 
\begin{eqnarray}
d\sigma^{\nu p}(x,Q^2)/dx &=& \int_0^1 dy\,
d^2 \sigma^{\nu p}(x,y)/dx\,dy   \nonumber \\
G^2 &=& { (\pi \alpha)^2\over 2 M_W^4 \sin^4\theta_W } 
\end{eqnarray}

From Eqs.\ \ref{F1csv} and \ref{F3rel}, we see that if charge 
symmetry is exact, and if the strange quark and antiquark  
distributions are equal at all $x$, then by taking the difference between 
cross sections for the appropriate charged current cross sections 
for neutrinos and antineutrinos, the relevant $F_2$ structure functions
will cancel, leaving just the $F_3$ structure functions.  This
follows from Eq.\ \ref{F1pn}.  In particular, the difference between 
the charged current cross section from neutrino scattering on an isoscalar 
target, and the cross section from antineutrinos on that target, is just 
equal to the sum of the structure functions $xF_3$ for neutrinos and
antineutrinos on the isoscalar target.  

If we don't require charge symmetry and equality of strange and 
antistrange quark distributions, then Eq.\ \ref{F3rel} becomes 
\begin{eqnarray}
{3\pi \over 2 G^2M_NE} \left( d\sigma^{\nu p}/dx - 
  d\sigma^{\bar\nu n}/dx \right) &=& 
  2x\left[ \delta d(x) + \delta \bar{u}(x) + s(x) - \bar{s}(x) 
  \right] \nonumber \\ &+& 
  {1\over 2}\left( xF_3^{W^+ p}(x,Q^2) + xF_3^{W^- n}(x,Q^2) 
  \right) \nonumber \\  {3\pi \over 2 G^2M_NE} \left( 
  d\sigma^{\nu n}/dx - d\sigma^{\bar\nu p}/dx \right) &=& 
  2x\left[ -\delta u(x) - \delta \bar{d}(x) + s(x) - \bar{s}(x) 
  \right] \nonumber \\ &+& 
  {1\over 2}\left( xF_3^{W^+ n}(x,Q^2) + xF_3^{W^- p}(x,Q^2) 
  \right) \nonumber \\  {3\pi \over 2 G^2M_NE} \left( 
  d\sigma^{\nu N_0}/dx - d\sigma^{\bar\nu N_0}/dx \right) &=& 
  x\left[ \delta d_{\rm v}(x) - \delta u_{\rm v}(x)+ 2\left(s(x) - 
  \bar{s}(x)\right) \right] \nonumber \\ &+& {1\over 2}\left( 
  xF_3^{W^+ N_0}(x,Q^2) + xF_3^{W^- N_0}(x,Q^2) \right) , 
\label{F3rcsv}
\end{eqnarray}
There are additional terms arising from charm
quark distributions; they are given in Eq.\ \ref{F1csv}, but 
we have not included them in Eq.\ \ref{F3rcsv}.   

From Eq.\ \ref{F3rcsv} we see that {\em the difference between 
neutrino and antineutrino cross sections on an isoscalar target
contains not only the $F_3$ structure functions, but two  
residual terms -- one of which depends on the quark CSV amplitudes, 
and the other depending on the difference between strange and
antistrange quark distributions.}   These additional terms 
then make a contribution at various $x$ values to the sum of
the $F_3$ structure functions on isoscalar targets.  From Eq.\ 
\ref{F1Nzero}, without these additional terms the difference
between charge-changing reactions induced by neutrinos and
antineutrinos gives the sum of up plus down valence quark 
distributions in the nucleon.  

The other relation between structure functions which allows an
experimental test of charge symmetry is the so-called ``charge ratio''
or the ``5/18$^{th}$ rule.''   Neglecting for the moment the
longitudinal/transverse ratio $R$ (which will cancel if we form the
ratio of the two structure functions), we have from Eq.\ \ref{F1Nzero}
\begin{eqnarray}
\overline{F_2}^{W\, N_0}(x,Q^2) &\equiv&  {F_2^{W^+\, N_0}(x,Q^2) + 
  F_2^{W^-\, N_0}(x,Q^2)\over 2} = x\left[ \overline{Q}(x) \right. 
  \nonumber \\  &-& \left.
  {1\over 2} \left(\delta u(x) + \delta \bar{u}(x) + \delta d(x) + 
  \delta \bar{d}(x)\right) \right] , \nonumber \\ 
  F_2^{\gamma \,N_0}(x, Q^2) &=& x\left[ {5\over 18}\overline{Q}(x) 
  - {1\over 6}\left(s(x) + \bar{s}(x) - c(x) - \bar{c}(x)\right) 
  \right. \nonumber \\ &-& \left. 
  {2\over 9}\left( \delta d(x) + \delta \bar{d}(x) \right) 
  - {1\over 18}\left( \delta u(x) + \delta \bar{u}(x) \right)\right]~~, 
  \nonumber \\ 
  \overline{Q}(x) &\equiv& \sum_{j=u,d,s,c} q_j(x) + \bar{q}_j(x)
\label{chgrat}
\end{eqnarray}
From Eq.\ \ref{chgrat}, if we take the ratio of the two $F_2$ structure
functions we obtain  
\begin{eqnarray} 
{F_2^{\gamma\, N_0}(x,Q^2) \over \overline{F_2}^{W\, N_0}(x,Q^2) }
  &=&  {5\over 18}\left[ 1- {3 \left( s(x) + \bar{s}(x)\right)\over 
  5 \overline{Q}(x)} - { 4\delta d(x) + 4\delta \bar{d}(x) + \delta u(x) 
  + \delta \bar{u}(x)\over 5 \overline{Q}(x)} \right] 
  \nonumber \\ &/& \left[ 1 - 
   {\delta d(x) + \delta \bar{d}(x) + \delta u(x) 
  + \delta \bar{u}(x)\over 2 \overline{Q}(x)}\right] ~~, \nonumber \\ 
  &\approx& {5\over 18}\left[ 1- {3 \left( s(x) + 
  \bar{s}(x)\right)\over 5 \overline{Q}(x)} + {3 \left( \delta u(x) + 
  \delta \bar{u}(x) - \delta d(x) -\delta \bar{d}(x)\right) \over 10 
  \overline{Q}(x)} \right]  \nonumber \\ 
  \,&{CS\atop =}&\, {5\over 18}\left[ 1- {3 \left( s(x) + 
  \bar{s}(x)\right)\over 5 \overline{Q}(x)} \right]  \nonumber \\ 
\label{Rcdef}
\end{eqnarray}
In Eq.\ \ref{Rcdef} we have dropped the small contribution from charm
quarks.  In the second of Eqs.\ \ref{Rcdef} we have expanded to 
lowest order in the (small) CSV terms.  The ratio $5/18$ in 
Eq.\ \ref{Rcdef}, the so-called ``charge ratio'' for these structure 
functions, occurs because the virtual photon couples to the
squared charge of the quarks, while the charged-current reactions
induced by neutrinos couple to the weak isospin mediated by $W$ 
exchange.  

In the charge symmetric limit, we can use the $F_2$ structure functions
from either neutrino or antineutrino induced reactions, since 
\begin{eqnarray*} 
 F_2^{W^+\, N_0}(x,Q^2) &{CS\atop =}& F_2^{W^-\, N_0}(x,Q^2) {CS\atop =}
\overline{F_2}^{W\, N_0}(x,Q^2).
\end{eqnarray*}
If we use the neutrino or
antineutrino $F_2$ structure functions instead of their average in
Eq.\ \ref{Rcdef}, the only thing which changes in the ratio is the 
weighting of the various CSV terms, plus an additional contribution 
if the strange and antistrange distributions are not identical.    
  
In order to get theoretical estimates for the structure function
relations, we need to know the quark CSV contributions, and we
also need to know the strange quark and antiquark distributions
in the nucleon.  The strange quark and antiquark distributions can
be obtained experimentally by measuring production of opposite
sign dimuons in reactions induced by neutrinos or antineutrinos.  
We review this in the following section.  
We obtain estimates for the charge-symmetry violating parton distributions 
using the model for quark CSV which we derive in Sect.\ 3.  
In Sect.\ 4, we
will estimate the contribution of the CSV terms to existing tests of 
charge symmetry at high energies.  In Sect.\ 6, we will discuss
possible CSV sea quark contributions to the various sum rules, in 
particular the Adler and Gross-Llewellyn
Smith sum rules.    

\subsection{Extraction of Strange Quark Distributions}

In Sect.\ 2.3 we introduced the assumption that
strange quark and antiquark distributions were identical at
all $x$, and were identical for protons and neutrons (a
similar assumption is made for charmed quark distributions).  In
Sect.\ 2.5 we saw that structure function tests of charge symmetry
also contained contributions from strange quark and antiquark
distributions.  Thus, to get accurate tests of parton charge
symmetry, we must have reliable measurements of strange quark
distributions.  

The strange quark distribution can be assessed in two ways.  
First, it can be obtained ``indirectly'' by comparing 
DIS reactions induced by charged leptons (muons or electrons) with
charge-changing currents
from neutrinos.  From Eq.\ \ref{chgrat}, we see that for an 
isoscalar target (assuming the Callan-Gross
relation and neglecting charm quark contributions), we have 
\begin{eqnarray}
\overline{F_2}^{W\, N_0}(x,Q^2) - {18\over 5}F_2^{\gamma N_0}(x,Q^2) 
  &=& {3x\over 10}\left[\delta u(x) +\delta\bar{u}(x) - \delta d(x) - 
  \delta\bar{d}(x) \right] \nonumber \\ &+& 
  {3x\over 5}\left[ s(x) + \bar{s}(x) \right]
\label{F2comp}
\end{eqnarray}
Eq.\ \ref{F2comp} is ``indirect'', in that we must compare
experiments with muon and neutrino beams, performed under different 
conditions and with different normalizations.  Furthermore, we
need to know the CSV contributions in order to extract the strange
quark distribution.  

The ``direct'' way of extracting strange quark distributions
is by measuring the yield of opposite sign dimuons produced in 
nuclear reactions induced by neutrinos.  In leading order, 
the incoming neutrino has a hard scattering with an $s$ or
$d$ quark, producing a charm quark which fragments into a
charmed hadron.  The subsequent semileptonic decay of the
charmed hadron produces an opposite sign muon, through
the process 
\begin{eqnarray*}
\nu_\mu + N \rightarrow \mu^- &+& c + X \nonumber \\
 &\rightarrow& \mu^+ + \nu_\mu \nonumber 
\end{eqnarray*}
The antineutrino process produces charm antiquarks from 
$\bar{d}$ and $\bar{s}$ sea quarks.  As the first muon
produced tends to come from the original scattering and 
to have larger transverse momentum with respect to the 
direction of the hadron shower, experimenters can tell whether 
the muon pair arose from a neutrino or antineutrino collision.   
In this section we
summarize the latest direct measurements and their results.  

\begin{figure}
\centering{\ \psfig{figure=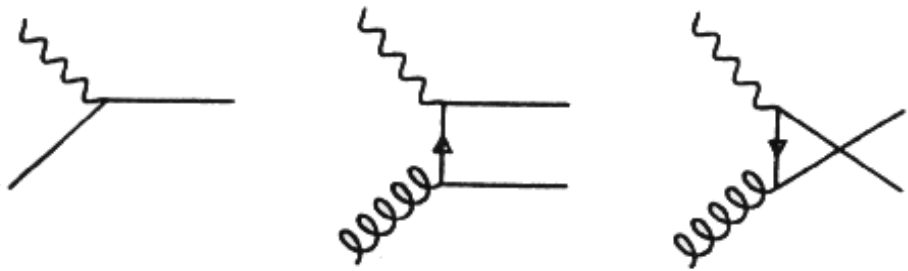,width=13cm}}
\vspace{0.15truein}
\caption{Contribution to production of an opposite sign dimuon pair
via intermediate charm quark production. Leading-order quark initiated
diagram, plus two NLO gluon-initiated diagrams.}
\vspace{0.1truein}
\label{fig27}
\end{figure}

In Fig.\ \ref{fig27} we show the leading order [LO] mechanism
for production of $\mu^+\mu^-$ pairs in neutrino-induced reactions.  
A virtual $W^+$ is absorbed on an $s$ or $d$ quark, producing a
charm quark, which then decays semi-leptonically.  Using the slow rescaling 
model \cite{Geo76} and assuming charge symmetry, the leading
order cross section for production of opposite
sign dimuons, for neutrino reactions on an isoscalar
target, has the form 
\begin{eqnarray}
{d^2\sigma (\nu N_0 \rightarrow \mu^+\mu^- X )\over d\xi_c dy}
  &=& {G^2 ME_\nu \over\pi}\left[ \left( \xi_c u(\xi_c,\mu^2) + 
\xi_c d(\xi_c,\mu^2)
  \right)|V_{cd}|^2 + 2\xi_c s(\xi_c,\mu^2)|V_{cs}|^2 \right] 
  \nonumber \\ &\times & \left[ 1 - {m_c^2\over 2 M E_\nu \xi_c} 
  \right] D(z) B_c
\label{smeas}
\end{eqnarray}
In Eq.\ \ref{smeas}, $D(z)$ is the fragmentation function 
for a charmed quark into a charmed hadron, and $B_c$ is the 
branching ratio for semileptonic decay of a charmed hadron. 
The quantity $\xi_c$ is given by the slow rescaling formula, 
Eq.\ \ref{rescal}, and the parton distributions depend through
QCD on the scale $\mu^2$.  
The corresponding cross sections for antineutrino interactions
are obtained by replacing $q_k(\xi_c,\mu^2)$ by 
$\bar{q}_k(\xi_c,\mu^2)$ for each quark flavor, in Eq.\ \ref{smeas}.   

Because the CKM matrix element $V_{cs}$ is substantially
larger than $V_{cd}$, i.e.\ $|V_{cs}|^2 \approx 0.95$ while 
$|V_{cd}|^2 \approx 0.05$, opposite sign dimuon production 
from neutrinos is most sensitive to the 
strange quark distribution in the nucleon, even 
though the $d$ quark content of the proton is roughly ten
times the strange quark density.  So the Cabibbo suppression
of the $d$ quark contribution to charm production makes
the strange quark contribution relatively more important.  This 
suppression factor is also present for reactions induced by
antineutrinos, but here the relevant antiquark distributions 
$\bar{s}$ and $\bar{d}$ are equal to within about a factor of
two.  

Because of the suppression of contributions from the
valence quarks, the next-to-leading [NLO] contributions
from gluon exchange turn out to be quite important 
\cite{Bar93}.  The most important such processes, $t$ and $u$ channel
diagrams initiated by gluons, are also 
shown in Fig.\ \ref{fig27} \footnote{We have not shown radiative-gluon
and self-energy diagrams which also occur in NLO.}.  
The extra factor of $\alpha_s$
which enters these diagrams is compensated for by the fact
that the gluon density is an order of magnitude larger than
the antiquark density.  It turns out to be crucial to
include the NLO contributions to this reaction, in this
kinematic region (relatively near charm threshold).  

\begin{figure}
\centering{\ \psfig{figure=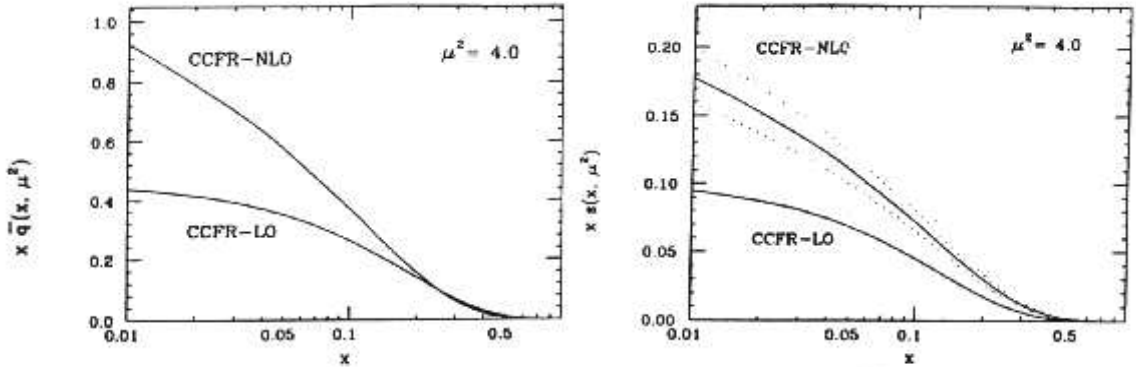,width=16cm}}
\vspace{0.15truein}
\caption{a) Sea quark distribution $x\sum_j \bar{q}_j(x)(x)$, 
$j=u,d,s$, from NLO analysis of CCFR Collaboration data, Ref.\ 
\protect\cite{Baz95}. b) Strange quark distribution $s(x)$ from same data.}
\vspace{0.1truein}
\label{fig29}
\end{figure}

Recent measurements by the CCFR collaboration (experiments
E744 and E770 at the Fermilab Tevatron with the quadrupole
triplet neutrino beam) \cite{Fou90,Rab93,Ste95} 
have amassed a substantial number of
events for both neutrinos and antineutrinos.  The following results 
have been obtained with the latest NLO analysis of the
CCFR data \cite{Baz95}.  In Fig.\ 
\ref{fig29}a we plot $x\bar{q}(x,\mu^2) = x\bar{u}(x,\mu^2)
 + x\bar{d}(x,\mu^2) + xs(x,\mu^2)$ extracted from
these measurements \cite{Baz95}, for a scale $\mu^2 = 4$ GeV$^2$.  
In Fig.\ \ref{fig29}b we plot the strange quark distribution 
$xs(x, \mu^2)$ at the
same scale.  Both the LO results, and the NLO results, are 
plotted.  We see that the NLO curves differ dramatically
from the LO results, with the NLO curves much softer than
the LO results (they are significantly larger at small $x$ and fall off
faster) \footnote{However, see the paper by Gl\"uck {\it et al.} 
\cite{Glu96}, who claim that a consistent treatment of acceptance 
corrections gives NLO results, for the same CCFR data, which are much 
closer to the original LO results.}.  

Second, the strange/nonstrange 
antiquark fraction can be extracted.  If we define
\begin{eqnarray}
  \kappa &=& {S + \bar{S}\over (\bar{U} + \bar{D})}, \quad {\rm where} 
  \nonumber \\ 
  S &=& \int_0^1 xs(x) dx, \quad {\rm the~CCFR~group~obtain}
  \nonumber \\ 
  \kappa &=& 0.477 {+ 0.046\atop -0.044} {\rm (stat)} 
  +{ +0.023\atop -0.024} {\rm (syst)}
\end{eqnarray}

The extracted value of $\kappa$ shows a substantial violation
of SU(3) symmetry in the nucleon sea.  However, 
the value of $\kappa$ obtained in the NLO analysis \cite{Ste95} 
is substantially larger than the result obtained in a LO  
analysis of the same data \cite{Rab93}, which produced a 
value $\kappa_{LO} \approx 0.37$.  This arises because the 
CCFR structure functions extracted from the NLO analysis 
give larger values at small $x$ for 
both the nonstrange sea and the strange sea.  The importance
of the NLO analysis is quite striking here.  

Next, the CCFR group compared the $x$ dependence of the
strange and nonstrange sea.  They parameterized the strange
quark distribution as 
\begin{equation}
xs(x,\mu^2) = A_s (1-x)^\alpha \left[ {x\bar{u}(x,\mu^2) + 
  x\bar{d}(x,\mu^2) \over 2} \right] 
\end{equation}
They obtained the best fit value $\alpha = -0.02 + {+0.60\atop 
 -0.54} {\rm (stat)} + {+0.28\atop -0.26}$ (syst).  This value is 
 consistent with
zero, so the $x$ dependence of strange and nonstrange sea appears
to be identical.  The LO analysis \cite{Rab93} appeared to show a
much softer $s$ distribution than the nonstrange sea.  

Fourth, this data can be used to extract the CKM matrix
element $V_{cd}$.  At present, the uncertainty in this
parameter gives the greatest contribution to the uncertainty
in the Weinberg angle, $\sin^2(\theta_W)$.  

Finally, the CCFR group tested whether the
strange antiquark distribution $\bar{s}(x)$ differed in
shape from the strange quark distribution $s(x)$.  The LO
analysis of this group \cite{Rab93} had suggested some
difference between the two distributions.  There have
been several theoretical suggestions that this might
be the case \cite{Sig87,Ji95,Hol96,Bro96,Mel97}.  If this should prove
to be the case, then our formulae for CSV need to be
modified to include differences between quark
and antiquark distributions for heavy quark flavors.  This
will turn out to be particularly important in tests of
charge symmetry involving charge-changing weak currents
on isoscalar nuclei, as is discussed in Sect.\ 5.1. 

The CCFR group analyzed their strange quark distributions
assuming that $xs(x,\mu^2) = (1-x)^{\beta}x\bar{s}(x,\mu^2)$.  They
obtained the value $\beta = -0.46 \pm 0.42 \pm 0.36 \pm 0.65 
\pm 0.17$; the quoted errors (from left to right) are 
statistical, systematic, from the uncertainty in the 
semileptonic charged hadron branching ratios, and the
uncertainty in $\mu^2$ scale.  The value obtained is 
consistent with zero, i.e.\ no difference in the shape of
the strange and antistrange distributions.     

Let us review the outline for subsequent sections of
this paper.  
In Sect.\ 3, we review model calculations which give 
order of magnitude estimates for charge symmetry violating 
contributions to both valence and sea quark distributions.  
In Sect.\ 4 we review existing experiments which allow us to
put upper limits on the magnitude of charge symmetry violation
in the parton distributions.  We will show that present upper
limits on CSV in parton distributions depend on the $x$ region
in question.  In the region $0.1 \le x \le 0.4$, the comparison 
of CCFR neutrino structure functions to the structure functions
from the NMC muon measurements suggests an upper limit on CSV
of a few percent.  At larger values of $x$, there is much less 
experimental data, so the upper limits on parton CSV are at
least 10\% in the parton distributions.  For smaller values 
$0.01 \le x \le 0.1$, the NMC and CCFR results currently disagree 
at a level between $10-20$\%.  

We will also review recent experimental tests of flavor 
symmetry in the proton sea.  We will show that these ``tests'' 
demonstrate large breaking of either flavor symmetry or
charge symmetry, and that at the present time it is difficult 
to rule out significant breaking of charge symmetry in the
nucleon sea (although we certainly do not expect CSV effects
of the magnitude necessary to agree with recent Drell-Yan
experiments).     
In Sect.\ 5  we suggest several new experiments
which can differentiate between charge symmetry and flavor
symmetry.  These experiments, if carried out, could either demonstrate 
the existence of CSV terms
in parton distributions, or provide more stringent 
upper limits on quark CSV.

\section{Theoretical Estimates of Charge Symmetry
Violation in Parton Distributions}
\mb{.5cm}

In this section, we will derive theoretical estimates for charge 
symmetry violation in parton distributions.  First, we will
review theoretical calculations of CSV for valence quark distributions.  
Next, we will give an estimate of the size of CSV effects in antiquark 
distributions.  We will discuss the robustness of these estimates.  In 
later sections, these calculations 
will be used to provide estimates of the magnitude of CSV effects which
could be expected in various experiments.  In absence of more detailed 
calculations of CSV effects, and lacking firm upper limits from experiment, 
these estimates are the best we have at present.  As we will show, we
believe that our estimates of both the sign and magnitude 
of valence quark CSV effects should be rather well determined.  
Neither the sign nor magnitude of sea quark CSV contributions is
well known; however, our model calculations predict very small
CSV effects from the sea.  

\subsection{Charge Symmetry Violation for Valence Quarks}

Theoretical
investigations of parton CSV for valence quark distributions by Sather 
\cite{Sat92}, Rodionov, Thomas and Londergan \cite{Rod94} and Benesh 
and Goldman \cite{Ben97a} concluded 
that one could make reasonably model-independent
estimates of the size of these effects.  Here we follow the 
method for calculating twist-two parton distributions with proper
support, which has been developed by the Adelaide group 
\cite{Sig89a,Sig89b,mst}.  The starting point is 
the evaluation of quark distributions through the relation
\begin{equation}
q(x,\mu^2) = M\sum_X\,|<X|\psi_+(0)|N>|^2 \,\delta(M(1-x) - p_X^+)
\label{pardis}
\end{equation} 
In Eq.\ \ref{pardis}, $\psi_+ = (1+\alpha_3)\psi/2$, $X$ 
represents a complete set of eigenstates of the Hamiltonian, and 
$\mu^2$ represents the starting scale for the quark distribution.  

The advantage of this method is that the resulting parton distribution 
$q(x,\mu^2)$ is guaranteed to have proper support, {\it i.e.} it 
vanishes for $x > 1$, regardless of the model used for the matrix 
element $<X|\psi_+(0)|N>$ in Eq.\ \ref{pardis}.  Parton distributions 
calculated from quark models generally lack this support, and this
can lead to serious problems in obtaining reliable results.  
Thus, Thomas and collaborators showed that reasonable
parton distributions could be obtained from models such as the MIT 
bag \cite{Sig89a}.  The
other advantage of this method is that it is often possible to
obtain reasonable results taking into account only the lowest-energy
spectator states in the sum over states of Eq.\ \ref{pardis}. 

We want to use the relation in Eq.\ \ref{pardis} to calculate
differences in parton distributions due to violation of charge symmetry.  
Thus, we wish to estimate the difference between, say, the up
quark distribution in the proton and the down quark distribution
in the neutron.  From Eq.\ \ref{pardis}
we see that CSV contributions will have four sources: first, 
from electromagnetic effects which break charge symmetry; second, 
from charge symmetry violating mass differences of the struck quark; 
third, from mass 
differences in the spectator multiquark system; and fourth from 
charge symmetry violation in the quark wavefunctions.  
In model calculations, it was found that the quark wavefunctions 
are almost invariant under small mass changes typical of CSV.  
At high energies, electromagnetic effects should also be small,
and we neglect these.  Consequently, parton charge symmetry violation
will arise predominantly through mass differences of the struck quark, 
and from mass differences in the spectator multiquark system.  

\begin{figure}
\centering{\hbox{ \hspace{-0.2in} \psfig{figure=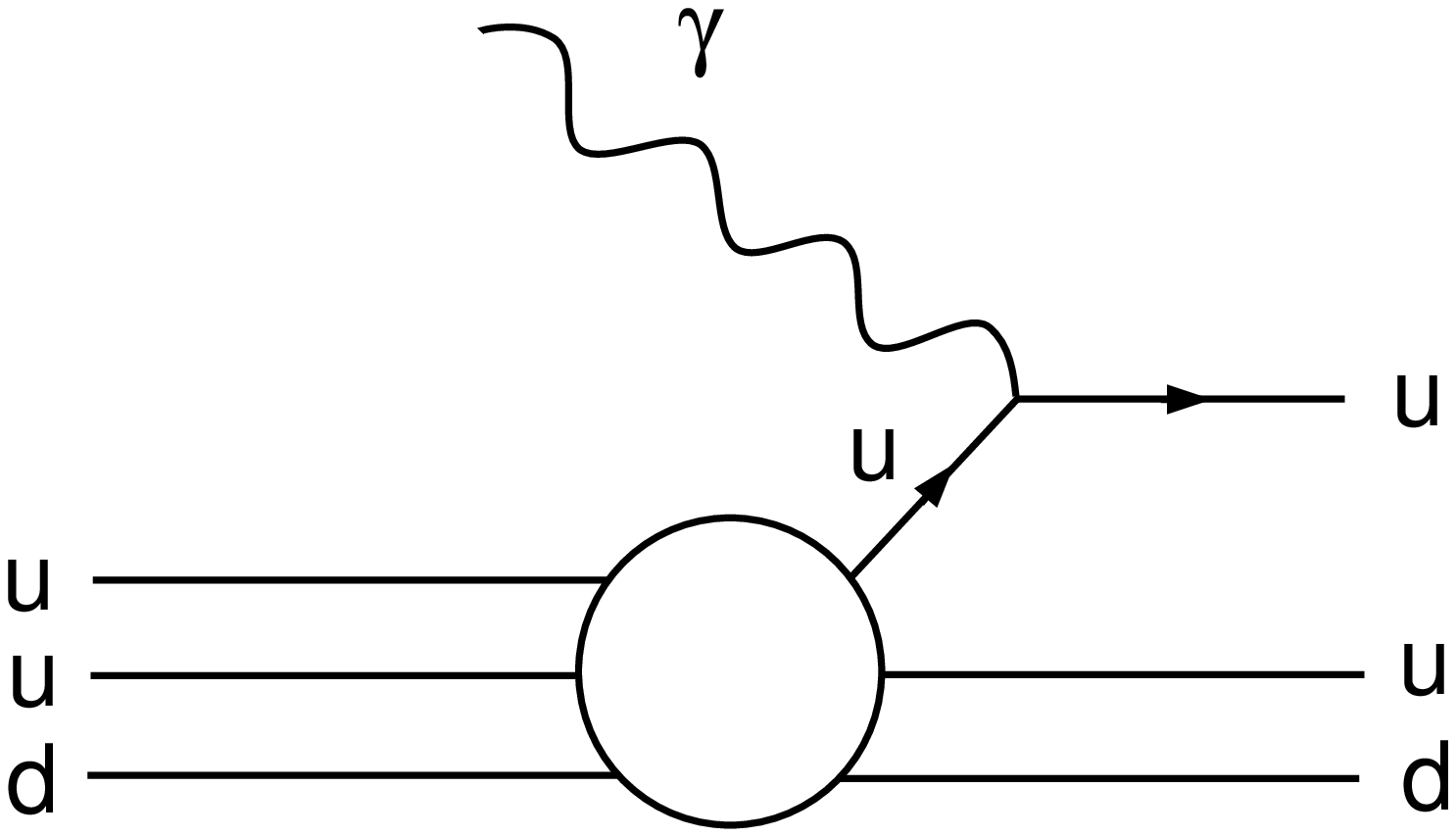,height=4.5cm}
\hspace{0.1in} \psfig{figure=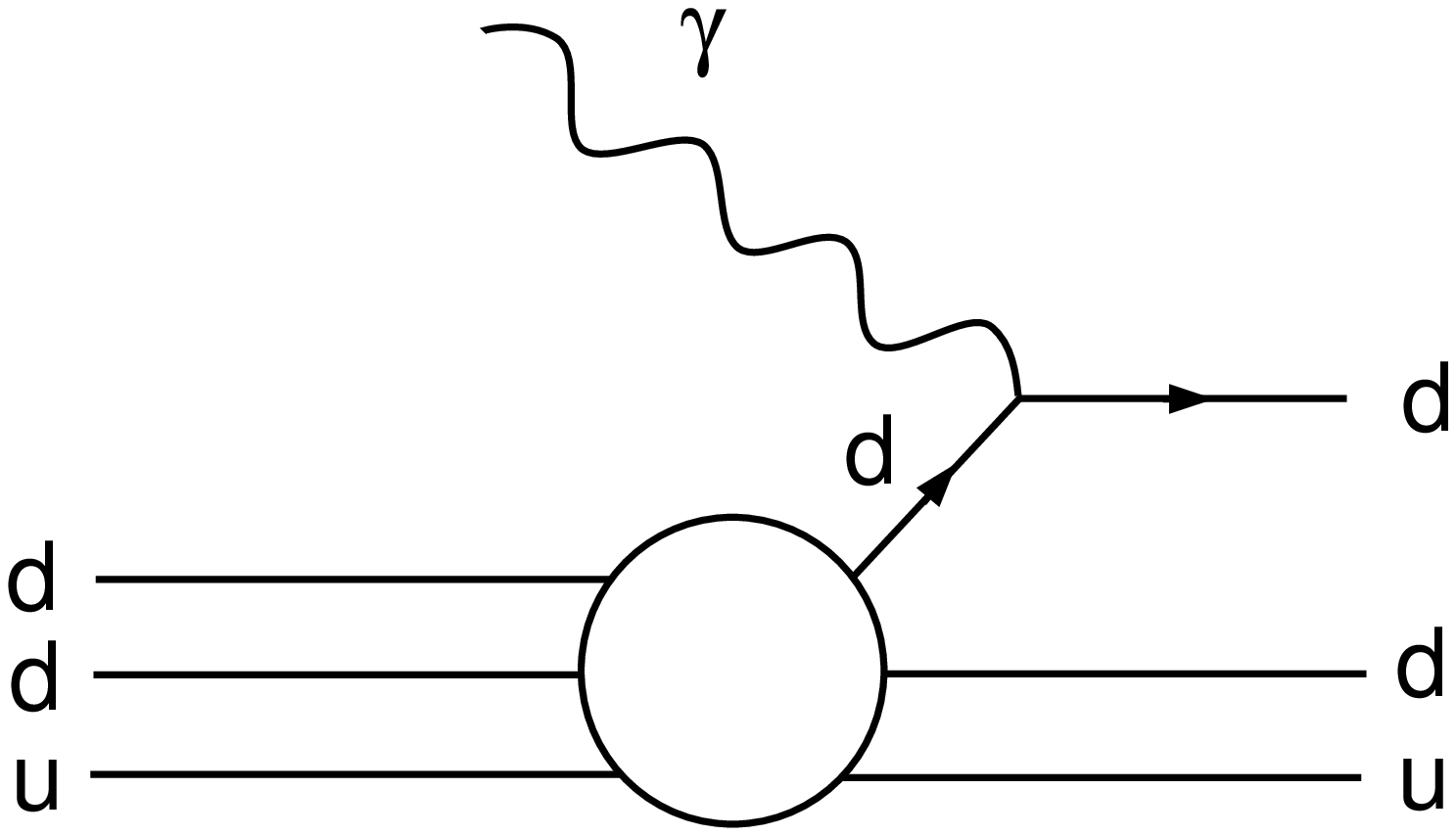,height=4.5cm}}}
\vspace{0.5truein}
\caption{Schematic picture of lowest energy contributions to  
``majority'' quark distributions: struck quark plus spectator diquark. 
(a) $u^p(x)$ for proton. (b) $d^n(x)$ for neutron.}
\vspace{0.1truein}
\label{fig31}
\end{figure}

As an example, we consider valence quark CSV where for the 
intermediate states $X$ in Eq.\ \ref{pardis} we include the
lowest two-quark spectator states from the MIT bag model 
\cite{Rod94,Cho74,Sig89a,Sig89b}.  
There are more sophisticated quark models available but the 
similarity of the results obtained by Naar and Birse \cite{Naa93} using 
the color dielectric model suggests that similar results would be 
obtained in any relativistic model based on confined current quarks.  
In Fig.\ \ref{fig31} we show schematically the lowest contributions
to the ``majority valence quark'' distributions, {\it i.e.} $u^p_{\rm v}(x)$ 
and $d^n_{\rm v}(x)$.  The majority quark CSV term is as defined in Eq.\ 
\ref{deluv}, $\delta u_{\rm v}(x) = u_{\rm v}^p(x)- d_{\rm v}^n(x)$.  
From Fig.\ \ref{fig31} we see that the only contribution to the
``majority'' quark CSV is the up-down mass difference $m_u - m_d$; 
the intermediate spectator diquark is the same ($ud$) for both 
proton and neutron.  

\begin{figure}
\centering{\hbox{ \hspace{-0.2in} \psfig{figure=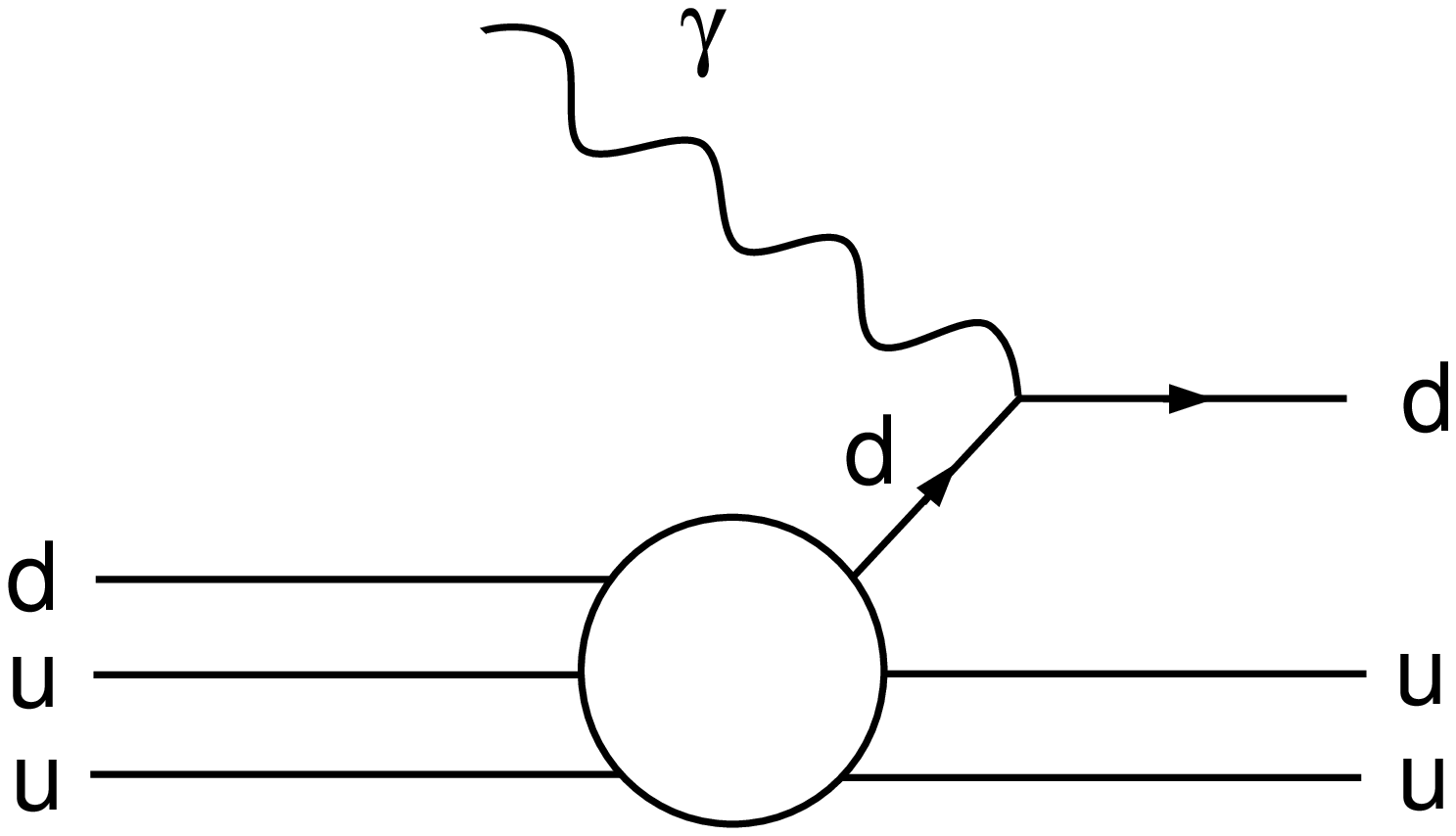,height=4.5cm}
\hspace{0.1in} \psfig{figure=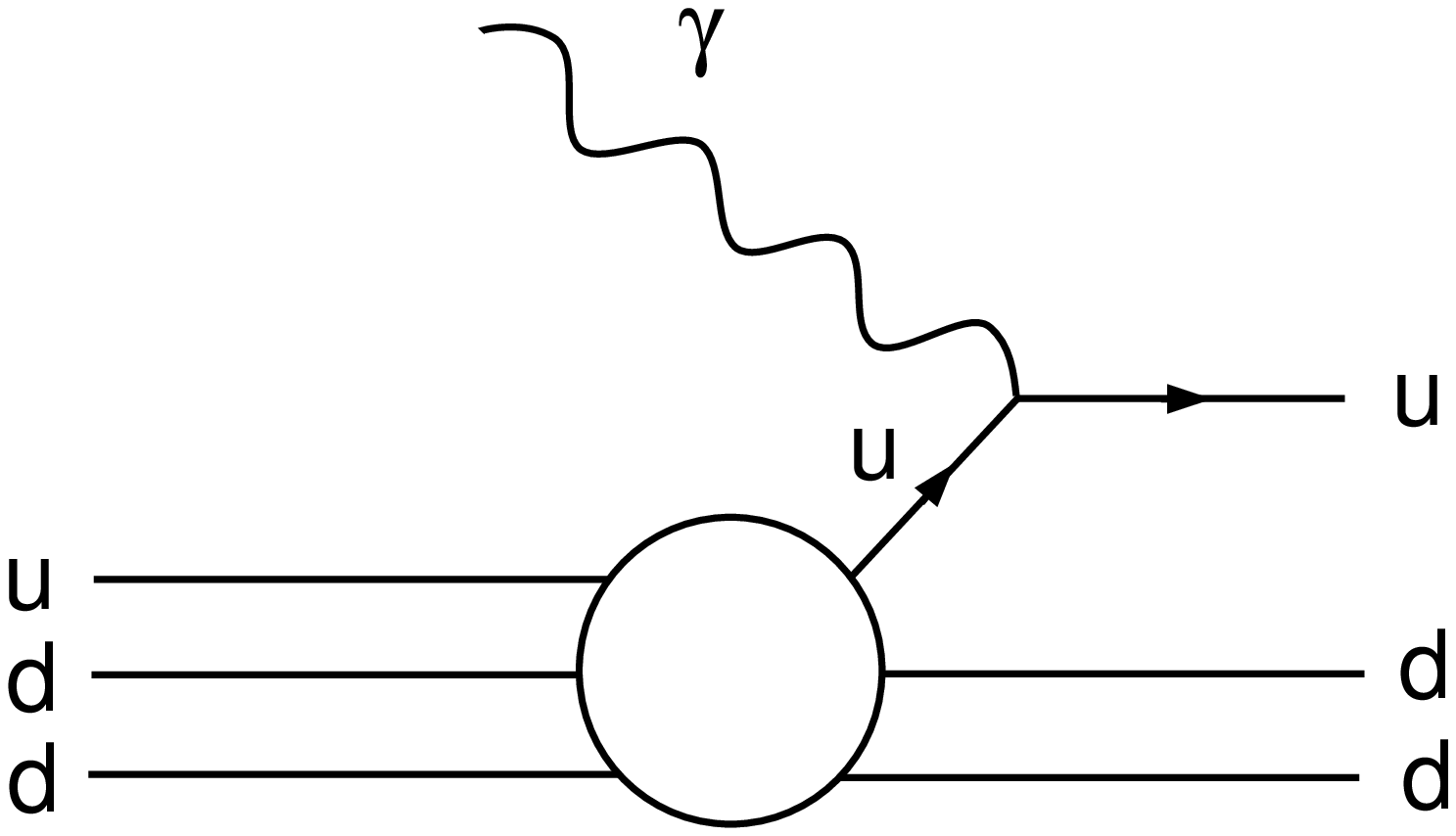,height=4.5cm}}}
\vspace{0.5truein}
\caption{Schematic picture of lowest energy contributions to  
``minority'' quark distributions. 
(a) $d^p(x)$ for proton. (b) $u^n(x)$ for neutron.}
\vspace{0.1truein}
\label{fig32}
\end{figure}

\begin{figure}
\centering{\ \psfig{figure=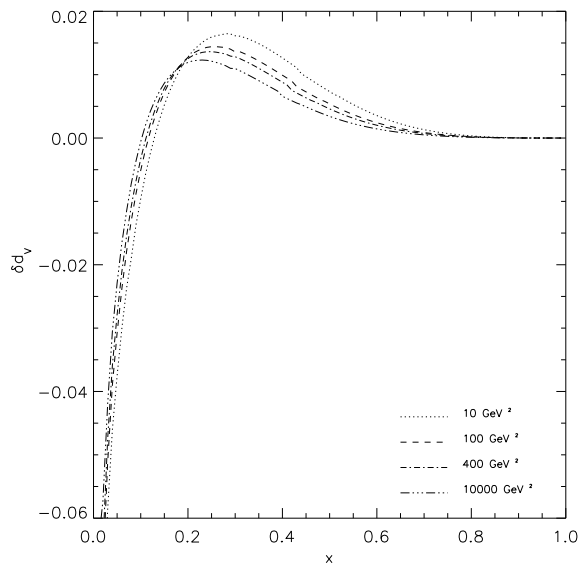,width=8cm}}
\vspace{0.05truein}
\caption{ Calculation of minority valence quark CSV 
contribution, $\delta d_v(x)$ vs.\ $x$. Dotted curve: $Q^2 = 10$
GeV$^2$; dashed curve: $Q^2 = 100$ GeV$^2$; dot-dashed curve: $Q^2 = 400$
GeV$^2$; dash-triple dot curve: $Q^2 = 10,000$ GeV$^2$. From Ref.\ 
\protect\cite{herathy}.}
\vspace{0.1truein}
\label{fig33}
\end{figure}

\begin{figure}
\centering{\ \psfig{figure=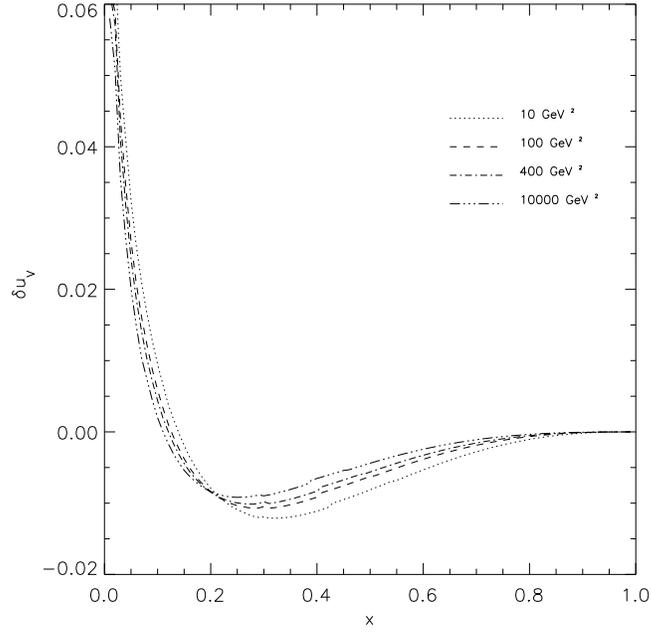,width=9cm}}
\vspace{0.05truein}
\caption{Majority valence quark CSV contribution, $\delta u_v(x)$. 
Notation is that of Fig.\ \protect\ref{fig33}.}
\vspace{0.1truein}
\label{fig33b}
\end{figure}

In Fig.\ \ref{fig32} we show the lowest contributions
to the ``minority valence quark'' distributions, {\it i.e.} 
$d^p_{\rm v}(x)$ and $u^n_{\rm v}(x)$.  From Fig.\ \ref{fig32} we see 
that there are two
contributions to minority quark CSV.  One is the up-down mass 
difference $m_u - m_d$; the second source of charge symmetry violation
comes from the intermediate spectator 
diquark mass, which is $uu$ for the proton and $dd$ for the neutron.  
In Fig.\ \ref{fig33} we show the calculated minority valence quark CSV 
term, $\delta d_{\rm v}(x) = d^p_{\rm v}(x) - u^n_{\rm v}(x)$.  Fig.\ 
\ref{fig33b} shows the majority valence quark CSV term, 
$\delta u_{\rm v}(x) = u^p_{\rm v}(x) - d^n_{\rm v}(x)$.  These are 
calculated from Eq.\ \ref{pardis} using quark wave functions 
from the MIT bag model.  The 
contributions are calculated at the bag scale, then evolved upwards
in $Q^2$.  At small $x$, $\delta d_{\rm v}(x)$ is negative, while for 
larger $x$ it is positive.  The majority quark CSV term has exactly
the opposite sign.  As a result, theoretical calculations suggest that 
\begin{equation} 
|\delta u_{\rm v}(x) + \delta d_{\rm v}(x)| << |\delta u_{\rm v}(x) -
 \delta d_{\rm v}(x)| ~~, 
\label{csvdff}
\end{equation}
so that experimental quantities which depend on the sum of the majority 
and minority valence quark CSV terms should be substantially smaller 
than those which depend on the difference between the majority and
minority CSV terms.  
 
\begin{figure}
\centering{\ \psfig{figure=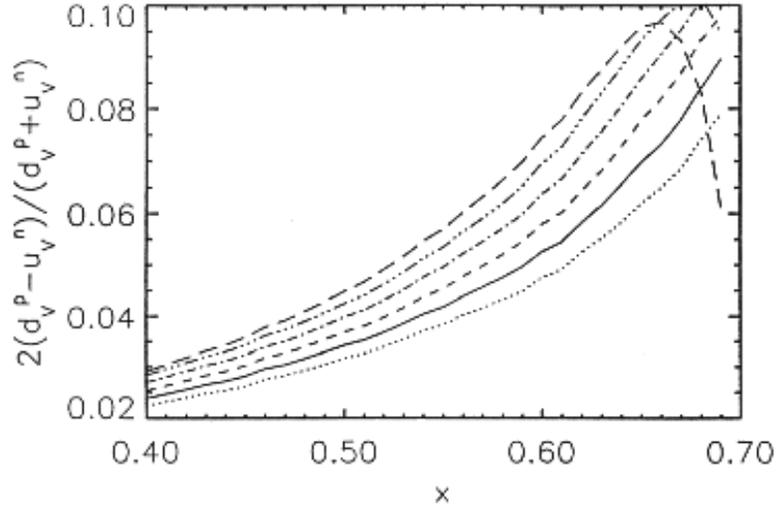,width=11cm}}
\vspace{0.05truein}
\caption{Model calculation of percent minority valence quark CSV 
term, $\delta d_v(x)/d_v(x)$ vs.\ $x$, evolved to $Q^2 = 
10$ GeV$^2$. From Ref.\ \protect\cite{Lon94}.}
\vspace{0.1truein}
\label{fig34}
\end{figure}

Because the integral over $x$ of the valence quark distributions is 
normalized (one for the minority valence quark distribution, two for
the majority distribution), the integral of the CSV distributions
must be zero, {\it i.e.} 
\begin{equation}
\int_0^1 \delta u_{\rm v}(x)\,dx = \int_0^1 \delta d_{\rm v}(x)\,dx 
  = 0
\label{csvint}
\end{equation}
In Fig.\ \ref{fig34} we show the percent CSV contribution, 
$\delta d_{\rm v}(x)/d_{\rm v}^p(x)$ vs.\ $x$.  For large $x$, 
{\it i.e.} $x \ge 0.5$, the minority valence quark CSV is predicted to 
be between $5-10$\%.  
This is an extremely large violation of charge symmetry, especially
since at low energy scales (e.g., low energy nuclear physics) 
charge symmetry is generally valid to at least 1\%.  
Compared with the minority quark CSV contributions, the majority quark 
term has roughly the same magnitude, as can be seen from Figs.\ 
\ref{fig33} and \ref{fig33b}. However, for large $x$ the percent 
majority quark CSV 
is predicted to be much smaller than the minority quark CSV fraction.  

Why should the minority valence quark CSV, $\delta d_{\rm v}(x)$,  
be a much larger percentage at large $x$ than the majority CSV term, 
$\delta u_{\rm v}(x)$? We can understand this qualitatively because 
the dominant source of CSV is the mass difference of the residual 
diquark pair when one quark is hit in the deep-inelastic process. 
For the majority quark distribution, the diquark pair is $ud$ for 
both proton and neutron, as can be seen from Fig.\ \ref{fig31}.  
Consequently there is no CSV contribution from this term.  
For the minority quark distribution the residual diquark is $uu$ in the 
proton, and $dd$ in the neutron. Thus, in the difference, 
$d^p_{\rm v}(x) - u^n_{\rm v}(x)$, the up-down mass difference enters 
twice.  Including 
electromagnetic effects, we know that the diquark mass difference
is $\delta M_2 = M_2^{dd} - M_2^{uu} \approx 4$ MeV.  From Eq.\ 
\ref{pardis}, we
can see that the valence quark distribution $q_{\rm v}(x)$ will peak at 
roughly $x_{peak} \sim 1- M_2/M$ where $M$ is the nucleon mass.  For 
the minority quark
distribution, say, $d_{\rm v}^p(x)$, the lowest diquark state for
the $uu$ pair will be $S=1$ by the Pauli principle.  From the 
$N-\Delta$ mass splitting we know that an $S=1$ diquark pair will
have an effective mass $M_2(S=1) \sim 800$ MeV, while the 
$S=0$ diquark (which is present only for the majority quark term 
$u_{\rm v}^p(x)$) has a mass $M_2(S=0) \sim 600$ MeV.  

As was pointed out by Close and Thomas \cite{close}, this explains why 
the down quark distribution $d_{\rm v}^p(x)$ peaks 
at a value $x \sim 0.1$, while the up valence quark distribution
$u_{\rm v}^p(x)$ peaks at $x \sim 0.3$.  It also predicts that 
$d_{\rm v}^p(x) << u_{\rm v}^p(x)$ at large $x$, as is observed 
experimentally.  
We can obtain an estimate of the magnitude of minority quark 
charge symmetry breaking, 
\begin{equation}
{\delta d_{\rm v}(x) \over d_{\rm v}^p(x)} \approx {\delta M_2 
  \over M - M_2} 
\label{csvapp}
\end{equation}
Eq.\ \ref{csvapp} gives an estimate of about 4\% for minority quark
CSV.  This can be compared with the 3-7\% estimates of CSV by 
Sather \cite{Sat92}, Rodionov {\it et al.} \cite{Rod94}, and
Benesh and Goldman \cite{Ben97a}.  

Fig.\ \ref{fig34} shows the fractional minority quark
CSV term, $2(d^p_{\rm v} - u^n_{\rm v})/(d^p_{\rm v} + u^n_{\rm v})$ vs.\ 
$x$ for several values of
the intermediate mean diquark mass.  Although the precise value of
the minority quark CSV changes with mean diquark mass, the size is
always roughly the same and the sign is unchanged.  This shows that
``smearing'' the mean diquark mass will not dramatically diminish the
magnitude of the minority quark CSV term (the mean diquark mass must be
roughly 800 MeV in the $S=1$ state to give the correct $N-\Delta$
mass splitting).

We reiterate that the magnitude of charge symmetry violation, predicted
for the ``minority'' valence quarks at large $x$, is both extremely large 
and surprising.  First, since the integral over $x$ of the valence
quark CSV term is zero, as given in Eq.\ \ref{csvint}, large CSV
contributions to valence quark parton distributions do not necessarily
imply large charge symmetry violation at low energies.  Second, 
experimental verification of these CSV effects is not a simple 
matter.  Since we predict that charge symmetry is well obeyed for the 
majority valence quarks, finding CSV effects requires experiments
which are sensitive to the minority quark distributions, in a region 
(large Bjorken $x$) where the minority quark distributions are much
smaller than the majority quark terms.  In Sects.\ 4 and 5, we
will review the current experimental limits on parton charge
symmetry, and suggest experiments which would be sensitive to our
predicted effects.  

However, as we have tried to emphasize, our predictions of charge
symmetry violation depend on rather simple assumptions, which 
have been shown to produce reasonable parton distributions 
\cite{Sig89a,Sig89b}.  Furthermore, all theoretical calculations
of parton charge symmetry obtain the same qualitative result: 
that the percentage of CSV for minority valence quarks should be 
substantially larger than the fraction of majority quark CSV at large 
$x$ \cite{Sat92,Rod94,Ben97a}.  To state this another way, theory
predicts that $\delta d_{\rm v}(x) \approx -\delta u_{\rm v}(x)$; 
however, at large $x$ $d_{\rm v}(x) << u_{\rm v}(x)$, so the fraction 
of minority quark CSV will be substantially larger than for 
majority quarks.   
In particular, the calculation of Sather \cite{Sat92} gave what
was essentially a model-independent estimate of parton CSV.  
Sather also predicted that minority quark CSV effects should be, 
relatively, substantially larger than majority quark CSV.  This
gives us confidence that our predictions are robust.  

Finally, since all phenomenological fits to date assume parton
charge symmetry at the outset, existing parton distributions then
have CSV effects included implicitly.  As we pointed out in Sect.\ 2, a 
truly consistent treatment of these effects would begin at the
outset with charge asymmetric parton distributions and proceed to
fit experimental data, without making the prior assumption of charge
symmetry.  Then the CSV contributions could be deduced in a 
consistent fashion.  In the absence of such a consistent procedure,
our calculations of charge symmetry violation should be taken
as order of magnitude estimates only.  
This is a difficult process, and we assume
it will not take place until there is some definite experimental
evidence of charge symmetry violation in parton distributions.   

\subsection{Estimate of Charge Symmetry Violation for Sea Quarks}

In the preceding section, we made estimates of valence quark charge
symmetry violation, using a formalism for quark distributions 
which was guaranteed to produce the proper support.  We could
use the same formalism to calculate antiquark distributions, 
\begin{equation}
\overline{q}(x,\mu^2) = M\sum_X\,|<X|\psi^\dagger_+(0)|N>|^2 \,
\delta(M(1+x) - p_X^+)
\label{pardis2}
\end{equation} 
We would proceed with the same assumptions as for valence quarks: 
we would take light cone quark momentum wave functions from simple
models, and truncate the sum over states in Eq.\ \ref{pardis2} to
the lowest energy states which can contribute.  

Recently, such a procedure has been followed by Benesh and 
Londergan \cite{Ben97b}.  One is confident that reasonably 
model-independent estimates can be made for valence quark CSV.  This
is borne out by substantial agreement between various theoretical
results.  Calculations of sea quark CSV require some additional 
assumptions, and the model dependence of sea quark CSV is not
entirely clear.  However, in these calculations one predicts
that charge symmetry violation for antiquarks should be very
small, probably at least an order of magnitude smaller than the
corresponding effects for valence quarks.  

We can give simple qualitative arguments why sea quark CSV effects
should be small.  The relative magnitude of CSV effects will be given 
approximately by 
\begin{equation}
{ \delta \bar{q}\over \bar{q}} \sim {\delta M \over M} ~~,
\end{equation}
where $M$ is the energy of the lowest intermediate state in 
Eq.\ \ref{pardis2}, and $\delta M$ is the mass difference for
intermediate states related by charge symmetry.  For antiquarks,
the lowest energy states which contribute are four-quark states, 
whose energy is roughly twice the energy of the lowest diquark
states which contribute for the valence quarks.  
We estimate the mass difference between charge symmetric four
quark states as $\delta M \approx 1.3$ MeV, or three times smaller
than the mass difference for minority valence quarks.  Thus, a naive
expectation would be that CSV effects for sea quarks would be
at least a factor of six smaller than for minority valence quarks.  

In their calculations Benesh and Londergan \cite{Ben97b} 
typically obtained
estimates for sea quark CSV at least an order of magnitude smaller
than those for valence quark CSV effects.  As a result, in 
subsequent sections we will occasionally neglect sea quark CSV effects 
relative to valence quark CSV terms.  Benesh and Londergan also 
obtain $\delta\bar{u}(x) \approx - \delta\bar{d}(x)$; therefore, 
one would expect that observables proportional to the difference
between up and down sea quark CSV distributions would be significantly
larger than observables which measure the sum of sea quark CSV terms.  
In Sects.\ 6 and 7 we review parton sum rules.  There we will
show that, in principle, it would be possible to get explicit
measurements of charge symmetry and/or flavor symmetry violation
in parton distributions, by measuring appropriate integrals of deep 
inelastic cross sections.   
 
\section{Present Experimental Limits on Parton Flavor Symmetry 
and Charge Symmetry}
\mb{.5cm}

In this section, we review the limits we can place on parton
flavor symmetry in the proton sea, and on charge symmetry 
of parton distributions.  In the following section, we propose
a series of experiments which could in principle sharpen the
limits on CSV in parton distribution, and which could also
discriminate between flavor symmetry and charge symmetry
violation.  

\subsection{Drell-Yan Tests of Flavor Symmetry in the Proton
Sea}

Recently, there has been much interest in the details of
proton sea quark distributions.  The NMC experiment 
\cite{Ama91,Arn94} measured the $F_2$ structure functions
for muons on proton and deuteron targets.  With these
data they were able to test the Gottfried Sum Rule 
\cite{Got67}, which requires the integral of the difference
between proton and neutron $F_2$ structure functions.  
The experimental value $S_G = 0.235 \pm 0.026$ is more
than four standard deviations below the ``naive'' expectation 
of 1/3.  Both the experimental and theoretical situations 
are summarized in detail in the recent review by Kumano 
\cite{Shunzo}.  We review the Gottfried
Sum Rule in Sect.\ 6.1. 

The most likely cause for the NMC result is a substantial
difference in the $\bar{d}$ and $\bar{u}$ distributions in
the proton.  The NMC experiment suggests that 
\begin{eqnarray*}
\int_0^1\,dx\,\left[ \bar{d}^p(x) - \bar{u}^p(x)\right] 
  \,&{CS\atop =}&\, 0.147 \pm 0.039
\end{eqnarray*}
There is no leading order QCD correction to the Gottfried
Sum Rule.  
Ross and Sachrajda \cite{sachrajda} showed that higher
order perturbative QCD contributions lead to a value 
much smaller than this.  Consequently, there has been much 
interest in experiments which might give a ``direct'' 
measurement of the sea quark distributions $\bar{d}^p(x)$
and $\bar{u}^p(x)$, and which might map out their $x$
dependence (the NMC experiment gives only the integral over
$x$ of this difference).  

Ellis and Stirling \cite{Ell91} suggested that this could
be measured by comparing Drell-Yan processes initiated by
protons, on proton and deuteron targets.  We review here
the information which could be obtained from these
measurements.

\begin{figure}
\centering{\hbox{ \hspace{0.1in} \psfig{figure=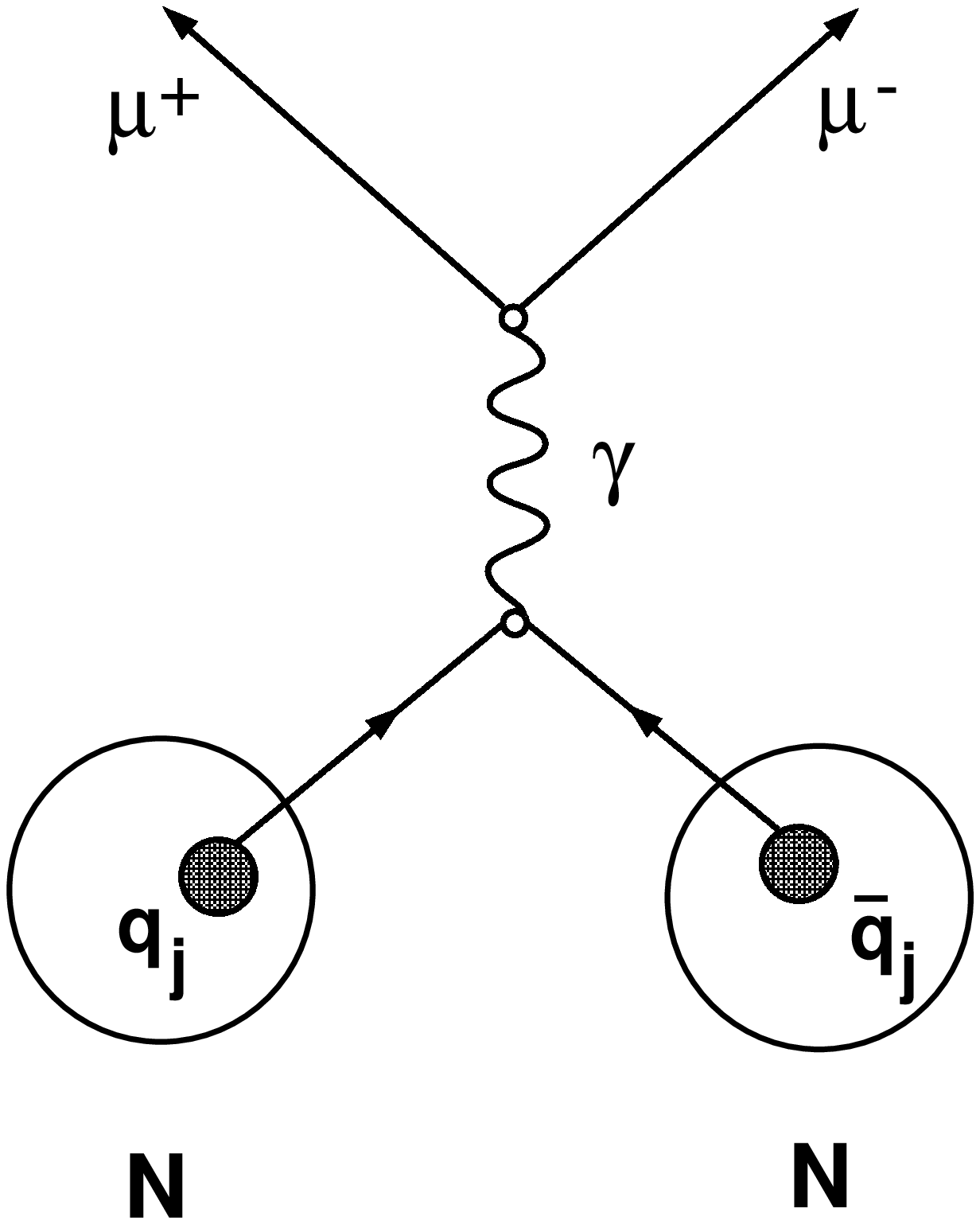,height=6.5cm}
\hspace{0.8in} \psfig{figure=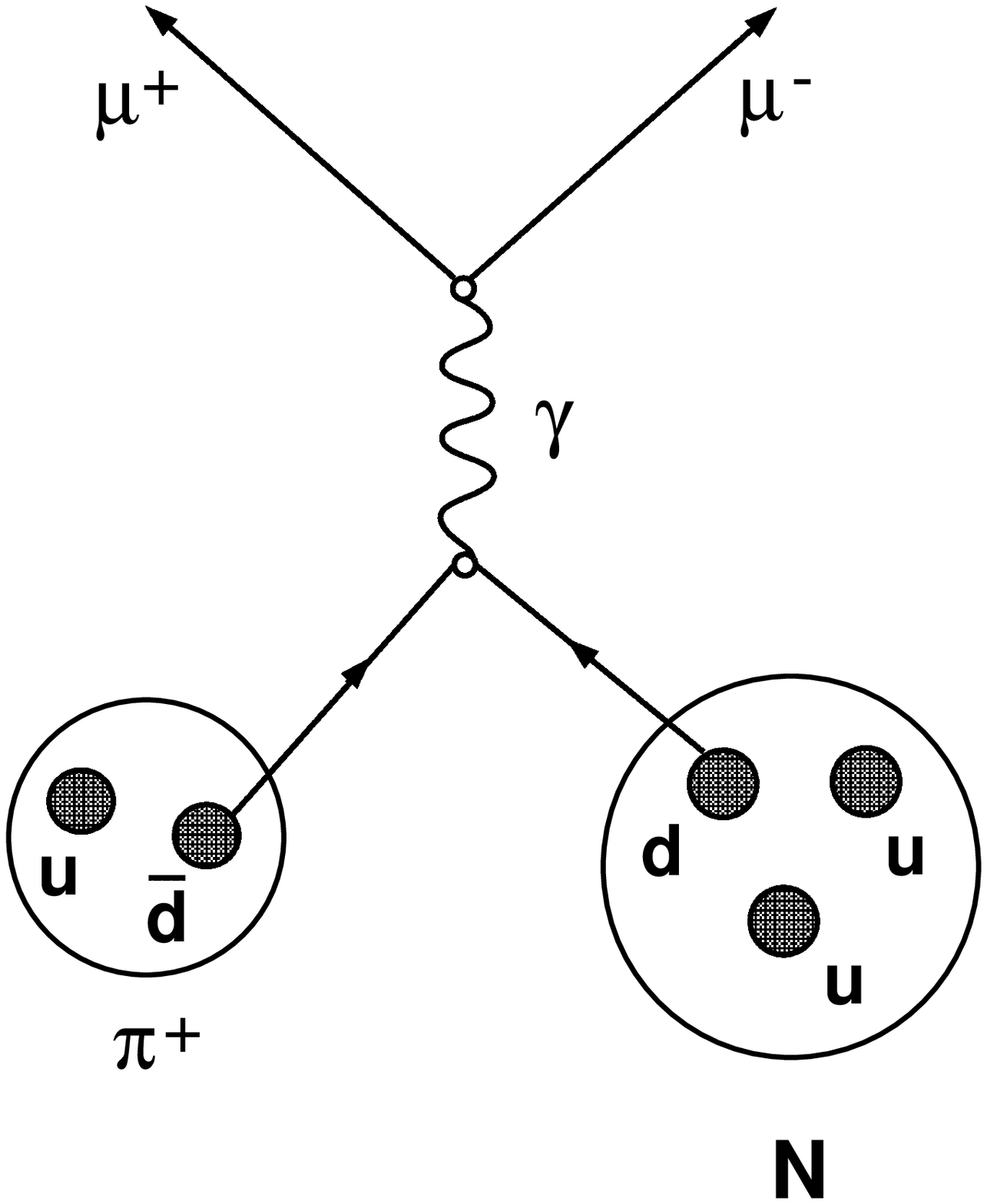,height=6.5cm}}}
\vspace{0.05truein}
\caption{Schematic picture of Drell-Yan [DY] process for two hadrons.  
(a) $NN$ DY is sensitive to antiquarks in one nucleon; 
(b) $\pi^+-p$ DY process, in the valence region for both
particles.}
\vspace{0.1truein}
\label{fig41a}
\end{figure}

In the Drell-Yan [DY] process \cite{drellyan} one observes 
lepton pairs of opposite charge and large invariant mass 
which arise from hadronic collisions.  This process 
occurs when a quark (antiquark) from the projectile 
annihilates an antiquark (quark) of the same flavor
from the target.  This produces a virtual photon which
subsequently decays into a pair of charged leptons.  The
process is shown schematically in Fig.\ \ref{fig41a}a, for $NN$ DY 
processes.  A quark in one nucleon annihilates an antiquark of the
same flavor in the other nucleon.  In Fig.\ \ref{fig41a}b we show 
the corresponding DY process for $\pi^+ +p$ reactions, in the valence
region for both particles.  

The Drell-Yan process for the interaction of hadron $A$ with 
hadron $B$ has the form
\begin{equation}
{d^2\sigma^{AB}\over dx_1\,dx_2} = {4\pi\alpha_s^2 \over
  9sx_1\,x_2}\,K(x_1,x_2) \sum_i\, e_i^2 \left[ 
 q^A_i(x_1)\bar{q}^B_i(x_2) 
 + \bar{q}^A_i(x_1)q^B_i(x_2) \right] 
\label{dyeq}
\end{equation}
In Eq.\ \ref{dyeq}, $s$ is the square of the CM energy, and $x_1$ 
and $x_2$ are, respectively, the longitudinal momentum fractions
carried by the target (projectile) quarks (or antiquarks) of flavor 
$i$ and charge $e_i$.  For example, the 
quantity $\bar{q}^B_i(x_2)$ is the antiquark distribution of the
target for quarks of flavor $i$ and momentum fraction $x_2$.  The 
factor $K(x_1,x_2)$ accounts for the higher-order
QCD corrections which enter the DY process.  Detailed reviews
of their form can be found in several articles 
\cite{Lea96,Ste95}.  

The values of $x_1$ and $x_2$ can be extracted from experiment
through the equations 
\begin{eqnarray}
M^2 &=& sx_1x_2 \approx 2 P_{\ell^+}P_{\ell^-}\left(1- 
  \cos\theta_{\ell^+\ell^-}\right) \nonumber \\
  x_F &\equiv& x_1 - x_2 = {2\left(P_{\ell^+}+ P_{\ell^-}
  \right)_L \over s} -1 \nonumber \\
  \tau &=& x_1\,x_2 \nonumber \\ 
  y &=& {\ln\left(x_1/x_2\right)\over 2} 
\label{DYdef}
\end{eqnarray}
In Eq.\ \ref{DYdef}, $P_{\ell^+}$ and $P_{\ell^-}$ are 
respectively the laboratory momenta of the outgoing leptons, 
$\left(P_{\ell^+}+ P_{\ell^-}\right)_L$ is the longitudinal
momentum of the lepton pair, and $\theta_{\ell^+\ell^-}$ 
is the angle between their momentum vectors.    

The experiments have compared Drell-Yan cross sections for
incident protons on proton and deuterium targets.  Taking
ratios of Drell-Yan cross sections avoids the necessity for precise
knowledge of the factor $K$ in Eq.\ \ref{dyeq}.  Assuming
the validity of the impulse approximation, the DY cross section
on the deuteron is just the sum of the DY cross sections on the 
free proton and neutron.  In that case the DY cross sections
are proportional to 
\begin{eqnarray}
\sigma_{DY}^{pp} &\sim& {4\over 9} \left[ u^p(x_1)\bar{u}^p(x_2) 
 + \bar{u}^p(x_1)u^p(x_2)\right] + {1\over 9} \left[ 
 d^p(x_1)\bar{d}^p(x_2) + \bar{d}^p(x_1)d^p(x_2)\right] \nonumber \\
\sigma_{DY}^{pD} &\sim& {4\over 9} \left[ u^p(x_1)\left( \bar{u}^p(x_2)
 + \bar{u}^n(x_2)\right) + \bar{u}^p(x_1)\left( u^p(x_2)+ 
 u^n(x_2)\right) \right] \nonumber \\ 
 &+& {1\over 9} \left[ d^p(x_1)\left( \bar{d}^p(x_2) + 
 \bar{d}^n(x_2)\right) \bar{d}^p(x_1)\left( d^p(x_2)+ d^n(x_2)
 \right) \right] 
\label{sigDYpd}
\end{eqnarray}
If we assume charge symmetry then Eq.\ \ref{sigDYpd} reduces
to 
\begin{eqnarray}
\sigma_{DY}^{pD} \,&{CS\atop\sim}&\,  \left( {4\over 9}u^p(x_1)
 + {1\over 9}d^p(x_1)\right) \left( \bar{u}^p(x_2)
 + \bar{d}^p(x_2)\right) \nonumber \\ &+& \left( 
 {4\over 9}\bar{u}^p(x_1) + {1\over 9}\bar{d}^p(x_1)\right) 
 \left( u^p(x_2)+ d^p(x_2)\right)
\label{sigDYcs}
\end{eqnarray}

The physics is most clear if we go to large $x_1$, i.e.\ 
large $x$ for the projectile proton, and substantially smaller
$x_2$ for the target.  In this regime the probability for finding 
antiquarks in the projectile is extremely small, so as a good 
approximation the DY process proceeds by quarks in the
projectile annihilating antiquarks in the target.  In this
region the ratio of DY processes on the deuteron, to those
from the proton, are given by 
\begin{eqnarray}
R_{sea}(x_1,x_2) = {\sigma^{pD}_{DY}(x_1,x_2)\over 
  2\sigma^{pp}_{DY}(x_1,x_2)} \,&{CS\atop =}&\, {1\over 2}
  { \left( 1 + \bar{R}(x_2)\right)\left(1 + r(x_1)/4 \right)
 \over \left(1+ r(x_1)\bar{R}(x_2)/4 \right)} \quad ,
\label{DYrat}
\end{eqnarray}
In Eq.\ \ref{DYrat}, we define the ratios
\begin{eqnarray}
 \bar{R}(x) &=& {\bar{d}^p(x)\over \bar{u}^p(x)} \nonumber \\     
 r(x) &=& {d^p(x)\over u^p(x)} 
\label{ptrat}
\end{eqnarray}

From Eqs.\ \ref{DYrat} and \ref{ptrat} we see that if 
$\bar{d}^p(x) = \bar{u}^p(x)$
for some $x=x_e$, then $R_{sea}(x_1,x_e) = 1$.   
For large $x$, the ratio $r(x)$ is small (the probability of
finding an up quark in the proton at large $x$ is significantly 
higher than for finding a down quark).  In that case, 
as $x_1 \rightarrow 1$, we have
\begin{eqnarray}
R_{sea}(x_1,x_2) \,&{CS\atop \longrightarrow}&\, {1\over 2}
   \left( 1 + \bar{R}(x_2)\right)\left(1 + {r(x_1)\over 4}
   \left[1- \bar{R}(x_2)\right] \right) \quad ,
\label{DYrapp}
\end{eqnarray}

From Eq.\ \ref{DYrapp}, we see that the ratio of DY cross
sections in this kinematic region should directly measure
the ratio of the down antiquark to up antiquark 
distributions in the proton, at a given value of 
$x_2$.  Neglecting terms of order $r(x_1)$, the quantity 
$R_{sea}(x_1,x_2)$ would be less than one if $\bar{d}^p(x_2) < 
\bar{u}^p(x_2)$, and greater than one if $\bar{d}^p(x_2) > 
\bar{u}^p(x_2)$.  There have been three recent experiments 
which enable us to extract the difference between down and
up antiquark distributions in the proton.  

The experiment E772 at FNAL \cite{garvey} measured Drell-Yan processes
for 800 GeV protons on a variety of nuclear targets.  The
targets employed were D, C, Ca, Fe, and W.  As there
was no proton target, the $D/p$ ratio could be inferred
by comparing the isoscalar targets with those for which
$N \ne Z$.  For non-isoscalar targets the excess neutron fraction
is proportional to $\epsilon = (N-Z)/A$.  As $\epsilon$ 
is rather small for these targets, it is difficult to measure 
neutron/proton differences, and hence hard to isolate any 
difference between down and up antiquarks in the proton.  
The experimental results appeared to show differences 
between $\bar{d}^p$ and $\bar{u}^p$ \cite{garvey}, and did 
disagree with some theoretical suggestions for 
$\bar{d}^p/\bar{u}^p$, but it was hard to draw any firm
conclusions regarding this question from the E772 experiment.  

Experiment NA51 at CERN \cite{Bal94} measured Drell-Yan
processes for 450 GeV protons on proton and deuteron
targets.  The NA51 data looked primarily at symmetric
kinematics $x_1 = x_2$.  The symmetric geometry is 
particularly good for minimizing errors in comparison of
different experiments.  For this geometry 
the approximations used to generate Eq.\ \ref{DYrat} are not
valid.  Instead, Ellis and Stirling \cite{Ell91} showed that
for symmetric kinematics the ratio of Drell-Yan cross
sections could be written as 
\begin{eqnarray}
R_{sea}(x_1=x_2=x) \,&{CS\atop =}&\, {\sigma^{pD}_{DY}(x,x)\over 
  2\sigma^{pp}_{DY}(x,x)} = {1\over 2}
   \left( 1 + { {5\over 8}\left( \bar{R}(x)+ r(x)\right) \over 
   \left(1+ r(x)\bar{R}(x)/4 \right) }\right) \quad ,
\label{DYeql}
\end{eqnarray}

The NA51 group obtained a ratio for a single averaged 
point $<x> = 0.18$.  From their measured asymmetry in
Drell-Yan processes, they extracted the value 
\begin{equation}
{\bar{u}^p\over\bar{d}^p}|_{<x> =0.18} = 0.51 \pm 0.04 
(stat) \pm 0.05 (syst) 
\end{equation}
The NA51 result, although for only a single average $x$ value, 
suggests that there are twice as many down antiquarks as up
antiquarks at $x= 0.18$.  

\begin{figure}
\centering{\ \psfig{figure=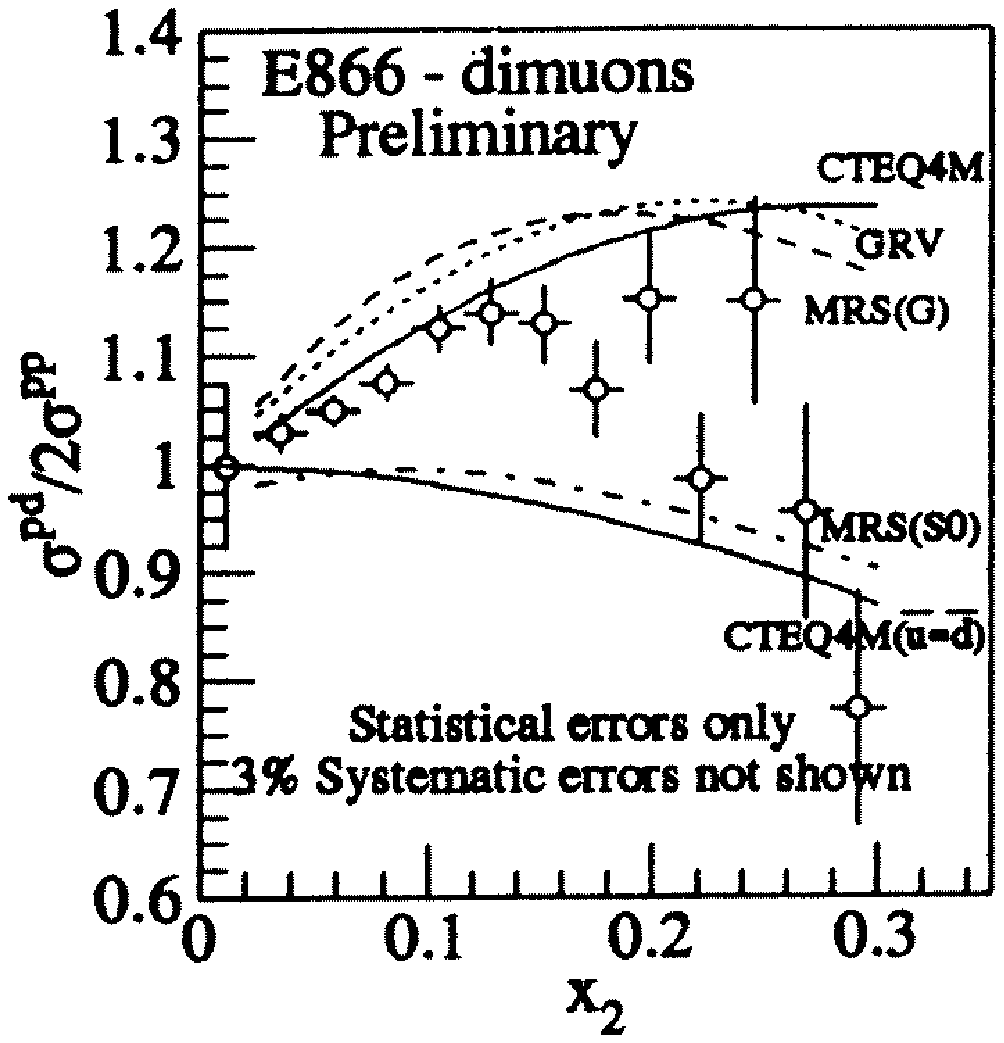,width=7cm}}
\vspace{0.15truein}
\caption{The ratio of pD to pp DY cross sections as a function of 
target $x_2$, preliminary results of the E866 Collaboration, Ref.\ 
\protect\cite{Haw97}. Curves are for phenomenological parton 
distributions.}
\vspace{0.1truein}
\label{fig42a}
\end{figure}

The E866 group at FNAL \cite{Haw97} compared Drell-Yan
processes for 800 GeV protons on liquid hydrogen and Deuterium
targets.  The E866 experiment has both the high statistics and
the wide kinematic range which made it difficult for prior 
experiments to investigate this issue.  In Fig.\ \ref{fig42a} 
we show preliminary results from E866 
for the ratio of DY cross sections $\sigma^{pD}/2\sigma^{pp}$, for 
positive $x_F = x_1- x_2$.  In this kinematic region we expect the 
antiquarks to come predominantly from the target.  For lower values
of $x_2$ the ratio is greater than one, and (with large errors) appears
to decrease at higher values of $x_2$.  
Where the ratio exceeds one, from Eq.\ \ref{DYrapp} this implies
$\bar{d}^p(x) > \bar{u}^p(x)$.  Furthermore, with this data 
one can map out the difference between $\bar{d}^p(x)$ and 
$\bar{u}^p(x)$ over a substantial region of $x$.  

The curves in Fig.\ \ref{fig42a} are from phenomenological parton
distributions \cite{Lai96,Mrs95,Grv95}.  The upper curves allow 
$\bar{d}^p(x) \ne \bar{u}^p(x)$, while the lower curves constrain 
the up and down sea quark distributions to be identical.  For 
$x_2 < 0.2$, the ratio of DY cross sections is clearly greater than
one.  In the following subsection we discuss the implications of the
E866 results, and we examine several theoretical models
which might generate large flavor symmetry violation in
the nucleon sea. 

\subsection{Implications of Large Flavor Symmetry
Violation from Drell-Yan Experiments} 
      
\begin{figure}
\centering{\ \psfig{figure=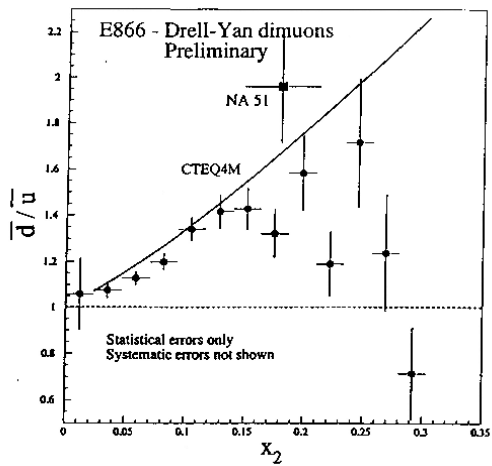,width=7cm}}
\vspace{-.1truein}
\caption{The ratio $\bar{d}^p(x)/\bar{u}^p(x)$ extracted 
from preliminary E866 results of Fig.\ \protect\ref{fig42a}.}
\vspace{0.1truein}
\label{fig42ab}
\end{figure}

The experimental Drell-Yan results appear to show a
substantial violation of flavor symmetry in the proton
sea.  The preliminary results of FNAL experiment E866 
\cite{Haw97} are most definitive in this regard, as they 
can extract $\bar{d}^p(x)/\bar{u}^p(x)$  
over a substantial range of $x$.  In Fig.\ \ref{fig42ab} 
we plot the ratio $\bar{d}^p(x)/\bar{u}^p(x)$ vs.\ $x$ which 
has been extracted from the (preliminary) E866 data \cite{Haw97}.  

The FSV contribution seen in the Drell-Yan experiments is
surprisingly large, as it is much larger than can be
accommodated by perturbative QCD.  Both NLO and NNLO
QCD calculations have been carried out, and predict 
very small FSV
effects.  Ross and Sachrajda \cite{sachrajda} showed that 
the flavor symmetry violating contribution which arises
from higher order QCD evolution is very small  
(for a comprehensive review of theoretical
estimates of FSV, see the review by Kumano \cite{Shunzo}).  
Consequently, we need a non-perturbative mechanism to
generate flavor-violating sea quark distributions which 
will reproduce the experimental result.  

Several authors (see the review by Kumano \cite{Shunzo} for
references)
have investigated mechanisms for producing a large
excess of $\bar{d}$ over $\bar{u}$ in the proton.  Since there
are two valence up quarks and one valence down quark in
the proton, the Pauli principle should make it easier to form
a $d\bar{d}$ pair than a $u\bar{u}$ pair in the presence of
the valence quarks (one should be somewhat careful of 
statements like this: a recent paper by Steffens and Thomas 
\cite{steffens} suggests that $u\bar{u}$ pairs may in fact
be favored if all antisymmetrization terms are considered).  
In Feynman and Field's early paper on
parton distributions \cite{feyfie}, they assumed an excess of 
$\bar{d}$ quarks in the proton on these grounds.  

\begin{figure}
\centering{\ \psfig{figure=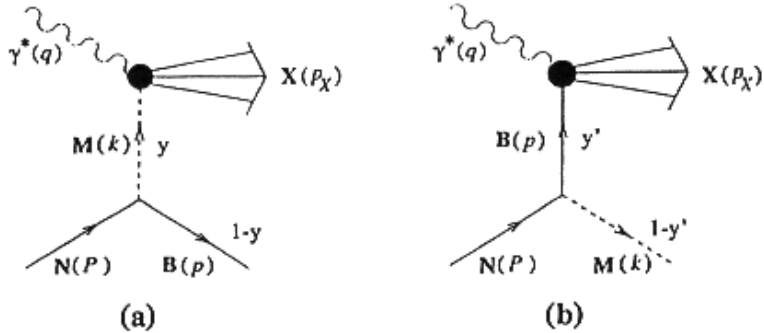,width=11cm}}
\vspace{-.1truein}
\caption{Schematic picture of ``mesonic'' contributions to DIS for
a nucleon. a) virtual meson contribution; b) recoil baryon contribution.}
\vspace{0.1truein}
\label{fig42c}
\end{figure}

Another mechanism for generating additional $\bar{d}$ quarks
in the proton is due to the ``Sullivan Effect'' \cite{sullivan},
whereby the virtual photon can couple to a meson created
by a quark.  Thomas \cite{tony83} pointed out in 1983 that 
``pionic'' effects would produce an excess of $\bar{d}$ over 
$\bar{u}$ in the proton.  The basic mechanism is shown in 
Fig.\ \ref{fig42c}.  A proton fragments into a neutron and a $\pi^+$.  
If the virtual photon scatters from the down antiquark in the $\pi^+$, 
it will produce an excess of $\bar{d}$ over $\bar{u}$.  

Several authors have shown that effects given by the ``Sullivan''
process can produce an excess of $\bar{d}$ quarks in the 
proton sea.  Detailed reviews of this process, and of
the literature on this subject, are given by Kumano \cite{Shunzo},
and by Speth and Thomas \cite{STadv}.   
The original papers \cite{tony83} considered the contributions
to this process from excess pions in the nucleus.  Kumano
and Londergan \cite{Kum91} calculated a model which included 
contributions from pions, nucleons and $\Delta$ isobars (the
$\Delta$ contribution tends to cancel the nucleon-only
contribution at larger $x$).   
This relatively simple model
has been expanded by the Adelaide \cite{mst} and J\"ulich
\cite{Szczurek} groups to include all of the meson and baryon
states normally associated with ``meson-exchange'' models.   

The ``mesonic'' models look promising, in that both the
magnitude of this effect, and the $x$-dependence of the 
predicted $\bar{d}^p/\bar{u}^p$ difference, are in qualitative
agreement with experiment.  In Fig.\ \ref{fig42ab}, the 
solid curve is the CTEQ4M phenomenological parton distribution 
\cite{Lai96}.  This is quite similar to the `mesonic' model result 
of the J\"ulich group \cite{Szczurek} for the ratio of down to up
antiquark distributions.  The model of Kumano
and Londergan \cite{Kum91}, which has only $\pi$ and $\Delta$ in 
addition to the nucleon, is similar to the CTEQ4M prediction at small
$x$ but gives a smaller ratio at large $x$.  Both mesonic calculations 
correctly predict
the excess $\bar{d}^p(x) > \bar{u}^p(x)$, and both get the
general shape of the down antiquark excess as a function of
$x$.  At larger $x$ the error bars are sufficiently large 
that detailed comparisons are difficult.  Furthermore, the
mesonic models are very sensitive to small
changes in the $\pi N\Delta$ coupling constant, and to
the shapes of the $NN\pi$ and $N\Delta\pi$ form factors.    

There are also other theoretical models which predict an
excess of $\bar{d}$ antiquarks in the proton.  
Eichten, Hinchliffe and Quigg \cite{ehq} have investigated the 
contribution from a model in which quarks couple chirally to pions 
\footnote{We note that this tends to overestimate the asymmetry as it
overlooks constraints on the quark states available in the hadron 
spectator.}.  
Dong and Liu \cite{kfliu} estimate the contributions
from mesons in lattice gauge calculations.  They try to 
separate the contributions from ``cloud'' antiquarks and 
``sea'' antiquarks in a lattice calculation.  It is not possible
to make a precise separation on the lattice, but their calculations 
also suggest an excess of down antiquarks relative to up antiquarks.  
All of this work is summarized in Kumano's review article 
\cite{Shunzo}.  One additional possibility is that instanton 
condensates in the nucleon \cite{Shu93,Sch94} might produce an excess of 
down sea quarks relative to up sea quarks in the proton.  

The Drell-Yan experiments appear to show conclusively a
large violation of flavor symmetry in the proton sea.  
However, it is important to note that {\em all these results
depend on the assumption of parton charge symmetry}.  If one
relaxes this assumption, {\em one could in principle reproduce 
the Drell-Yan results even if flavor symmetry is exact}.  From 
Eq.\ \ref{sigDYpd}, let us assume exact flavor symmetry in
the proton sea, i.e. 
\begin{eqnarray}
\bar{u}^p(x) &=& \bar{d}^p(x) \equiv \bar{q}^p(x) \nonumber \\ 
\bar{u}^n(x) &=& \bar{d}^n(x) \equiv \bar{q}^n(x) \nonumber \\ 
\bar{q}^p(x) &\ne & \bar{q}^n(x) 
\label{DYncs}
\end{eqnarray}
The parton distributions of Eq.\ \ref{DYncs} are completely
flavor symmetric but not charge symmetric.  If we go to 
the region $x_F > 0$ we find that Eq.\ \ref{DYrapp} now becomes
\begin{eqnarray}
R_{sea}(x_1,x_2) \,&{FS \atop \rightarrow}&\, {1\over 2}
   \left( 1 + {\bar{q}^n(x_2)\over \bar{q}^p(x_2)} \right) \quad .
\label{DYnovp}
\end{eqnarray}
The Drell-Yan experiments could thus be reproduced, even if
flavor symmetry was exact, with a sufficiently large violation
of charge symmetry in the parton distributions.  
It would require an astonishingly large CSV contribution in 
the nucleon sea to
reproduce the E866 results: this would be a factor 25-50 larger
than our estimates in Sect.\ 3.  Alternatively, the E866 
results could be due to a linear combination of FSV and CSV effects 
in the nucleon sea.  

In the next subsections, we will review the experimental
constraints on charge symmetry in parton distributions.  In
Sect.\ 5, we will suggest a number of new experiments which
might provide more stringent tests of parton 
charge symmetry.   

\subsection{The ``Charge Ratio:'' Comparison of Muon with 
Neutrino Induced Structure Functions}

In Sect.\ 2.5, we derived the relation between the structure
function $F_2$ measured in neutrino induced charged current reactions, 
and the $F_2$ structure function for charged lepton DIS, both measured
on isoscalar targets.  From Eq.\ \ref{chgrat}, at sufficiently high 
energies the structure functions have the form
\begin{eqnarray}
F_2^{\gamma N_0}(x)  &=& x \left[ {5\over 18}\,\overline{Q}(x) - 
 {1\over 6} \left(s(x) + \bar{s}(x)\right) - {4\delta d(x) + 
 4\delta \bar{d}(x) + \delta u(x) + \delta \bar{u}(x) \over 18}
 \right] \nonumber \\ \overline{F_2}^{W N_0}(x) &=& x\left[ 
 \overline{Q}(x) - {\delta d(x) + 
 \delta \bar{d}(x) + \delta u(x) + \delta \bar{u}(x) \over 2} \right]  
 \nonumber \\ \overline{Q}(x) &=& \sum_{j=u,d,s,c} q_j(x) + 
 \overline{q}_j(x) \nonumber \\ 
 \widetilde{Q}(x) &\equiv& \overline{Q}(x) - {3\left( s(x) + 
 \bar{s}(x) \right) \over 5} 
\label{F2gamm}
\end{eqnarray}
In Eq.\ \ref{F2gamm} we have neglected the charmed quark
contribution to the structure functions, and for the moment we have set 
$R=0$.  The function $\overline{F_2}^{W N_0}(x)$ is the average of
the neutrino and antineutrino induced charged current structure
functions.  

From Eq.\ \ref{F2gamm} we see that there is a simple relation between 
the two structure functions, in the limit of exact charge symmetry.  The 
ratio of the two structure functions in Eq.\ \ref{F2gamm}, 
when corrected for the strange quark contribution and the factor
$5/18$ (which reflects the fact that the virtual photon couples to
the squared charge of the quarks while the weak interactions couple
to the weak isospin), is defined as 
the ``charge ratio'' $R_c(x,Q^2)$ or, as it is sometimes termed, the 
``5/18$^{th}$ rule.''  This quantity should
be one, independent of $x$ and $Q^2$, in the naive parton
model.  If we expand the ratio $R_c$ to lowest order in the 
presumably small charge symmetry violating terms, we obtain 
\begin{eqnarray}
R_c(x) &\equiv& 
 {F_2^{\gamma N_0}(x) \over {5\over 18} \overline{F_2}^{W N_0}(x)
 - x\left( s(x) + \bar{s}(x) \right)/6 } \nonumber \\ 
 &\approx& 1 + {3 \left( \delta u(x) + \delta \bar{u}(x) - \delta d(x) 
 - \delta \bar{d}(x)\right) \over 10\widetilde{Q}(x) }
\label{Rc}
\end{eqnarray}
As we pointed out in Section 2.6, the strange quark
distributions can be obtained independently by measuring
opposite sign dimuon events in neutrino DIS from nuclei.  
Using these strange quark distributions in Eq.\ \ref{Rc}, 
and comparing the $F_2$ structure functions for lepton-induced processes 
with the $F_2$ structure functions from weak processes mediated by
$W$-exchange, we can in principle measure parton charge symmetry 
violation and determine its $x$ dependence.  By measuring
$R_c(x)$ we can place upper limits on parton CSV as a function of $x$.  
The longitudinal/transverse ratio $R$ can be included in forming
the structure functions, and will cancel when the ratio $R_c$ is 
taken.  

Eq.\ \ref{Rc} requires averaging the $F_2$ structure functions for 
neutrino and antineutrino cross sections.  If we instead take the
ratio using only neutrino-induced reactions, it is straightforward
to obtain 
\begin{eqnarray}
R_c^\nu(x) &\equiv& 
 {F_2^{\gamma N_0}(x) \over {5\over 18} F_2^{W^+ N_0}(x)
 - x\left( s(x) + \bar{s}(x) \right)/6 } \nonumber \\ 
 &\approx& 1 - {s(x)-\bar{s}(x)\over \widetilde{Q}(x)} 
 + {\left( 4\delta u(x) - \delta \bar{u}(x) - 4\delta d(x) 
 + \delta \bar{d}(x)\right) \over 5\widetilde{Q}(x) }
\label{Rc2}
\end{eqnarray}
Eq.\ \ref{Rc2} differs from Eq.\ \ref{Rc} since it has a term 
proportional to the difference between strange and antistrange quark
distributions, and also in the relative weighting of the CSV terms which
enter.  The $s-\bar{s}$ term is absent if one is able to average 
neutrino and antineutrino cross sections.   

The charge ratio test allows us to place the strongest limits to
date on parton CSV.  There should be no additional QCD corrections to this
relation so it should be independent of $Q^2$, {\em provided} that 
the structure functions are calculated in the so-called ``DIS scheme.''  
In this scheme, the $F_2$ structure functions are defined so that
they have the form $F_2(x) = x\sum_i e_i^2 [q_i(x) + \bar{q}_i(x)]$ to 
all orders, where 
$e_i$ is the quark charge appropriate for either the electromagnetic
or weak interactions.  For example, the CTEQ4D parton distributions 
\cite{Lai96} were determined in the DIS scheme. Despite the robustness of 
the charge ratio test, it also depends on a
large number of assumptions and corrections, which must be taken into
account to obtain limits on CSV terms.  Among these corrections
are: 
\begin{itemize}
\item{} Relative normalization between leptonic and neutrino cross
sections. 
\item{} Corrections due to strange quarks.  As outlined in Sect.\ 
2.6, $s(x)$ ($\bar{s}(x)$) can be independently extracted from the
cross section for opposite sign
dimuons from reactions induced by neutrinos (antineutrinos)
\item{} Corrections from excess neutrons.  Eq.\ \ref{Rc} was derived
for isoscalar targets.  In order to obtain reasonable cross sections, 
neutrino reactions are now measured on iron targets.  This requires
a correction for the excess neutrons in the target.  
\item{} Heavy target corrections.  If the leptonic structure functions
are obtained from light targets and neutrino reactions performed
on heavy targets, it is necessary to correct the neutrino $F_2$ 
structure functions for heavy target effects.  At low and intermediate
$x$, heavy target structure functions are decreased because of shadowing 
and EMC effects, respectively; at very large $x$ Fermi motion effects
increases the structure functions for heavy targets. 
\item{} Higher twist effects on parton distributions. 
\item{} Heavy quark threshold effects.  At sufficiently low energies, 
heavy quark threshold effects will modify structure functions, as 
we reviewed in Sect.\ 2.2.  
\end{itemize}

\begin{figure}
\centering{\ \psfig{figure=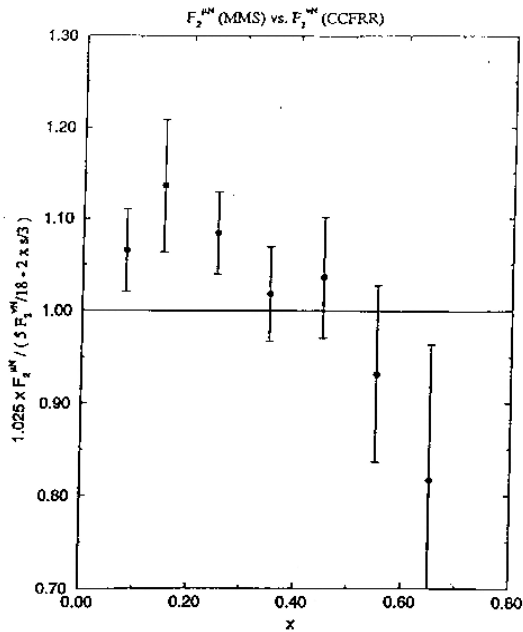,width=9cm}}
\vspace{-.1truein}
\caption{Charge ratio $R_c^\nu(x)$ of Eq.\ \protect\ref{Rc2} vs.\ $x$.  
Muon measurements of Meyers et al, Ref.\ \protect\cite{Meyers} 
[using muons on Fe at FNAL], with CCFRR neutrino measurements on iron, 
Ref.\ \protect\cite{Macfar}.}
\vspace{0.1truein}
\label{fig45}
\end{figure}

In Fig.\ \ref{fig45} we plot the charge ratio $R_c^\nu$, 
i.e.\ the ratio of muon
$F_2$ structure functions measured by Meyers {\it et
al.} on iron \cite{Meyers} to the value of $F_2$ extracted
from the CCFRR neutrino measurements \cite{Macfar}.  
The muon measurements were taken at FNAL with 93 and 215 
GeV muons, using the multimuon spectrometer at FNAL.  The CCFRR 
neutrino measurements were made with the
FNAL narrow-band neutrino beam.  

In comparing the muon and neutrino measurements, the following 
corrections were made by Meyers {\it et
al.}.  First, the $F_2$ structure functions were 
modified by including the strange quark contribution, determined 
as described in Section 2.6 \cite{Baz95}.  Second, corrections were made 
for the excess neutrons in iron.   
Third, there was a discrepancy in the extraction of the
$F_2$ structure functions.  The muon data were analyzed 
assuming longitudinal/transverse ratio $R=0$, while the
neutrino data assumed $R=0.1$.  Meyers {\it et al.} 
corrected the muon data to make them consistent.  The muon 
data have been renormalized by the factor 1.025.  

Within the errors (two standard deviations), 
$R_c^\nu$ is consistent with unity, except possibly at
the largest value of $x$ where $R_c^\nu(x  = 0.65) = 0.82 \pm 
0.09$.  From these experiments, the upper limits on the CSV contribution 
to $R_c$ are generally no better than about 10\%, and at large values of 
$x$ the errors are significantly larger.  The experimental data is 
consistent with zero charge symmetry violation and certainly rules out 
any extremely large violation of parton charge symmetry.  
From Eq.\ \ref{Rc2} and the theoretical calculations of parton CSV 
given in Sect.\ 3, we expect that the CSV contribution
to the charge ratio will not exceed 1-2\% at any value of $x$.  
Consequently, any deviation of the charge ratio from unity, at any value
of $x$, would be surprising and very interesting.  

\begin{figure}
\centering{\ \psfig{figure=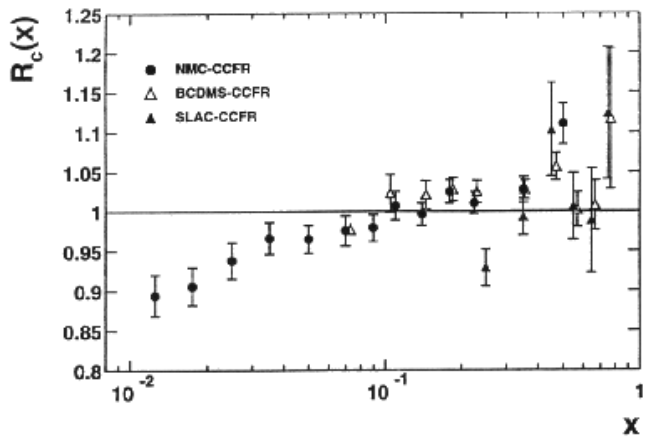,width=10cm}}
\vspace{-.1truein}
\caption{Charge ratio $R_c^\nu(x)$ of Eq.\ \protect\ref{Rc2} vs.\ $x$.  
CCFR neutrino measurements on iron with the FNAL wide-band 
neutrino beam, Ref.\ \protect\cite{CCFR}. Circles: 
$\mu +D$ measurements of the NMC group, 
Ref.\ \protect\cite{Ama91}; open triangles, muon measurements from
the BCDMS group on deuterium, Ref.\ \protect\cite{Ben90} and carbon, 
Ref.\ \protect\cite{BCDMS}; solid triangles: SLAC electron scattering
data, Ref.\ \protect\cite{Whi90,Whi90b}.}
\vspace{0.1truein}
\label{fig45a}
\end{figure}

More recent data for both muons and neutrinos allows us to make
substantially more precise tests of parton charge symmetry.  The
NMC group \cite{Ama91} measured the $F_2$ structure function for
muon interactions on deuterium at energy $E_\mu = 90$ and 280 GeV.  
The CCFR group \cite{Sel97} has extracted the $F_2$ structure
function for neutrino and antineutrino interactions on iron using
the Quadrupole Triplet Beam at FNAL.  The CCFR measurements provide
the most copious sample of neutrino events, and allow the most
precise limits on parton CSV.  In Fig.\ \ref{fig45a} we plot the 
charge ratio $R_c$ of Eq.\ \ref{Rc} vs.\ $x$.  The circles are
the NMC/CCFR ratio.  The open triangles are the BCDMS/CCFR charge
ratio, where BCDMS represents the muon scattering results of the
BCDMS group on deuterium \cite{Ben90} and carbon \cite{BCDMS}.  The
solid triangles are the SLAC/CCFR charge ratio, where SLAC denotes 
electron scattering results of the SLAC group \cite{Whi90,Whi90b}. 

The charge ratio has been calculated by C. Boros \cite{Bor97}.  
The results differ somewhat from those produced by Seligman {\it et al.} 
in their calculation of the charge ratio \cite{Sel97,Sel97a}.  In 
comparing the
data sets, Boros takes only those points with the same $x$ value and 
sums over overlapping $Q^2$ values, while Seligman interpolates
between measured values of the structure functions.  In addition, 
in Fig.\ \ref{fig45a} there is no correction for strange quarks.  

In the region $0.1 \le x \le 0.4$, the charge ratio test is 
consistent with unity, and the data gives an upper limit to CSV effects 
in the charge ratio at about the 3\% level.  For larger values of
$x$ the upper limit on CSV effects is more consistent with the
5-10\% level, due mainly to the poorer statistics and, as we will
see, on the large Fermi motion corrections needed for the heavy 
target at large $x$.  Both the new muon and neutrino data are 
more precise than the older measurements.  In addition, the more recent
phenomenological parton distributions are better determined.  
Relative normalizations of lepton and neutrino cross sections appear to 
be well understood.  All data is analyzed with consistent assumptions
about the longitudinal/transverse ratio $R$.  Heavy  
quark threshold effects should also be under control.  

\begin{figure}
\centering{\ \psfig{figure=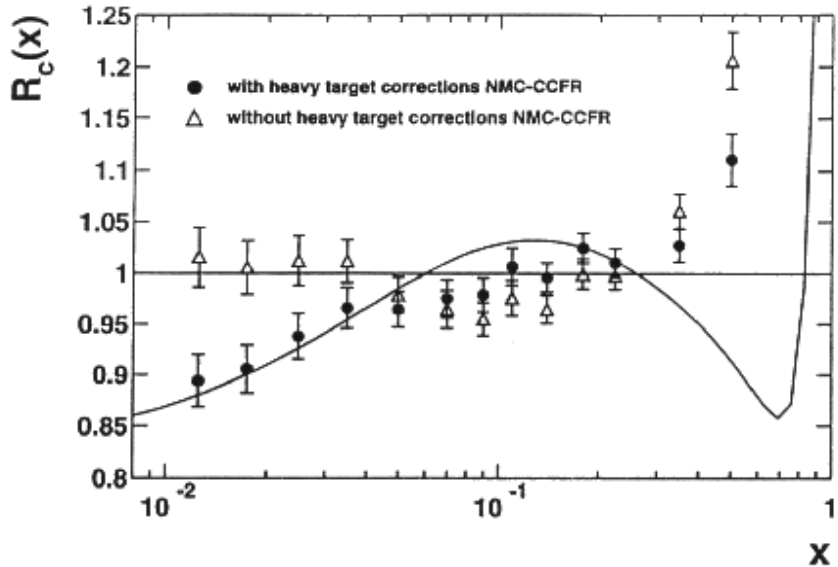,width=13cm}}
\vspace{0.15truein}
\caption{Effects of heavy target correction on charge ratio of Fig.\ 
\protect\ref{fig45a}.  Triangles: NMC-CCFR data without heavy target
correction; circles: NMC-CCFR data including heavy target correction 
(solid curve).}
\vspace{0.1truein}
\label{fig45ab}
\end{figure}

Probably the most significant correction is the heavy target correction, 
necessary because we are comparing muon data on deuterium, where the
correction is presumably very small, to neutrino data on iron.  In 
Fig.\ \ref{fig45ab} we show the same charge ratio as in Fig.\ 
\ref{fig45a} for the NMC-CCFR comparison, but here we explicitly show the 
heavy target corrections.  The open triangles show the ratio without
heavy target corrections, and the solid circles show the ratio after
applying these corrections.  The solid line is the iron target
correction factor as a function of $x$.  
  
After including the heavy target correction, there appears to be a
significant deviation of the charge ratio from one at the smallest
values $x < 0.1$; the discrepancy approaches 15\% at the smallest
values of $x$, with the electromagnetic structure functions being
smaller than the neutrino ones.  Several suggestions have been
made to explain this discrepancy.  We summarize the explanations 
listed by Seligman \cite{Sel97}.  First, the lack of agreement could
result from difficulties in analyzing the low-$x$ neutrino events.  
This will be tested with the next generation FNAL neutrino experiment E815.  
Second, it is conceivable that the disagreement arises from effects
at small $Q^2$ which differ between leptonic and neutrino induced
reactions \cite{Don94}.  However, these effects appear to be quite small 
for $Q^2 > 1$ GeV$^2$ \cite{Zhu96}. 

The discrepancy increases monotonically at small $x$, where the strange
quark effects are largest.  One intriguing possibility is that strange
quark effects might account for all of the apparent discrepancy.  In
this case it is possible that the present phenomenological analysis of 
both strange quark and antiquark distributions need to be modified 
substantially, as has recently been argued by Brodsky and Ma \cite{Bro96}. 
In any case, the recent NMC-CCFR comparison allows us to put rather
tight upper limits on parton CSV contributions, and focuses our
attention on the low-$x$ region where there is currently a discrepancy
between the $F_2$ structure functions extracted from the two
reactions. 

\subsection{Comparison of Neutrino and Antineutrino Cross Sections
on Isoscalar Targets}

On an isoscalar target, the differential cross sections
for charged current reactions induced by neutrinos or
antineutrinos can be written in the general form
\begin{eqnarray}
{d\sigma^{\nu N_0}\over dx} &=& {G^2 M_N E\over \pi}
  \left[ \left({1\over 2} + {1\over 6(1+R)} \right)
  F_2^{W^+ N_0}(x,Q^2) + {1\over 3} xF_3^{W^+ N_0}(x,Q^2)
  \right] \nonumber \\
{d\sigma^{\bar\nu N_0}\over dx} &=& {G^2 M_N E\over \pi}
  \left[ \left({1\over 2} + {1\over 6(1+R)} \right)
  F_2^{W^- N_0}(x,Q^2) - {1\over 3} xF_3^{W^- N_0}(x,Q^2)
  \right] .  
\label{signnbar}
\end{eqnarray}
In Eq.\ \ref{signnbar}, the quantity $R$ is the 
longitudinal/transverse ratio, 
\begin{equation}
 R(x, Q^2) = {\sigma_L\over \sigma_T} = {F_2(x,Q^2) - 
  2x\,F_1(x,Q^2)\over 2x\,F_1(x,Q^2)} .
\label{Rdeff}
\end{equation}

From Eq.\ \ref{F1Nzero} we see that the structure functions 
$F_2$ and $F_3$ per nucleon for an isoscalar target 
can be written as
\begin{eqnarray}
  F_2^{W^+ N_0}(x) &=& x\left[ u^p(x) + d^p(x) + \bar{u}^p(x) 
  + \bar{d}^p(x) + 2s(x) - \delta u(x) - \delta\bar{d}(x)
  \right] \nonumber \\ 
  F_2^{W^- N_0}(x) &=& x\left[ u^p(x) + d^p(x) + \bar{u}^p(x) 
  + \bar{d}^p(x) + 2\bar{s}(x) - \delta d(x) - \delta\bar{u}(x)
  \right] \nonumber \\ 
  xF_3^{W^+ N_0}(x) &=& x\left[ u^p(x) + d^p(x) - \bar{u}^p(x) 
  - \bar{d}^p(x) + 2s(x) - \delta u(x) + \delta\bar{d}(x)
  \right] \nonumber \\ 
  xF_3^{W^- N_0}(x) &=& x\left[ u^p(x) + d^p(x) - \bar{u}^p(x) 
  - \bar{d}^p(x) - 2\bar{s}(x) - \delta d(x) + \delta\bar{u}(x)
  \right] .
\label{isoF}
\end{eqnarray}
In Eq.\ \ref{isoF}, we assume that the momentum transfers
are sufficiently high that threshold effects can be
neglected.  In this equation, we have neglected the contribution
from charmed quarks in the nucleon, and for the moment we have set $R=0$.  
In this limit, the $F_2$ structure
functions from neutrinos and antineutrinos are identical except
for CSV contributions.  In addition, the $F_2$ and $F_3$ structure
functions are identical except that the antiquark contributions
have different signs.  Consequently, if we go to large $x$
where the sea quark contributions become small with respect to 
the valence quark terms, then both $F_2$ and $F_3$ structure
functions for both neutrinos and antineutrinos should become
equal.  From Eq.\ \ref{signnbar} we see that $F_2$ and $F_3$ 
will add together in the neutrino cross section, but will 
cancel in the antineutrino cross section.  

We thus define the ratio of antineutrino to neutrino charged
current cross sections on an isoscalar target, 
\begin{equation}
r^{\nu/\bar\nu}(x) \equiv {d\sigma^{\bar\nu N_0}(x)/dx \over
  d\sigma^{\nu N_0}(x)/dx}
\label{nurat}
\end{equation}
We will focus on this relation at reasonably large values of 
$x \geq 0.3$.  For these values of $x$ the sea quark 
distribution will be small relative to the valence quark
distributions.  In this region we can expand the ratio 
of Eq.\ \ref{nurat} to lowest order
in small quantities, and we obtain
\begin{eqnarray}
r^{\nu/\bar\nu}(x) &\approx& {1\over 3} + {8 \left( \bar{u}^p(x) 
  + \bar{d}^p(x)\right) \over 9 \left( u^p_v(x)+  d^p_v(x) 
  \right)}+ { \delta u(x) - \delta d(x) \over 3 
  \left( u^p_v(x)+  d^p_v(x) \right)} + {4 s(x)\over 3 
  \left( u^p_v(x)+  d^p_v(x) \right)}
    \quad [R=0] \nonumber \\ 
r^{\nu/\bar\nu}(x) &\approx& {21\over 65} + {3784 \left( \bar{u}^p(x) 
  + \bar{d}^p(x)\right) \over (65)^2 \left( u^p_v(x)+  d^p_v(x) 
  \right)}+ {21 \left(\delta u(x) - \delta d(x) \right) \over 65 
  \left( u^p_v(x)+  d^p_v(x) \right)} \nonumber \\ &+& {88 s(x)\over 65 
  \left( u^p_v(x)+  d^p_v(x) \right)}   \quad [R=0.1] 
\label{rnudef}
\end{eqnarray} 
In Eq.\ \ref{rnudef}, we treat $R$ as a constant (we use the
value averaged over $x$).  If $R=0$, the ratio 
$r^{\nu/\bar\nu}(x)$ is predicted to approach the value 
1/3 at large values of $x$; for $R= 0.1$, the ratio should 
approach 21/65.  $x$-dependent deviations from this constant value 
will arise from either sea quark or CSV contributions.  

\begin{figure}
\centering{\ \psfig{figure=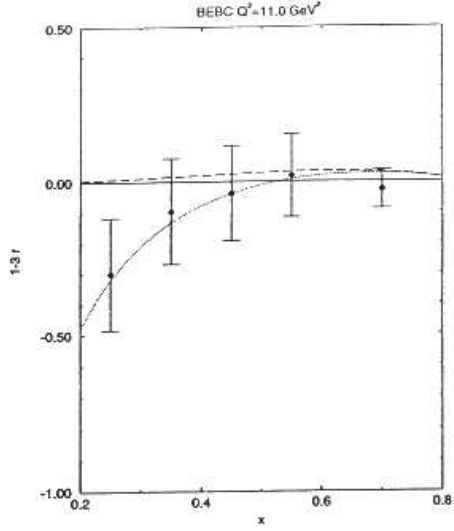,width=7cm}}
\vspace{0truein}
\caption{The quantity $1- 3r^{\nu/\bar{\nu}}$ of Eq.\ 
\protect\ref{nurat}, vs.\ $x$ for neutrino and antineutrino data on BEBC 
D bubble chamber. Data from WA25 experiment, Ref.\ \protect\cite{WA25}.}
\vspace{0.1truein}
\label{fig43}
\end{figure}

At present, this ratio can
be obtained from experimental data on deuterium and iron
targets.  In Fig.\ \ref{fig43} we plot $1- 3r^{\nu/\bar\nu}$ 
vs.\ $x$ for the WA25 data \cite{WA25}.  This consists of neutrino and 
antineutrino cross sections taken in the BEBC D bubble chamber at 
CERN.  The experimental points
plotted have an average momentum transfer $Q^2 = 11$ 
GeV$^2$.  They were analyzed assuming $R=0$.  The dotted 
curve in Fig.\ \ref{fig43} is the total contribution from 
both sea quarks and our model predictions from valence quark
CSV (see Section 3), while the dashed curve is the 
model contribution from CSV alone.  The model prediction 
is small in absolute value, and small relative to the sea
quark contribution, except at large values $x \geq 0.6$, 
where the CSV contribution is predicted to dominate.  
In the quantity $1- 3r$, the sea quarks are weighted
with a factor 8/3 relative to the CSV contribution.  
In this ratio, we predict that the sea quarks will give a
negative contribution to the quantity $1- 3r$, while the CSV
contribution is predicted to be positive.  

The experimental error bars range from about 20\% 
at small $x$ to 50\% at large $x$.  The errors are 
substantially larger than the 
theoretical CSV contribution, for all measured values of 
$x$.  If it were possible
to obtain precise neutrino and antineutrino data on
isoscalar targets such as D, for large values of $x$, the 
quantity $r^{\nu/\bar\nu}(x)$ could in principle give a
sensitive upper limit on parton charge symmetry violation.  
Unfortunately, measurement of these cross sections is
notoriously difficult.  We know of no current plans for
precision measurements of charged current 
cross sections for neutrino and antineutrino beams 
on isoscalar targets.  From Fig.\ \ref{fig43}, we see that 
the experimental error bars would have to be at least an
order of magnitude smaller than their current values to reach 
our predicted CSV signal.  Further, it is doubtful that the
neutrino results could attain the limits on CSV already
reached by the ``charge ratio'' comparison between $F_2$
measurements from muons and neutrinos; this was reviewed in 
Sect.\ 4.3.  

This is unfortunate, since the neutrino comparisons have
fewer implicit assumptions than the charge ratio -- if
accurate structure functions were available for neutrino
and antineutrino bombardment of isoscalar
targets, one would be comparing data taken in the same
experiments, one would not have to make corrections for
excess neutrons, nor would one have to correct for the
strange quark distributions, extracted as described in Sect.\ 2.6.   

\begin{figure}
\centering{\ \psfig{figure=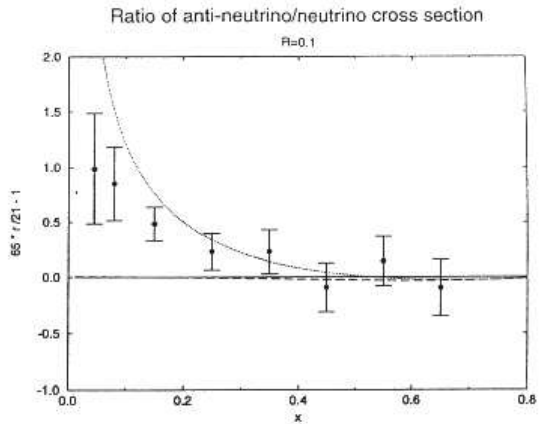,width=8cm}}
\vspace{-.1truein}
\caption{The quantity $1- 65r^{\nu/\bar{\nu}}/21$, vs.\ $x$ for 
CCFRR neutrino and antineutrino data on iron, Ref.\ 
\protect\cite{Macfar}.}
\vspace{0.1truein}
\label{fig43b}
\end{figure}

In Fig.\ \ref{fig43b} we plot $1- 65r^{\nu/\bar\nu}/21$ vs.\ $x$ for 
the CCFRR data, neutrino and antineutrino cross sections on
iron, taken at FNAL \cite{Macfar}.  The plotted points correpond to 
an average momentum transfer $Q^2 = 12.69$ GeV$^2$.  
They were analyzed assuming $R=0.1$.  The dotted curve in 
Fig.\ \ref{fig43b} is the total contribution from 
both sea quarks and our model predictions from valence quark
CSV, while the dashed curve is the model contribution from CSV alone.  
The CCFRR data is similar to the WA25 results shown in Fig.\ \ref{fig43}, 
in that the experimental error bars are much larger than
the theoretical CSV contribution, and the data is 
consistent with no charge symmetry violation.  

The experimental error bars range from about 20\% 
at small $x$ to 100\% at large $x$.  The errors are 
again substantially larger than the 
theoretical CSV contribution, for all measured values of 
$x$.  Since iron is not an isoscalar target, it is necessary
to make corrections from the excess neutrons in iron.  
These corrections have been taken into account in 
Figs.\ \ref{fig43} and \ref{fig43b}.    
There is more recent experimental data for
neutrino and antineutrino cross sections from the 
CCFR collaboration \cite{Sel97}.  We are unaware of a systematic 
study of the quantity $r^{\nu/\bar\nu}(x)$ by this group.  

\section{Proposed New Experimental Tests of Parton Charge
Symmetry}
\mb{.5cm}

In the preceding Section, we reviewed existing experiments
and showed the limits they placed on charge symmetry
and flavor symmetry violation in parton distributions.  The
latest Drell-Yan data for protons on proton and deuteron 
targets appear to show clear evidence of substantial flavor
symmetry violation in the proton sea.  However, it is 
conceivable (although unlikely) that this result could also
be due to charge symmetry violation in parton distributions, 
or by some combination of flavor symmetry and charge symmetry
violation.  In this Section we propose a series of experiments,
all of which are chosen specifically to set limits on
CSV contributions to parton distributions.  

\subsection{Test of Weak Current Relation $F_1^{W^+ N_0}(x) = 
F_1^{W^- N_0}(x)$}

In Sect.\ 2.5, we reviewed the high-energy limiting form
to the electroweak structure functions.  We showed that, at
sufficiently high energies, the charge-changing structure
functions on an isoscalar target are equal except for
contributions from valence quark CSV, and possible strange
or charmed quark terms, i.e.\ 
\begin{equation}
 F_1^{W^+ N_0}(x,Q^2) - F_1^{W^- N_0}(x,Q^2) =
 {\delta d_{\rm v}(x) 
  - \delta u_{\rm v}(x) \over 2}+  s(x) - \bar{s}(x) - 
  c(x) + \bar{c}(x) ,
\label{f1diff}
\end{equation}
as shown in Eq.\ \ref{F1Nzero}.   

At the enormous values of $Q^2$ that can be probed at HERA, weak
interaction processes such as $e^- p \rightarrow \nu_e X$
are not completely negligible with respect to the electromagnetic process
$e^- p \rightarrow e^- X$.  Furthermore, we are well above
heavy quark production thresholds, so threshold effects on the CKM 
matrix elements and issues like ``slow rescaling'' which are 
critical at lower energies, are unimportant in this regime.  
The $(e^-,\nu_e)$ reaction picks out the positively charged partons 
in the target, so
that if one could accelerate deuterons in the HERA ring, the
structure functions for this reaction would have the form 
\begin{equation}
F_1^{W^-D}(x) = \left[ u^p(x) + \bar{d}^p(x) +u^n(x) + 
 \bar{d}^n(x) + 2\bar{s}(x) + 2c(x) \right].
\label{Wmin}
\end{equation}
As in Sect.\ 2, we denote the $(e^-,\nu_e)$ reaction by the
charge of the virtual $W$ absorbed by the target.  We neglect
differences in strange and charm quark distributions between proton 
and neutron.  On the 
other hand, if we have a positron beam the $(e^+,\bar{\nu}_e)$ 
deep inelastic reaction measures only the negatively charged partons, 
so for a deuteron target this would have the form:
\begin{equation}
F_1^{W^+D}(x) = \left[ d^p(x) + \bar{u}^p(x) +d^n(x) + 
 \bar{u}^n(x) + 2s(x) + 2\bar{c}(x) \right].
\label{Wpl}
\end{equation}
Taking the difference of the $e^+$ and $e^-$ charge-changing weak 
interaction cross sections one therefore obtains Eq.\ \ref{f1diff}.  
The difference between the structure functions $F_1^{W^+D}$ and 
$F_1^{W^-D}$ has been studied recently by Londergan, 
Braendler and Thomas \cite{herathy}.  We summarize their
results here.  
 
To indicate the size of expected differences in the $e^\pm D$ 
charge-changing cross sections, we construct the ratio 
\begin{eqnarray}
R_W(x) &\equiv& { 2\left( F_1^{W^+D}(x) - F_1^{W^-D}(x)\right) \over 
  F_1^{W^+D}(x) + F_1^{W^-D}(x)} \cr &=& { \delta d_{\rm v}(x) 
  - \delta u_{\rm v}(x) + 2\left[s(x) - \bar{s}(x)\right] \over 
  \sum_{j=p,n} \left[ u^j(x) + \bar{u}^j(x) + d^j(x) + \bar{d}^j(x) 
  \right] + 2(s(x) + \bar{s}(x)) } \nonumber \\ &=& R_{CSV}(x) + R_s(x) 
\label{Rcsv}
\end{eqnarray}
Direct comparison of the $F_1$ structure functions for
charge-changing weak interactions on an isoscalar target
is a strong test of charge symmetry in parton distributions.  
As we have shown, the only source of difference is either
a difference between strange quark and antistrange quark 
distributions, or charge symmetry violating components
of the valence quark distributions.  Note that Eq.\ \ref{Rcsv} is
unchanged if we use the $F_2$ structure functions rather than
the $F_1$ structure functions: the additional factors proportional
to the longitudinal/transverse ratio $R$ cancel in forming the
ratio.  Eq.\ \ref{Rcsv} is also true for any isoscalar nuclear
target, if we replace the nucleon parton distributions by their 
nuclear counterparts.   

\begin{figure}
\centering{\hbox{ \hspace{-0.1in} \psfig{figure=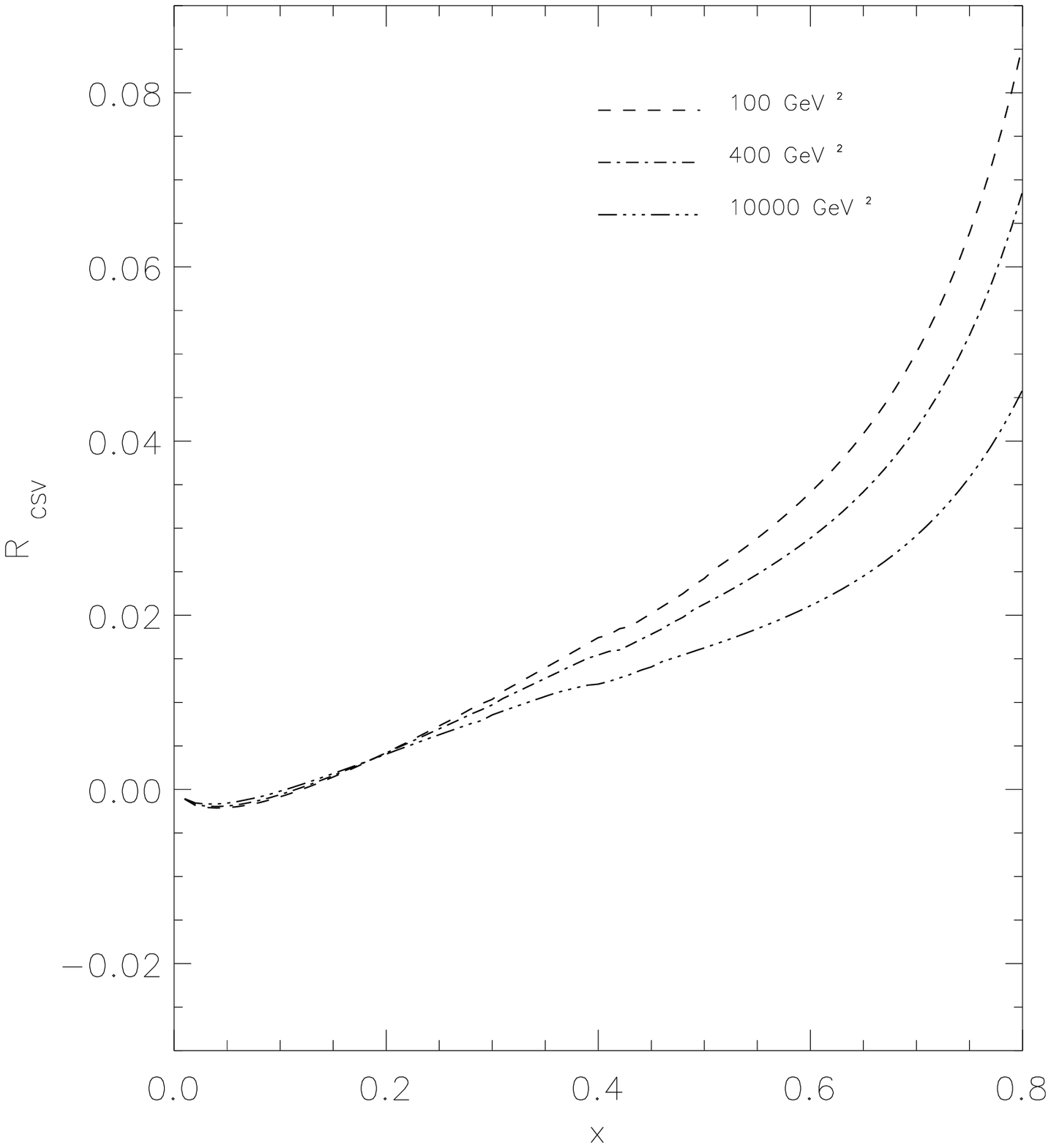,height=7.8cm}
\hspace{0.5in} \psfig{figure=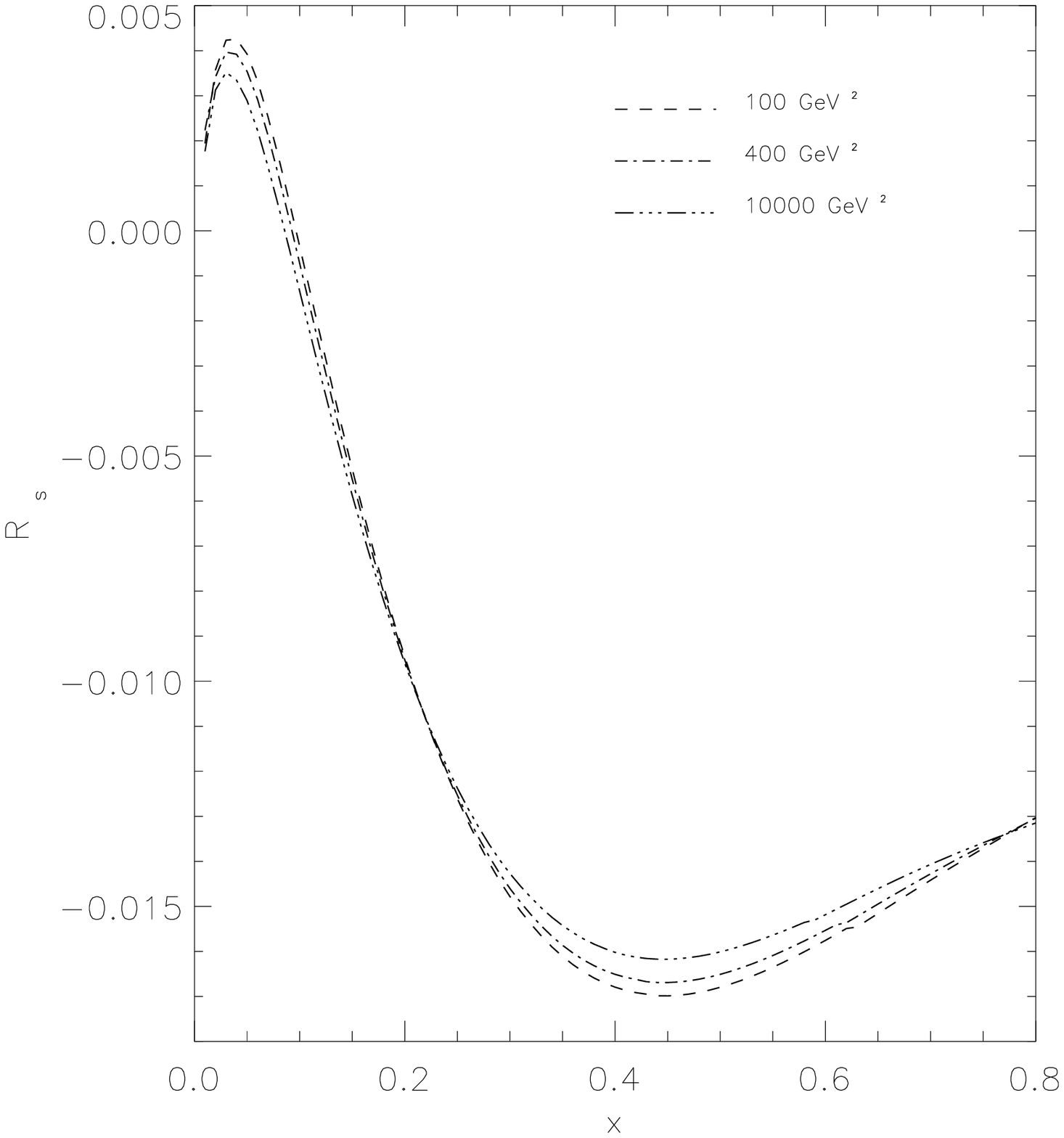,height=7.8cm}}}
\vspace{0.05truein}
\caption{a) Theoretical estimates of CSV contribution $R_{CSV}$ to 
the ratio $R_W(x)$ of Eq.\ \protect\ref{Rcsv} vs.\ $x$, for various 
values of $Q^2$.  b) Theoretical estimates of $s-\bar{s}$ difference, 
$R_s(x)$ of Eq.\ \protect\ref{Rcsv}. From Ref.\ \protect\cite{herathy}.}
\vspace{0.1truein}
\label{fig42cc}
\end{figure}

In order to illustrate the size and shape of the effect expected we
plot in Fig.\ \ref{fig42cc}a the theoretical CSV contribution, 
$R_{CSV}(x)$ from Eq.\ \ref{Rcsv}.  The dashed curve is calculated for 
$Q^2 =100$ GeV$^2$, the dot-dashed curve for $Q^2 =400$ GeV$^2$, and the 
dash-triple dot curve for $Q^2 =10,000$ GeV$^2$.  The CSV quantity 
$R_{CSV}(x)$ is predicted to be greater than 0.02 provided $x > 0.4$.  
The valence quark CSV terms are taken from the work of Rodionov 
{\it et al.}, Ref.\cite{Rod94}, as discussed in Sect.\ 3.1.  
For values $x > 0.1$ we predict that 
$\delta d_{\rm v}(x)$ will be positive and $\delta u_{\rm v}(x)$ negative, 
so their effects should add.  As we mentioned previously, estimates of
valence quark CSV have little model dependence, so we would expect
to see differences of this magnitude and sign at large Bjorken $x$.  
We predict several percent effects at the largest values of $x$, if
the structure functions could be determined in this region.    

The term $R_s$ of Eq.\ \ref{Rcsv} is shown in Fig.\ \ref{fig42cc}b.  
The term $R_s(x)$ is proportional
to the difference between strange quark and antiquark distributions.  
There has been quite a lot of interest recently 
\cite{Sig87,Ji95,Hol96,Bro96,Mel97} in the possibility that 
$s(x) - \bar{s}(x)$ might be non-zero.  The ``mesonic models,'' 
which have had success in reproducing the experimental 
values for $\bar{d}(x)/\bar{u}(x)$, as discussed in Sect.\ 4.1 
and 4.2, naturally give rise to differences in these
distributions.  In such models, the $\bar{s}$ arises from 
virtual kaon production, while the $s$ comes from the residual 
strange baryon ($\Lambda$ or $\Sigma$); this was first pointed
out by Signal and Thomas \cite{Sig87}.  There was some suggestion 
of experimental support for the idea \cite{Fou90}, based on LO
analysis of the CCFR neutrino data, but subsequent NLO analysis
of the same data \cite{Baz95} saw no difference between 
strange and antistrange quark distributions, within the experimental 
errors.  We reviewed the
current experimental situation regarding strange/antistrange quark 
distributions in Sect.\ 2.6. 

For an estimate of the difference between strange and antistrange
quarks, we have used the mesonic model calculation of Melnitchouk 
and Malheiro\cite{Mel97}.  
For this estimate we took the (poorly determined) 
$NKH$ vertex function to be a monopole form factor of mass 1 GeV, the
largest value consistent with the latest CCFR data\cite{Fou90,Baz95}. 
Fig.\ \ref{fig42e} shows the ratio $R_s$ of Eq.\ \ref{Rcsv} 
from this model.  The order of magnitude of the $s-\bar{s}$ 
difference is comparable to that arising from CSV.  As the two
effects have the opposite sign, we predict some cancellation between
the two contributions.  However, the predicted shapes are completely
different, and one should be able to separate the two contributions
on the basis of the measured $x$-dependence of $R_W$.  

We emphasize that even the sign of the $s-\bar{s}$ difference is 
not well determined, so the theoretical ``error bars'' associated
with the curves of Fig.\ \ref{fig42e} are large.  As we mentioned in Sect.\ 
2.6, the quantity $s-\bar{s}$ can be independently extracted by
measuring opposite sign dimuons arising from scattering of
neutrinos or antineutrinos from nuclei.  

The structure functions $F_1^{W^+D}$ and $F_1^{W^-D}$ should be nearly
identical at all $x$.  If they are not, this would be quite surprising.  
{\it Any 
deviation from zero, at any value of $x$, would be extremely 
interesting}, whether its origin lies in parton CSV or intrinsic
strangeness.  
This is a strong test of the validity of
charge symmetry in parton distributions.  Since the ratios require
comparison of charge-changing reactions induced by electrons and
positrons, it is important to have very accurate calibration
of the relative reaction rates.  Detector efficiencies should not
be a major problem as the signal involves prominent jets on the
hadron side and very large missing energy and momentum on the lepton
side.  Clearly it will be important to determine
experimentally whether or not $F_1^{W^+D} - F_1^{W^-D}$
is non-zero. The
interpretation in terms of CSV, $s \neq \bar{s}$ or possibly both, 
can then be pursued in detail.

\subsection{Drell-Yan Processes Initiated by Charged Pions
on Isoscalar Targets}

In Sect.\ 4.1 and 4.2, we showed the dramatic results obtained
by comparing pp and pD Drell-Yan [DY] processes.  The preliminary
results from the E866 experiment at FNAL \cite{Haw97} 
appear to show substantial flavor symmetry violation in the proton sea.  
We can also use DY processes as a specific 
test for charge symmetry violation in parton distributions.  For this
we want to differentiate between, say, up quarks in the proton
and down quarks in the neutron.  We will show that this could
be accomplished by comparing DY processes induced by charged pions 
on isoscalar targets.  

The crucial element here is that at large momentum fraction
$x$, the nucleon distribution is dominated by its three valence 
quarks, while at similar large $x$ the pion is predominantly 
a valence quark-antiquark pair.  If one uses beams of pions, and 
concentrates on the region where Bjorken $x$ of
the target quarks is reasonably large, then the 
annihilating quarks will predominantly come from the nucleon and the
antiquarks from the pion.  Comparison
of Drell-Yan processes induced by $\pi^+$ and $\pi^-$ in this kinematic
region provides in principle a sensitive method for
comparing $d$ and $u$ quark distributions in the nucleon, since the 
$\pi^+$ contains a valence $\bar{d}$ (and will annihilate a $d$ quark 
in the nucleon) and $\pi^-$ a valence $\bar{u}$ (and will annihilate
a nucleon $u$ quark).  

As an example, we consider reactions on the deuteron, although our
results will be true for any isoscalar nuclear target.  
Consider the DY process for a $\pi^+$ on a deuteron.  
In Fig.\ \ref{fig41a}b, we showed a schematic picture of the
dominant process (in a kinematic region dominated by valence
quarks for the meson and nucleon), for charged pion DY processes
on a proton.  Provided 
that $x,x_\pi \geq 0.3$, to minimize the contribution from 
sea quarks, the dominant process will be the annihilation of 
a $\bar{d}$ in the $\pi^+$, with momentum fraction $x_\pi$, with
a down quark in the deuteron with momentum fraction
$x$.  Neglecting for the moment sea quark
effects, the $\pi^+-D$ ($\pi^-$D) DY cross sections will be proportional 
to:
\begin{eqnarray}
\sigma_{\pi^+D}^{DY}(x,x_\pi) &\sim& {1 \over 9}\left( d^p(x) 
 + d^n(x) \right) 
 \overline{d}^{\pi^+}(x_\pi)~~, \nonumber \\  
  \sigma_{\pi^-D}^{DY}(x,x_\pi) &\sim& {4\over 9} \left( u^p(x) + 
 u^n(x) \right)\overline{u}^{\pi^-}(x_\pi)~~. 
\label{eq:pipd}
\end{eqnarray}
If we construct the ratio, $R^{DY}_{\pi D}(x,x_\pi)$:
\begin{equation}
 R^{DY}_{\pi D}(x,x_\pi) =  \frac{4 \sigma_{\pi^+D}^{DY}(x,x_\pi) - 
 \sigma_{\pi^-D}^{DY}(x,x_\pi)}
 {\left(4 \sigma_{\pi^+D}^{DY}(x,x_\pi) + 
 \sigma_{\pi^-D}^{DY}(x,x_\pi) \right) /2},  
\label{eq:R}
\end{equation}
we expect the ratio $R^{DY}_{\pi D}$ to be 
quite sensitive to charge symmetry violating (CSV) terms in the 
nucleon valence parton distributions. 

From our model calculations of Section 3, the
CSV contributions should be at most a few percent.  Consequently 
we must include sea quark contributions for both nucleon and
pion in defining this ratio.  There will also be a contribution
from charge symmetry violating effects in the pion parton
distributions.  Including these contributions, 
the Drell-Yan ratio for pions can be written: 
\begin{eqnarray}
R^{DY}_{\pi D}(x,x_\pi) &\approx & \left( \frac {\delta d(x) - 
\delta u(x)}{u^p_v(x) + d^p_v(x)} 
 \right) \nonumber \\ &+& \frac{15\left[ 2\pi_v(x_\pi)u^p_s(x)
 + \pi_s(x_\pi)\left( u^p_v(x) + d^p_v(x) \right) \right]} 
 {4\,\overline{d}^{\pi}_v(x_\pi)\left[ u^p_v(x) + d^p_v(x) 
 \right]}; \nonumber \\
           &\equiv& R^N_{\pi D}(x) + R^{SV}_{\pi D}(x,x_\pi) ,
\label{eq:Rfinal}
\end{eqnarray}
Eq.\ \ref{eq:Rfinal} is valid at sufficiently large $x$ and $x_\pi$, 
where sea quark probabilities are small relative to valence quarks.
We have expanded it to lowest order in both sea quark and CSV terms  
\footnote{in Eq.\ \ref{eq:Rfinal} we neglected a pion CSV term 
$\delta \overline{d}^\pi(x_\pi) = \overline{d}^{\pi^+}(x_\pi) 
- \overline{u}^{\pi^-}(x_\pi)$.  This term was estimated using a
Nambu-Jona Lasinio (NJL) \cite{njl} model employed
recently by Toki and collaborators \cite{toki1,toki2}, which 
predicted a very small pion CSV effect \cite{Lon94}.}. 

The nucleon CSV term $R^N_{\pi D}(x)$ in Eq.\ (\ref{eq:Rfinal}) is 
a function only of $x$.  
It is not necessary to know absolute fluxes of charged pions to 
obtain an accurate value for $R^{DY}$.  The yield of $J/\psi$'s from 
$\pi^+-D$ and $\pi^--D$ should be identical to within 1\%, 
so this can be used to normalize the relative fluxes.  
Because $R^{DY}_{\pi D}(x,x_\pi)$ is a ratio of cross sections, a 
number of systematic errors should
cancel.  In particular, Eq.\ (\ref{eq:Rfinal}) is not sensitive to 
differences between the parton
distributions in the free nucleon and those in the deuteron 
\cite{bodek,fs,bicker,mst}.  Provided that both the neutron and proton 
parton distributions are modified in the same way, then the ratio  
in Eq.\ (\ref{eq:Rfinal}) will be unchanged.   

\begin{figure}
\centering{\ \psfig{figure=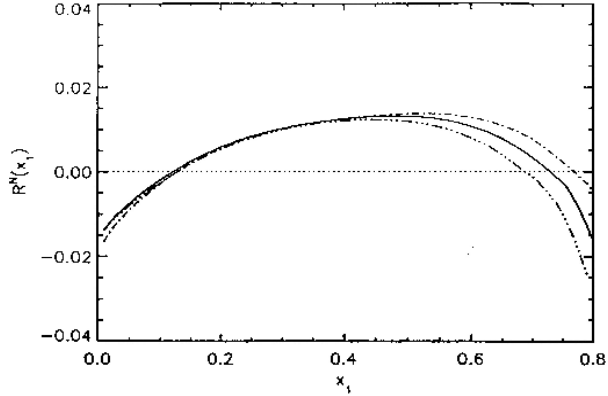,width=9cm}}
\vspace{0truein}
\caption{Theoretical estimate of nucleon CSV term $R^N_{\pi D}$ of 
Eq.\ \protect\ref{eq:Rfinal}.  From Ref.\ \protect\cite{Lon94}.}
\vspace{0.1truein}
\label{fig47}
\end{figure}

In Fig.\ \ref{fig47} we show the nucleon CSV contribution, 
$R^N_{\pi D}(x)$, using the bag model predictions for CSV, 
evolved to $Q^2 = 10$ GeV$^2$.  
As the main uncertainty in our calculation is the mean diquark 
mass, the results are shown for several values of this parameter. 
In the region $0.4 \leq x \leq 0.7$, we predict $R^N$ will always be 
positive, with a maximum value of about 1.7\%.  

 We predict that the contributions
from pion and nucleon CSV will all be the same sign and will
add constructively.  

Since the Drell-Yan ratios arising from CSV are expected to 
be very small, even small contributions from sea quarks
could make a substantial effect.  The dominant contribution will arise
from interference between one sea quark and one valence quark.  
We concluded that for sufficiently large $x$, e.g.\ $x \geq 0.4$, 
one should be able to separate the CSV ``signal'' from the 
sea-valence interference \cite{Lon94}.  
Unlike the CSV contributions of Eq.\ (\ref{eq:Rfinal}), the sea-valence
interference term $R^{SV}_{\pi D}(x,x_\pi)$ does not separate, so 
one could exploit the very different
dependence on $x$ and $x_\pi$ of the background and CSV terms.  
We
conclude that the CSV terms could be extracted even in the presence
of a sea-valence ``background''.     

Despite our prediction in Sect.\ 3 that the fractional ``minority 
quark'' CSV term, $\delta d(x)/d_{\rm v}(x)$, should be between 
3 and 7\% (c.f.\ Fig.\ \ref{fig34}), the nucleonic CSV 
ratio $R^N_{\pi D}$, shown in Fig.\ \ref{fig47}, is predicted 
to be more like 1-2\%.  This is because $\delta d$ in Eq.\ 
(\ref{eq:Rfinal}) is divided by $u^p + d^p$, and at large $x$  
$d^p(x) << u^p(x)$.  A much larger ratio could 
in principle be obtained
by comparing the $\pi^+ -p$ and ``$\pi^- -n$'' Drell-Yan processes
through the ratio : 
\begin{equation}
R^{DY}_{\pi N}(x,x_\pi) =  \frac{4 \sigma_{\pi^+p}^{DY}(x,x_\pi) 
 + \sigma_{\pi^-p}^{DY}(x,x_\pi)
 - \sigma_{\pi^-D}^{DY}(x,x_\pi)}
 {\left(4 \sigma_{\pi^+p}^{DY}(x,x_\pi) - 
 \sigma_{\pi^-p}^{DY}(x,x_\pi)
  + \sigma_{\pi^-D}^{DY}(x,x_\pi) \right) /2}.  
\label{eq:Rpin}
\end{equation}
In principle, the advantage of this measurement is that it isolates
the minority quark CSV term -- in fact, the dominant term in 
Eq.\ \ref{eq:Rpin} is the term $\delta d_{\rm v}(x)/d_{\rm v}(x)$ so
we expect CSV effects at the 3-7 \% level.  

We conclude, however, that this quantity is unlikely to 
provide unambiguous information regarding parton charge
symmetry violation.  First, to form the ratio in Eq.\ \ref{eq:Rpin}
one must know the relative normalization of DY cross sections on protons and 
deuterons.  This should be feasible by bombarding both hydrogen 
and deuterium targets simultaneously with charged pion beams.  
In order to extract the minority quark CSV term, it is necessary to 
know the precise relation between deuteron parton distributions 
and those for free nucleons.  If we include ``EMC'' changes in the 
deuteron structure functions relative to free proton and neutron
distributions, we find that 2-3\% changes in the parton distributions 
can produce 10-30\% changes in the ratios of Eq.\ 
\ref{eq:Rpin}.  In view of the sensitivity of this ratio to the EMC 
term, we conclude that information regarding CSV effects could not
be extracted from comparing DY cross sections for $\pi^- p$ 
with those for ``$\pi^+ -n$'' unless the  
Fermi motion and binding corrections for the deuteron were known
to great accuracy.  

Comparing the Drell-Yan yield 
for $\pi^+$ and $\pi^-$ on deuterons may provide a means  
to extract the charge symmetry violating [CSV] part of the nucleon 
parton distribution.  As the
$x$ and $x_\pi$ values of interest for the proposed measurements are
large ($x > 0.5$), a beam of 40-50 GeV pions will produce sufficiently
massive dilepton pairs that the Drell-Yan mechanism is applicable.  
A flux of more than $10^9$ pions/sec.\ is desirable, so 
these experiments might be feasible when 
the new FNAL Main Injector becomes operable.  

\subsection{Charged Pion Leptoproduction from Isoscalar Targets}

In a recent paper \cite{Lon96}, we pointed out that semi-inclusive 
pion production, from lepton DIS on nuclear targets, 
could also be a sensitive probe of CSV effects in the 
valence parton distributions for the nucleon.  

In the quark/parton model, the semi-inclusive production of 
hadrons in deep inelastic lepton scattering from a nucleon 
is given by
\begin{equation}
{1\over \sigma_N(x)}{d\sigma^h_N(x,z)\over dz}=
{N^{Nh}(x,z) \over \sum_i e^2_i q^N_i(x)}. 
\label{cs}
\end{equation}
The quantity $N^{Nh}$ in Eq.\ (\ref{cs}), the yield of hadron 
$h$ per scattering from nucleon $N$, has the form  
$N^{Nh}\equiv \sum_i e^2_i q^N_i(x)D^h_i(z)$, 
where $D^h_i(z)$ is the fragmentation function
for a quark of flavor $i$ into hadron $h$, which depends on the 
quark longitudinal momentum 
fraction $z=E_h/\nu$, where $E_h$ and $\nu$ are the energy of 
the hadron and the virtual photon respectively.

For pion electroproduction on an isoscalar target, (such as the 
deuteron) charge symmetry relates the ``favored'' production of
charged pions from valence quarks, by  
\begin{eqnarray}
N^{D\pi^+}_{fav}(x,z) &=& 4\,N^{D\pi^-}_{fav}(x,z) .
\label{csvlep}
\end{eqnarray}
In Eq.\ \ref{csvlep}, $N^{D\pi^+}_{fav}(x,z)$ represents
the yield of $\pi^+$ per scattering from the deuteron, via
the ``favored'' mode of production, e.g.,   
for $\pi^+$ ($\pi^-$) production, the ``favored'' mode of
charged pion production is from the target up (down) quarks.  
Since the semi-inclusive reactions are proportional to the square
of the quark charge, there is a relative weighting of 4 for 
$\pi^+$ production.  

Deviations from Eq.\ \ref{csvlep} will arise from effects due
to sea quarks, CSV effects in the parton distributions, and
contributions from the ``unfavored'' fragmentation functions.  
The HERMES collaboration at HERA \cite{HERA} 
is currently taking experimental data on semi-inclusive 
pion production from hydrogen and deuterium \footnote{Our description 
of fragmentation is correct only in the high energy limit, where 
hadron production is so copious that the
leading quark fragmentation and target fragmentation completely 
decouple.  It is likely that at HERMES energies, sufficiently few 
hadrons are produced that this picture is inaccurate.  Monte Carlo 
simulations of data at these energies could reveal the breakdown of 
this naive factorization picture, in which case the arguments presented
here would be applicable only at higher energies}.  

Assuming charge conjugation invariance and charge symmetry
for the fragmentation functions 
allows us to write the yield for leptoproduction of a charged 
pion from a proton as
\begin{eqnarray}
N^{p \pi^\pm}(x,z) &=& D^{\pi^\pm}_u(z) \left[ {4\over 9}u^p(x)  
  + {1\over 9}\bar{d}^p(x) \right] + 
  D^{\pi^\mp}_u(z)\left[ {4\over 9}\bar{u}^p(x) + 
  +{1\over 9} d^p(x) \right]  \nonumber \\ &+& 
  {1\over 9}D^{\pm}_s(z)\left[ s^p(x) + \bar{s}^p(x) \right]~~. 
\label{genfor}
\end{eqnarray}
The fragmentation functions have been extracted by the EMC
group \cite{emcfrag,emcfrag89}, and are being 
independently measured by the HERMES collaboration \cite{HERA}.  

We proposed measuring the quantity 
$\widetilde{R}^D(x,z)$, defined by 
\begin{eqnarray}
\widetilde{R}^D(x,z) &\equiv& {1- \Delta (z)\over 
  1+ \Delta (z)}\left[ {4\,N^{D\pi^-}(x,z) -  N^{D\pi^+}(x,z)
  \over N^{D\pi^+}(x,z) -  N^{D\pi^-}(x,z) } \right] \nonumber \\
  &\approx& {5 \Delta (z)\over 1+ \Delta (z)} +
  {4\left[ \delta d(x) - \delta u(x)\right]  \over 
  3\,\left[u^p_{\rm v}(x) + d^p_{\rm v}(x)\right]} \nonumber \\ 
  &+& { 5\left( \bar{u}^p(x) +  \bar{d}^p(x) \right) +  
  \Delta_s(z) \left[ s(x) + \bar{s}(x)\right]/(1+ \Delta (z)) 
  \over 
  \left[u^p_{\rm v}(x) + d^p_{\rm v}(x)\right]}  \quad . 
\label{sigtil}
\end{eqnarray}

In Eq.\ (\ref{sigtil}), we expand to first order in
``small'' quantities.  These are: the CSV nucleon terms,
$\delta d(x)$ and $\delta u(x)$, and the sea quark distributions   
(Eq.\ (\ref{sigtil}) is only valid at large $x$ where the
ratio of sea/valence quark distributions is small).  We have
neglected the CSV part of the fragmentation function; in Ref.\ 
\cite{Lon96} we estimated that this term
would be quite small.  

The quantity $\widetilde{R}^D(x,z)$ in Eq.\ \ref{sigtil} separates 
into three pieces.  The first piece depends only on $z$.  This term 
is a function only of the experimentally measured quantity 
$\Delta(z)$, the unfavored/favored ratio of fragmentation 
functions.  It decreases roughly 
monotonically as $z$ increases.  The second term  
depends only on $x$, and is proportional to the nucleon CSV fraction 
(relative to the valence quark distributions).  The final term 
in Eq.\ \ref{sigtil} depends on both $x$ and $z$.  It is proportional 
to the sea quark contributions, so it becomes progressively 
less important at large $x$.  It also contains a term which is
proportional to the strange/favored ratio of fragmentation functions
$\Delta_s(z)$: we estimate that this term is always negligible.    

Experimentally, one needs to measure accurately the  
$x$-dependence of $\widetilde{R}^D(x,z)$ for fixed $z$; 
in this case the $z$-dependent term will be
large (of order one) and constant.  The sea quark 
contribution will be large at small $x$, but should fall
off monotonically and rapidly with $x$.  
So, at sufficiently large $x$, the sea
quark contribution will be negligible relative to the CSV
term.  One then has to extract the small, $x$-dependent
term in Eq.\ (\ref{sigtil}) from the large term independent 
of $x$.  As a general rule, the larger the values of $x$ 
and $z$ at which data can be taken, the larger the CSV term will
be relative to the $z$-dependent term.  

\begin{figure}
\centering{\ \psfig{figure=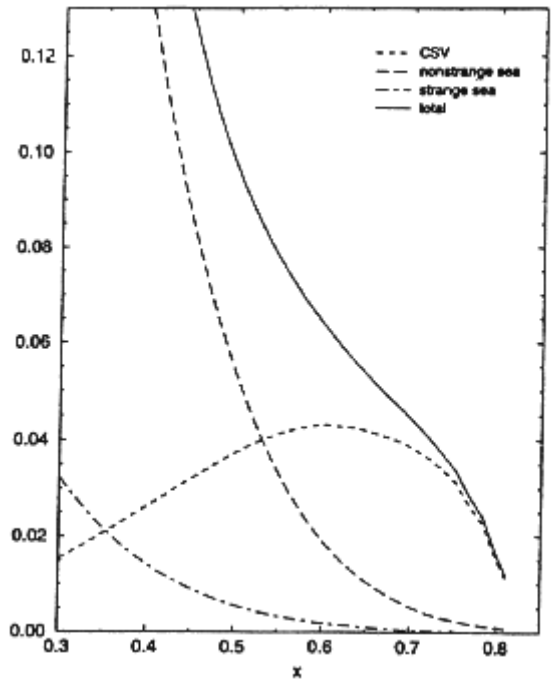,width=8cm}}
\vspace{0truein}
\caption{$x$-dependent contributions to charged pion leptoproduction,
from Ref.\ \protect\cite{Lon96}.}
\vspace{0.1truein}
\label{fig412}
\end{figure}

In Fig.\ \ref{fig412} we plot our predictions for the $x$-dependent
terms in $\widetilde{R}^D(x,z)$, at $Q^2 = 10$ GeV$^2$.  
The long dashed curve is the contribution
from nonstrange sea quarks to $\widetilde{R}^D_{sea}(x,z)$.  
This depends only on $x$, and is calculated using the CTEQ3M parton
distributions \cite{cteq}.  The short dashed curve 
is our prediction for the parton charge symmetry violating term, 
$\widetilde{R}^D_{CSV}(x)$; this uses the CTEQ3M parton distributions, 
plus the bag model prediction for valence quark CSV from Londergan 
{\it et al.} \cite{Lon94}, as discussed in Sect.\ 3.  The 
dot-dashed curve is our estimate 
of the strange quark contribution.  The solid 
curve is the sum of the three terms.  

For $x \approx 0.5$, the CSV term is as 
large as the sea quark contribution, and with increasing $x$ (e.g., 
for $x \geq 0.55$), the CSV term dominates the $x$ dependent 
terms.  We predict the maximum CSV contribution will be of order 
$0.02 - 0.04$.  The $x$ dependent contribution shown in
Fig.\ \ref{fig412} will sit on a large and constant $z$ 
dependent term.  This term is predicted to vary
between 1.5 and 1, as $z$ goes from 0.4 to 0.8 \cite{Lon96}.  
So the CSV term is expected
to be between 1-4\% of the $z$-dependent term.  

In the HERMES experiment at HERA, the goal is to make 
precision measurements of the spin structure functions, 
so the prospect for obtaining very
accurate spin-averaged charged pion leptoproduction data is
excellent.  Only data from deuterium targets is required; efficient
detection of both signs of charged pions is important, but absolute
yields are not required as overall normalizations cancel out in
the ratio of Eq.\ (\ref{sigtil}).

\section{Charge Symmetry and Flavor Symmetry Contributions to Sum Rules}
\mb{.5cm}

Sum rules can provide extremely useful information on 
parton distributions.  For example, from the quark 
model the integrals of up and down valence quark distributions
obey the quark number sum rules, given by 
the normalization conditions on the quark
distributions, see Eq.\ \ref{GSRnrm} in Sect. 2.   

Two sum rules, the Adler sum rule \cite{Adler} and 
Gross-Llewellyn Smith sum rule \cite{GLS}, can be directly 
related to linear combinations
of quark normalization integrals.  The Adler sum rule is 
obtained by integrating the 
difference between the $F_2$ structure functions for 
charged-current interactions of antineutrinos on protons,
and that for neutrinos on protons.  The Gross-Llewellyn
Smith sum rule is obtained by summing the $F_3$ structure
functions for charged current interactions of neutrinos,
and antineutrinos, on a proton target (the same
result is obtained for a neutron target). 
 
The Gottfried sum rule (GSR) \cite{Got67} is obtained by integrating the
difference between the $F_2$ structure functions for 
(photon mediated) neutral current interactions on protons
and neutrons.  Unlike the Adler or Gross-Llewellyn Smith
sum rules, the ``naive'' Gottfried sum rule expectation 
$S_G = 1/3$ is obtained only if we assume both charge symmetry 
for parton distributions, and what is frequently called ``SU(2) flavor 
symmetry'' in the proton sea.  That is, in addition to charge symmetry 
we assume that $\bar{u}^p(x) = \bar{d}^p(x)$. 
  
There has been much recent interest in the Gottfried sum rule, 
sparked by the rather precise measurements of the NMC group 
\cite{Ama91}.  These show rather
conclusively that $S_G$ is substantially less than 1/3.  
If charge symmetry is valid, this provides
information on SU(2) flavor symmetry violation [FSV] in the proton
sea.  In fact, the level of FSV needed to agree with the
NMC result is surprisingly large.  However, as we will see, 
the experimental GSR measurements are 
actually sensitive to a combination of FSV and CSV in the
nucleon sea.  

As we have argued in this paper, we would expect CSV effects 
in parton distributions to be no greater than about one percent 
for sea quark distributions, and for the ``majority'' valence
quark distribution.  However, we expect the ``minority'' valence
quark distribution to exhibit CSV effects of several percent at
large $x$.  As we discussed in Sect. 4, the current experimental upper 
limits on CSV are of the order of a few percent for $0.1 \le x \le 0.4$; 
upper limits for CSV are no better than 10\% for larger values of $x$, 
and for $x < 0.1$ the best experiments to date suggest a violation of
parton charge symmetry.  
It would therefore be useful to construct sum rules which
could in principle distinguish between CSV and FSV effects.  
In this Section, we will review the Adler, Gross-Llewellyn
Smith and Gottfried sum rules, clarifying the various assumptions 
implicit in their derivation (particularly the role of charge symmetry
in the sum rules).  Next, we will propose sum rules which
can clearly differentiate between charge symmetry violation
and flavor symmetry violation in nucleon sea quark
distributions.  Finally, we will present existing deep
inelastic structure functions in the context of these
new sum rules, to examine the degree to which existing
data can constrain limits on charge symmetry and/or
flavor symmetry violating effects.    

The discussion of new sum rules follows rather closely the 
prior theoretical work of Ma \cite{Ma92,Ma93}.  The sum rules
we introduce are linear combinations of those proposed by Ma.  He
also pointed out the potential confusion in the
literature on the question of charge symmetry.      

We first review existing sum rules (Gottfried, Gross-Llewellyn
Smith, Adler) without making the usual assumptions of
charge symmetry in quark distributions.  The NMC measurements of 
the Gottfried Sum Rule \cite{Ama91} are seen as 
strongly suggesting large flavor symmetry violation [FSV]
in the proton sea quark distributions. The Drell-Yan measurements 
carried out by the NA51 \cite{Bal94}
and E866 \cite{Haw97} groups are regarded as more or less definitive
proof of large FSV effects.  We point out that all three results
could in principle be explained by large 
charge symmetry violation in the nucleon sea quark parton distributions
(alternatively, a linear combination of FSV and CSV effects could
be responsible for these results).  

In Sect.\ 7, we introduce
``flavor symmetry'' and ``charge symmetry'' sum rules,
and discuss how they can separate CSV and FSV effects in
nucleon sea quark distributions.  We 
discuss what (if any) experimental limits on CSV and FSV can
be drawn from existing deep inelastic neutrino scattering
experiments.  

\subsection{Review of Gottfried Sum Rule}

Here we review the Gottfried Sum Rule \cite{Got67}.  We
go through this in considerable detail so that the underlying 
assumptions in its derivation 
are clear throughout.  For a comprehensive review of 
both experimental and theoretical aspects of the Gottfried Sum Rule, 
and the related question of the flavor symmetry of the proton
sea, see the recent work by Kumano \cite{Shunzo}.  

The Gottfried Sum Rule is given by
\begin{eqnarray}
S_G &\equiv& \int_0^1 \,{dx\over x}\, \left[ F_2^{\mu p}(x) - 
 F_2^{\mu n}(x) \right] \quad .  
\label{GSRi}
\end{eqnarray} 
Because the $F_2$ structure function from electron
and muon deep inelastic scattering depends on the
squared charge of the quarks, the up quark distributions are weighted
by a factor of four relative to the down quark distributions.  

Rewriting the structure functions in terms of quark 
distributions, we obtain the result  
\begin{eqnarray}
\int_0^1\,{dx\over x} \left[ F_2^{\mu p}(x) - F_2^{\mu n}(x) 
 \right] &=& {1\over 9}  
 \int_0^1\,dx\, \left[ 4\,u^p(x) + 4\,\bar{u}^p(x) - 
 4\,u^n(x)- 4\,\bar{u}^n(x) \right. \nonumber \\ &+& \left. d^p(x)
 + \bar{d}^p(x)-  d^n(x) -  \bar{d}^n(x) \right] .
\label{GSRiii}
\end{eqnarray} 
In obtaining Eq.\ (\ref{GSRiii}), we assume the strange quark
contributions for neutron and proton cancel.  We can 
invoke the ``strong'' assumption that the strange parton 
distributions for proton and neutron are identical at each value
of $x$, i.e.\ $s^p(x) = s^n(x)$  
(and similarly for the antiquark distributions); 
alternatively, we can assume the ``weak'' condition that the 
parton distributions need not be identical at all $x$, 
but that the {\em integrals} over $x$ of the appropriate 
parton distributions are identical for proton and neutron.  
There is no QCD modification of this sum rule.  

Introducing valence quark distributions as in Section
2, {\it i.e.}  
$u_{\rm v}(x) \equiv u(x) - \bar{u}(x)$, we obtain
\begin{eqnarray}
\int_0^1\,{dx\over x} \left[ F_2^{\mu p}(x) - F_2^{\mu n}(x) 
 \right] &=& 
 {1\over 9} \int_0^1\,dx\, \left[ 
  4\,u^p_{\rm v}(x) + 8\,\bar{u}^p(x) - 4\,u^n_{\rm v}(x) 
  \right. \nonumber \\ &-& \left. 8\,\bar{u}^n(x) + d^p_{\rm v}(x)
 + 2\,\bar{d}^p(x)-  d^n_{\rm v}(x) -  2\,\bar{d}^n(x) \right] .
\label{GSRval}
\end{eqnarray} 
We now invoke the valence quark normalization conditions, defined
in Eq.\ \ref{GSRnrm}, and we 
obtain the Gottfried Sum Rule, 
\begin{eqnarray}
S_G &\equiv& \int_0^1\,dx\, \left[ {F_2^{\mu p}(x) - 
 F_2^{\mu n}(x)\over x} \right] 
\nonumber \\ &=& {1\over 3} + {2\over 9}\,\int_0^1\,dx\,
\left[ 4\,\bar{u}^p(x) + \bar{d}^p(x) - 4\bar{u}^n(x) - 
\bar{d}^n(x) \right] \quad .
\label{GSRfin}
\end{eqnarray}

If we make the additional (and customary) ``strong'' 
assumption of charge symmetry, e.g. 
\begin{eqnarray}
d^n(x) &=& u^p(x) \nonumber \\
u^n(x) &=& d^p(x) \nonumber \\
\bar{d}^n(x) &=& \bar{u}^p(x) \nonumber \\
\bar{u}^n(x) &=& \bar{d}^p(x) \quad ,
\label{csass}
\end{eqnarray}
then we obtain the ``normal'' formulation of the Gottfried Sum Rule
[GSR], 
\begin{eqnarray}
S_G \,&{CS\atop =}&\, \int_0^1\,dx\, \left[ {F_2^{\mu p}(x) - 
F_2^{\mu n}(x)\over x} \right] 
\nonumber \\ &=& {1\over 3} + {2\over 3}\,\int_0^1\,dx\,
\left[ \bar{u}^p(x) - \bar{d}^p(x) \right] . 
\label{GSRcs}
\end{eqnarray}
Note that the charge symmetric Gottfried Sum Rule, Eq.\ \ref{GSRcs}, 
does not require 
the ``strong'' assumption of charge symmetry: it follows also 
from the ``weak'' condition that the antiquark distributions are 
not charge symmetric at all points $x$, but that the {\em integral} 
over all $x$ is the same for, say, the up antiquark distribution 
in the proton and the down antiquark distribution in the neutron.  

With the assumptions we have listed, the Gottfried Sum Rule
will be equal to 1/3 if we have SU(2) flavor symmetry in the proton
sea, i.e.\ if $\bar{u}^p(x)= \bar{d}^p(x)$, or if the integrals
over $x$ of these distributions are equal.   

One additional point is that the Gottfried ``Sum Rule'' cannot 
be obtained from current algebra, 
that is, the GSR cannot be expressed in terms of equal-time 
commutators of some observable.  The Gottfried Sum Rule is
simply a relation which holds in the quark/parton model, with 
additional assumptions regarding equality of strange and charmed
quark expectation values in the proton and neutron.  This is not the
case for the Adler sum rule, which can
be derived either from current algebra relations or the
quark/parton model.   

\begin{figure}
\centering{\ \psfig{figure=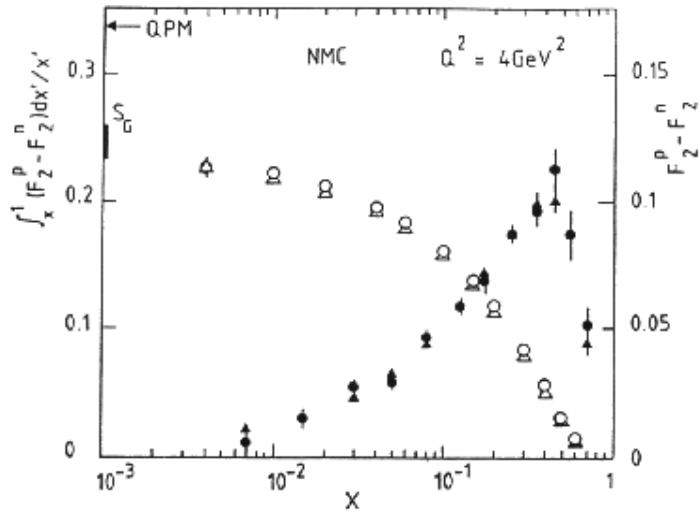,width=10cm}}
\vspace{0truein}
\caption{Experimental results for Gottfried sum rule from NMC
group, Ref.\ \protect\cite{Ama91}. Solid dots: $F_2^p - F_2^n$; open
dots: $\int_x^1\,[F_2^p(x') - F_2^n(x')]dx'/x'$.}
\vspace{0.1truein}
\label{fig51}
\end{figure}

For almost twenty years there have been indications that 
the GSR seems to be less than 1/3.  However, interest in
the Gottfried Sum Rule has intensified with 
the very accurate measurements in 1991 by the NMC group 
\cite{Ama91}; the data was re-analyzed in 1994 \cite{Arn94}.  
Their results are shown in Fig.\ \ref{fig51}.  The 
solid circles show $F_2^p(x) - F_2^n(x)$, while the open circles
plot $S_G(x) \equiv \int_x^1 [F_2^p(x)- F_2^n(x)]\,dx/x$.  From
Eq.\ \ref{GSRi} we see that the difference in $F_2$ structure
functions is multiplied by $1/x$.  This means that measurements
at low $x$ play a critical role in determining $S_G$.  Fig.\ 
\ref{fig51} shows that roughly half the contribution to the
GSR integral comes from the region $x \le 0.1$.  The 
neutron results, $F_2^n$, were inferred from reactions on deuterium.  

Earlier measurements by the SLAC, EMC and BCDMS groups 
\cite{whitlow,SLAC,emc,BCDMS} gave results which were lower than 
1/3, but these values had errors in $S_G$ of roughly 20\%, so that 
the results were within one standard deviation of 1/3.  The 
statistical error in the previous experiments 
was dominated by the lack of data at sufficiently
small $x$.  The NMC group obtained data for $x$ down to 0.003.  
The final value from the NMC group was 
\begin{equation}
S_G(0.004 \leq x \leq 0.8) = 0.221 \pm 0.008~(stat) \pm 
0.019~(syst) , 
\end{equation} 
The NMC group then fitted the
difference in $F_2$ structure functions by a power law and
extrapolated from $x=0.004$ to $x=0$ (the extrapolation to $x=1$ 
produces no measurable contribution to $S_G$).  Their extrapolated
result was
\begin{equation}
S_G = 0.235 \pm 0.026 . 
\end{equation} 
This is more than four standard deviations lower than the
``naive'' expectation of 1/3.  

If we use the structure functions for electromagnetic interactions,
Eqs.\ \ref{F1gamm} and \ref{F1gammn}, but do not invoke charge symmetry, 
we obtain
\begin{eqnarray}
\left[ {F_2^{\mu p}(x) - F_2^{\mu n}(x)\over x} \right] 
  &=& {1\over 3}\left[ u^p_{\rm v}(x) - 
   d^p_{\rm v}(x)\right] +  {2\over 3}\, \left[ \bar{u}^p(x) 
   - \bar{d}^p(x) \right] \nonumber \\ &+&  {1\over 9}\, \left[ 
   4\delta d(x) + \delta u(x) + 4\delta \bar{d}(x) + 
  \delta \bar{u}(x) \right] .  
\label{GSRcsv}
\end{eqnarray}
Integrating this over $x$ and normalizing
the valence quarks gives
\begin{eqnarray}
S_G &=& \int_0^1\,dx\, \left[ {F_2^{\mu p}(x) - 
F_2^{\mu n}(x)\over x} \right] 
\nonumber \\ &=& {1\over 3} + {2\over 3}\,\int_0^1\,dx\,
  \left[ \bar{u}^p(x) - \bar{d}^p(x) \right] \nonumber \\ 
  &+& {2\over 9}\,\int_0^1\,dx\,
  \left[ 4\delta\bar{d}(x) + \delta\bar{u}(x) \right] . 
\label{GSRcsv2}
\end{eqnarray}

There has been much speculation as to the cause of the
NMC result.  One possibility is that the Gottfried Sum Rule 
is, in fact, 1/3, and that the apparent deviation of the
GSR from 1/3 is an artifact of the procedure
for extrapolating the structure functions to $x=0$.  Martin, 
Roberts and Stirling \cite{Mar90} suggested that one might
have $S_G = 1/3$, where the ``missing''
contribution to $S_G$ comes from very small $x$ values, and
that the NMC power law extrapolation was in error.  

We discussed in Sect.\ 4 other theoretical suggestions 
for the origin of the excess of $\bar{d}^p$ over $\bar{u}^p$.  
Recently several groups have compared proton-induced Drell-Yan 
processes on protons and deuterons.  Recent results have been 
obtained for this process
by the NA51 group at CERN \cite{Bal94}, and 
preliminary results from the E866 group at FNAL \cite{Haw97}.  
Both experiments appear to confirm that $\bar{d}^p(x) > 
\bar{u}^p(x)$. This was discussed in
more detail in Section 4.

Assuming the NMC extrapolation is correct, from Eq.\ \ref{GSRcsv2} 
we see that deviation of the GSR from 
1/3 measures either charge symmetry violation [CSV],  
or flavor symmetry violation [FSV] in the nucleon sea quark
distributions (or a combination of the two effects).  If we assume 
the validity of charge symmetry,
then the NMC measurement implies a surprisingly large SU(2) 
flavor asymmetry in the proton antiquark distributions, namely
\begin{eqnarray}
  \int_0^1\,dx\, \left[ \bar{d}^p(x) - \bar{u}^p(x) \right] \ 
  &{CS\atop =}& \ 0.147 \pm 0.039
\label{fsvnmc}
\end{eqnarray}

The FSV contribution suggested by the NMC experiment is
surprisingly large, as it is much larger than can be
accommodated by perturbative QCD.  Both NLO and NNLO
QCD calculations have been carried out, and predict 
very small FSV effects \cite{sachrajda}.  
Consequently, we need a non-perturbative mechanism to
generate flavor-violating sea quark distributions which 
will reproduce the experimental result.  

The Pauli principle should make it easier to form
a $d\bar{d}$ pair than a $u\bar{u}$ pair in the presence of
the valence quarks.  
In Feynman and Field's early paper on
parton distributions \cite{feyfie}, they assumed an excess of 
$\bar{d}$ quarks in the proton on these grounds.  
A promising mechanism for generating additional $\bar{d}$ quarks
in the proton, first recognized in Ref.\ \cite{tony83}, is the 
``Sullivan Effect'' \cite{sullivan}.  
We discussed this briefly in Sect.\ 4, and refer the interested
reader to the review article of Kumano \cite{Shunzo}.  

It is important to note that in principle, {\em one could reproduce 
the NMC results even if flavor symmetry is exact}; this was pointed
out by Ma \cite{Ma92}.  From 
Eq.\ \ref{GSRcsv2}, if we assume exact flavor symmetry, but 
not charge symmetry, in the nucleon sea, then  
\begin{eqnarray}
\bar{d}^p(x) &=& \bar{u}^p(x) \equiv \bar{q}^p(x) \nonumber \\
 \bar{d}^n(x) &=& \bar{u}^n(x) \equiv \bar{q}^n(x) \nonumber \\
 \bar{q}^p(x) &\ne& \bar{q}^n(x)
\label{flsymm}
\end{eqnarray}
then the NMC measurement implies a substantial
charge symmetry violation in the nucleon sea, {\it i.e.}, 
\begin{equation}
  \int_0^1\,dx\, \left[ \bar{q}^p(x) - \bar{q}^n(x) 
  \right] = -0.088 \pm 0.023
\label{csvnmc}
\end{equation}
It would require a very sizable CSV contribution to
reproduce the NMC result.  The necessary parton sea CSV contribution 
is more than an order of magnitude larger than the theoretical
estimate we discussed in Sect.\ 3.2.  Alternatively, the NMC data  
could result from a 
linear combination of FSV and CSV effects in the nucleon sea.  

Eq.\ \ref{csvnmc} shows that the Gottfried Sum Rule is 
sensitive to charge symmetry violation in the nucleon sea.  
One can also have charge symmetry violation in the valence quark
distributions. However, the integral over $x$ of the charge symmetry 
violating pieces must vanish since CSV contributions cannot
change the valence quark normalizations. In principle, violation
of charge symmetry in the valence quark distributions makes
no contribution to the Gottfried Sum Rule.  However, as the
valence quark CSV contribution vanishes only when integrated
over all $x$, it is possible to obtain a contribution from
valence quark CSV if data is taken only over a finite range
of $x$.  It is important that extrapolations over
an unmeasured region properly account for these terms.   

\subsection{Adler Sum Rule}

The Adler Sum Rule is given by the integral of the $F_2$ structure
functions for charged current 
neutrino scattering.  The Adler Sum Rule, $S_A$, can be defined as   
\begin{eqnarray}
S_A &\equiv& \lim_{Q^2 \rightarrow \infty}\, 
\int_0^1\,dx\, \left[ {F_2^{W^- p}(x,Q^2) - 
F_2^{W^+ p}(x,Q^2)\over 2x} \right] 
\nonumber \\ &=& \int_0^1\,dx\,
\left[  u^p(x) - \bar{u}^p(x) - \left( d^p(x) - \bar{d}^p(x) 
 \right) \left(1- |V_{td}|^2 \right) 
 - s(x) +\bar{s}(x) \right] \nonumber \\
 &=&  1 \quad .
\label{Adler}
\end{eqnarray}
We obtain the result $S_A = 1$ if we neglect the term $|V_{td}|^2
\approx 1\times 10^{-4}$.  The Adler sum rule thus requires 
measuring the $F_2$ structure function for antineutrinos and
neutrinos on protons, dividing by $x$ (which emphasizes the
contribution from very small $x$), and subtracting them.  
The Adler sum rule then 
follows from the normalization of the quark distributions. 
As a consequence of the algebra of SU(2) charges,
the Adler sum rule has no QCD corrections.  

\begin{figure}
\centering{\ \psfig{figure=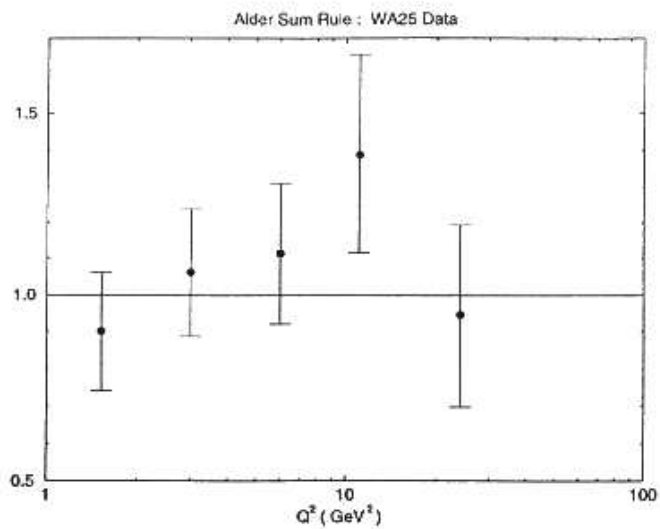,width=9.5cm}}
\vspace{0truein}
\caption{Experimental results for Adler sum rule, Eq.\ 
\protect\ref{Adler} from the WA25 group, Ref.\ \protect\cite{WA25}.}
\vspace{0.1truein}
\label{fig52}
\end{figure}

In Fig.\ \ref{fig52} we show the experimental situation regarding 
the Adler sum rule.  The experimental data are from the WA25
experiment \cite{WA25}, using the CERN-SPS wide
band neutrino and antineutrino beams in the
BEBC H and D bubble chambers.  The experimental data are
shown for several values of $Q^2$.  The average value 
is $S_A = 1.01 \pm 0.08 \pm 0.18$.  However, as pointed
out by Sterman {\it et al.} \cite{Ste95}, the total
$\nu N$ cross section used by the WA25 group is smaller than
the presently accepted value \cite{blair,berge}.  If the
WA25 value is readjusted to fit this total cross section
their result becomes $S_A = 1.08 \pm 0.08 \pm 0.18$.

Within the rather large errors,
the results are independent of $Q^2$.  The large errors arise
from the factor $1/x$ in the integral, Eq.\ \ref{Adler}.  
This gives a heavy weighting to the data at small $x$.  
The paucity of data in this region and the 
relatively large error bars there give a large uncertainty
in the sum rule value.  Because of the difficulties in
obtaining sufficient neutrino and antineutrino 
cross sections at small $x$ values
for light nuclear targets, it is unlikely that there
will be new experimental neutrino data in the near future
which would allow us to test the Adler sum rule.  

Note that the experimental points presented involve neutrino
measurements on neutrons (e.g., deuterons) and protons, 
and not antineutrinos and neutrinos on protons, as given
in the definition of the Adler sum rule, Eq.\ \ref{Adler}.  
This follows from the relation $F_2^{W^+ n}(x,Q^2) \,{CS\atop =} 
\, F_2^{W^- p}(x,Q^2)$, which follows if one assumes 
charge symmetry, as was discussed in detail in Section V.A.  
Thus, the WA25 group does not plot the integral $S_A$ of Eq.\ 
\ref{Adler}, but instead measures the quantity 
\begin{eqnarray}
\widetilde{S}_A &\equiv& \lim_{Q^2 \rightarrow \infty}\, 
\int_0^1\,dx\, \left[ {F_2^{W^+ n}(x,Q^2) - 
F_2^{W^+ p}(x,Q^2)\over 2x} \right] 
\nonumber \\ &=& \int_0^1\,dx\,
\left[  u^p_{\rm v}(x)- d^p_{\rm v}(x) - \delta u(x) 
  - \delta\bar{d}(x) \right] \nonumber \\
 &=&  1 -\int_0^1\,dx\, \left[ \delta\bar{u}(x)+ 
 \delta\bar{d}(x) \right] . 
\label{Adlercsv}
\end{eqnarray}
If charge symmetry is exact, then $S_A = \widetilde{S}_A$.  
If we assume that the quark normalization integral is indeed
one, then the experimental results allow us to place an upper
limit on the integral of the antiquark charge symmetry
violating amplitudes at the 20\% level.  We will discuss
this more in Sect.\ 7.2  in connection with the ``charge symmetry'' 
sum rule.

\subsection{Gross-Llewellyn-Smith sum rule}

The Gross-Llewellyn Smith [GLS] Sum Rule \cite{GLS} is derived from the
$F_3$ structure functions for neutrinos and antineutrinos, 
\begin{eqnarray}
S_{GLS} &\equiv&  \int_0^1\,{dx\over 2x} \left[ 
 xF_3^{W^+ N_0}(x) + xF_3^{W^- N_0}(x) \right] \nonumber \\ 
 &=& {1\over 2} \int_0^1\,{dx\over 2x} \left[ xF_3^{W^+ p}(x) 
 + xF_3^{W^- p}(x) + xF_3^{W^+ n}(x) + xF_3^{W^- n}(x)
 \right] \nonumber \\ &=& {1\over 2}\,\int_0^1\,dx\,
 [ u^p(x) - \bar{u}^p(x) + d^p(x) - \bar{d}^p(x) + 
 c(x) - \bar{c}(x) + s(x) - \bar{s}(x)  \nonumber \\
  &+& u^n(x) - \bar{u}^n(x) + d^n(x) - \bar{d}^n(x) + 
 c(x) - \bar{c}(x) + s(x) - \bar{s}(x) ] \nonumber \\
 &=& 3\left[\, 1- {\alpha_s(Q^2)\over \pi}-  
a(n_f)\left({\alpha_s(Q^2)\over \pi}\right)^2 - 
b(n_f)\left({\alpha_s(Q^2)\over \pi}\right)^3 \right] + 
\Delta HT \quad ,
\label{GLSdef}
\end{eqnarray}
which follows from the normalization of the quark valence 
distributions.  We have presented the sum rule for an 
isoscalar target.  An identical prediction is obtained for either 
a proton or neutron target, i.e.\ $xF_3^{W^+ p}(x) 
+ xF_3^{W^- p}(x)$ or $xF_3^{W^+ n}(x) + xF_3^{W^- n}(x)$, 
in the integrand of the sum rule.  Neither the
Adler nor Gross-Llewellyn Smith sum rules require any
additional assumptions regarding charge symmetry of quark
distributions.  The GLS sum rule does, however, acquire a 
QCD correction, which is represented by the term in square 
brackets in Eq.\ \ref{GLSdef}.  The higher order QCD corrections 
have been derived 
by Larin and Vermarseren \cite{larin}.  They depend on the 
strong coupling $\alpha_s(Q^2)$.  The terms $a$ and $b$ depend
on the number of quark flavors ($n_f$) available at a 
particular value of $Q^2$.  The quantity $\Delta HT$ 
represents a higher twist contribution \cite{braun}.   

\begin{figure}
\centering{\ \psfig{figure=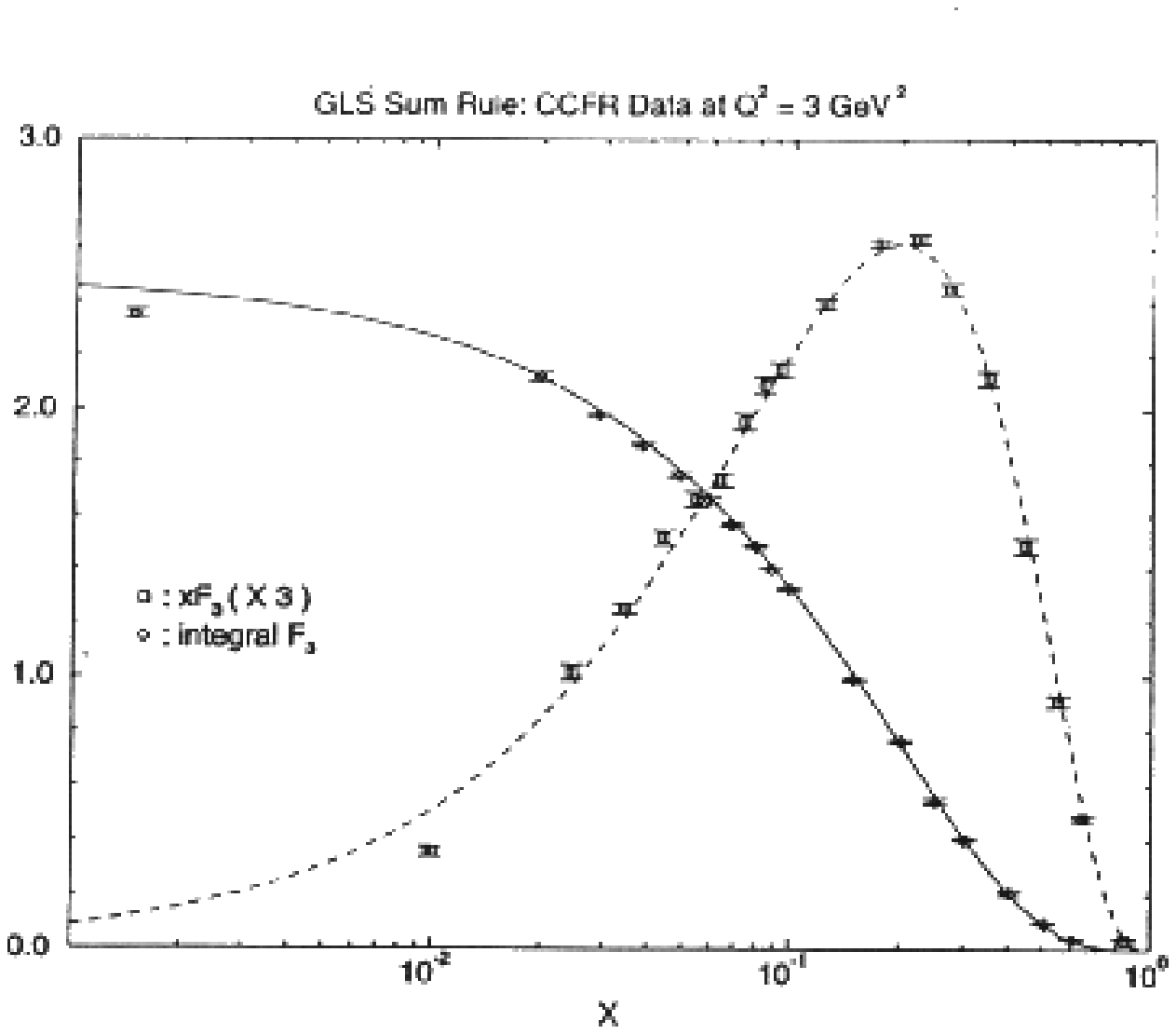,width=9.5cm}}
\vspace{0truein}
\caption{Experimental results for Gross-Llewellyn Smith sum rule, 
Eq.\ \protect\ref{GLSdef}, from CCFR group, Ref.\ \protect\cite{CCFR}.}
\vspace{0.1truein}
\label{fig53}
\end{figure}

As is the case for the Adler and Gottfried sum rules, the
Gross-Llewellyn Smith sum rule requires that the structure
function be divided by $x$ in performing the integral.  This
gives a strong weighting to the small-$x$ region, such that 
as much as 90\% of the sum rule comes from the region $x \leq 0.1$.  
The GLS sum rule is the most precisely known of the three
sum rules we consider.  The best (and most recent) value has
been obtained by the CCFR collaboration \cite{CCFR}, which 
measured neutrino and antineutrino cross sections on iron
targets, using the quadrupole triplet beam (QTB) at FNAL.  
In Fig.\ \ref{fig53} we show the CCFR measurements, the
experimental values of $xF_3$, and their integral, vs.\ 
$x$.  They obtain cross sections at several values of $x$
and $Q^2$; the final value for the sum rule is given for
$Q^2 = 3$ GeV$^2$.  Their reported value for the sum rule is
$S_{GLS} = 2.50 \pm 0.018 \pm 0.078$.  The GLS sum rule
is therefore known to 3\%.  

\begin{figure}
\centering{\ \psfig{figure=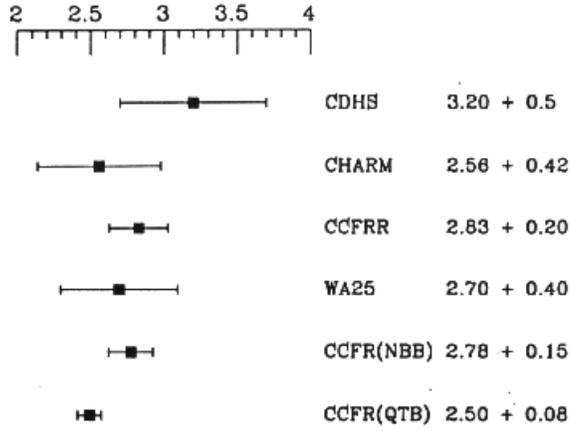,width=8.5cm}}
\vspace{0truein}
\caption{Experimental results for Gross-Llewellyn Smith sum rule, 
and their errors, for a series of experiments, in chronological order
from top to bottom.}
\vspace{0.1truein}
\label{fig54}
\end{figure}

A theoretical value for the Gross-Llewellyn Smith sum
rule requires evaluating the QCD corrections.  The most
recent calculations include next-to-leading order QCD
corrections.  They use a QCD scale parameter 
$\Lambda_{QCD} = 213 \pm 50$ MeV.  With this scale parameter
and NLO QCD corrections, one obtains a theoretical prediction 
$S_{GLS} = 2.63 \pm 0.04$.  
The theoretical prediction is just within two
standard deviations of the experimental value.  In 
Fig.\ \ref{fig54} we show the evolution over time of the GLS sum 
rule value.  

The errors on the GLS sum rule are now at a level where the value
of the strong coupling constant $\alpha_s$ is a major source of
error.  The CCFR group may now have data on $xF_3$ over a wide enough 
range of $Q^2$ that, together with renormalized data from several other 
experiments, they may be able to evaluate the GLS sum rule
without extrapolation for a large range of $Q^2$ values. 
This raises the hope that one can calculate the Gross-Llewellyn Smith 
sum rule as a function of $Q^2$, and to use the resulting $Q^2$ 
dependence of the sum rule to determine $\alpha_s(Q^2)$.  

The CCFR group has recently re-calculated both
the GLS sum rule, and the strong coupling constant $\alpha_s$  
\cite{harris}.  With data of this quality over a large $Q^2$ range, 
it may be possible to use the $Q^2$ dependence to put constraints on 
the strong coupling constant.  Additional information regarding this 
procedure can be found in the thesis by Seligman \cite{Sel97a}.  

The quantities $xF_3^{W^+ N_0}(x) + xF_3^{W^- N_0}(x)$, which
form the integrand for the GLS sum rule, are obtained by 
taking the difference between cross sections for neutrinos 
and antineutrinos on isoscalar targets.  This was discussed in Sect.\ 
2.5.  If charge symmetry is exact,
then the $F_2$ structure functions exactly cancel when we
take the difference of neutrino and antineutrino cross sections, 
and we
are left only with the $F_3$ structure functions.  However, 
if we allow charge symmetry violation, then {\em insofar as the 
sum of $F_3$ structure functions is defined as the difference
between neutrino and antineutrino cross sections on an isoscalar
target}, there are additional contributions to this integrand, 
{\it i.e.} using Eqs.\ \ref{F1Nzero} and \ref{F3rcsv}, 
\begin{eqnarray}
{3\pi \over 2 G^2M_NE} \left( 
  d\sigma^{\nu N_0}/dx - d\sigma^{\bar{\nu} N_0}/dx \right) &=& 
  {1\over 2}\left( xF_3^{W^+ N_0}(x,Q^2) + xF_3^{W^- N_0}(x,Q^2) 
  \right) \nonumber \\ 
  &+& F_2^{W^+ N_0}(x,Q^2) - F_2^{W^- N_0}(x,Q^2) 
  \nonumber \\ &=& {x\over 2}\left[ 2\left( u^p_{\rm v}(x) + 
  d^p_{\rm v}(x)\right) + 
  6\left(s(x)- \bar{s}(x)\right) \right. \nonumber \\ 
  &-& \left. 3\delta u_{\rm v}(x)  +  \delta d_{\rm v}(x) \right]  
\label{F3csv}
\end{eqnarray}

In addition to the valence quark distributions 
which we obtain in the charge symmetric limit, there is an
additional contribution of three times the difference between
strange and antistrange quarks.  There is an additional term
proportional to the valence quark 
CSV terms.  From Sect.\ 3, the CSV terms are predicted to 
have opposite signs so their contributions should add
coherently in Eq.\ \ref{F3csv}.  Although we expect charge
symmetry violating contributions of at most about two percent 
in the $F_2$ structure functions, the relative contribution
to the experimental values of $F_3$ could be somewhat  
larger.  This is because in the double differential 
neutrino cross sections (see, e.g., Eq.\ \ref{signuCC}), the 
$F_3$ structure functions are multiplied by a
coefficient $y - y^2/2$, while $F_2$ has a coefficient which is 
roughly 1.  Since the
average value of $y$ in these experiments is about 0.2, the
$F_2$ structure functions will on average 
be weighted by a factor 5 relative to the $F_3$ terms.  Thus, 
naively we expect CSV effects in $F_2$ to be magnified by a factor 
of about 5 in extracting $F_3$.  Thus a 2\% CSV amplitude
in the quark distributions could potentially change the extracted
value of $F_3$ by as much as 10\%.  

Despite the possibility that the $F_3$ structure functions
could be modified by quark CSV effects, {\em in principle, these 
CSV amplitudes should have no effect
on the GLS sum rule}.  Since the GLS sum rule involves
an integral over $x$, the net contribution to the sum rule
arising from the CSV effects is 
\begin{equation}
\delta S_{GLS}^{CSV} = \int_0^1\, dx \left[ \delta d_{\rm v}(x) - 
  \delta u_{\rm v}(x) \right] = 0
\label{GLScsv}
\end{equation}
The integral vanishes because the quark valence distributions
obey the normalization conditions, Eq.\ \ref{GSRnrm}.  Therefore
the integral over $x$ of both $\delta d_{\rm v}(x)$ and 
$\delta u_{\rm v}(x)$ must be exactly zero.  In practice, 
this requires having data over all $x$, or correctly 
performing an extrapolation over the unmeasured $x$ region, 
so that the contributions from valence quark CSV terms really 
average to zero.

\section{Flavor Symmetry and Charge Symmetry Sum Rules}
\mb{.5cm}

From the previous section, we see that both FSV and CSV
terms contribute to the Gottfried sum rule, and 
that what is conventionally called the ``Adler sum rule'' also 
contains a CSV contribution from sea quarks.  If 
sufficiently accurate experimental data can be obtained, it
would be useful to 
derive sum rules which could differentiate between
charge symmetry and flavor symmetry violation in the nucleon
sea.  Quantities like the electromagnetic interactions, and
Drell-Yan processes, which couple to virtual photons 
are proportional to the squares of the quark charges.  They 
will give the up quark (and antiquark) distributions a 
relative weighting four times that for the down quark 
distributions.  Neutrino deep inelastic structure
functions, which couple to the weak isospin, allows the possibility 
of separating these contributions.  

This process of defining sum rules which would differentiate between 
charge symmetry and flavor symmetry violation was originally carried 
out by Ma \cite{Ma92}.  Our charge symmetry
and flavor symmetry sum rules are linear combinations of the
integrals defined by Ma.    

\subsection{Charge symmetry sum rule}

We define a ``charge symmetry'' sum rule in terms of
the $F_2$ structure functions for charged current neutrino
interactions on the neutron and proton, 
\begin{eqnarray}
S_{CS} &\equiv& \lim_{Q^2 \rightarrow \infty} 
  \,\int_0^1 \,{dx\over 2x}\, \left[ F_2^{W^- p}(x) -
  F_2^{W^+ n}(x) \right]  \nonumber \\
  &=&  \int_0^1\,dx\, \left[ u^p(x) + \bar{s}(x) + \bar{d}^p(x) 
   -\left(d^n(x) + s(x) + \bar{u}^n(x) \right) \right] \nonumber \\
  &=& \int_0^1\,dx\, \left[ \delta u(x) + \delta \bar{d}(x) 
  + \bar{s}(x)- s(x)\right]  = \int_0^1\,dx\, \left[ 
  \delta\bar{u}(x)+ \delta\bar{d}(x) \right] .
\label{SCS}
\end{eqnarray}
In deriving $S_{CS}$, we assume that the strange and charmed 
contributions
for neutron and proton are identical.  We term this the ``charge
symmetry'' sum rule, since by inspection if either the strong
form or weak form of charge symmetry holds for the nucleon sea
quark distributions, then $S_{CS}$ will be identically equal to
zero, and any deviation from zero will be due to violation of 
charge symmetry.  (To reiterate, the ``strong form'' of charge symmetry
for sea quarks is the statement that $\bar{u}^p(x) = \bar{d}^n(x)$ 
for all $x$; the ``weak form'' would state that the distributions
might not be identical, but that their integrals over $x$ are
equal.  With either form of charge symmetry, the contribution 
to the integrals in the sum rule would vanish).  Just as for
the Adler sum rule, there are no QCD corrections to the
charge symmetry sum rule.  

We can relate the charge symmetry sum rule to the sum rules
listed in Sect. 6.2.  We can easily see that 
\begin{eqnarray}
S_{CS} &=& S_A - \widetilde{S}_A \quad\quad {\rm where} \nonumber \\
S_A &=& \lim_{Q^2 \rightarrow \infty}\, 
\int_0^1\,dx\, \left[ {F_2^{W^- p}(x,Q^2) - 
F_2^{W^+ p}(x,Q^2)\over 2x} \right] , 
\nonumber \\ \widetilde{S}_A &=& \lim_{Q^2 \rightarrow \infty}\, 
\int_0^1\,dx\, \left[ {F_2^{W^+ n}(x,Q^2) - 
F_2^{W^+ p}(x,Q^2)\over 2x} \right] 
\label{ScSa}
\end{eqnarray}
The Adler sum rule $S_A$ requires comparing the structure function 
$F_2$ for charged current weak interactions on protons with
antineutrino beams, and with neutrino beams.  We discussed this
at length in Sect.\ 6.2 (see Eq.\ \ref{Adlercsv}).  It is required to be
one from normalization of valence quark distributions, and has
no contribution from CSV terms.  

The sum
rule $\widetilde{S}_A$ requires subtracting the corresponding $F_2$ 
structure functions for neutrinos on neutrons (i.e., 
deuterons) and protons, respectively; neutrino beams on
different targets.  The ``charge symmetry'' sum rule
requires comparing antineutrinos on protons, with neutrinos
on neutrons.  As an alternative to the charge symmetry sum
rule one could measure $S_A$ and $\widetilde{S}_A$ and
compare them.  If $S_A =1$ and $\widetilde{S}_A \ne 1$, 
this would give clear evidence for charge symmetry violation 
in the nucleon sea quark distributions.  
 
One can straightforwardly construct sum rules using 
different linear combinations 
of $F_2$ structure functions for neutrinos or antineutrinos
on protons or neutrons, which contain the same information
as the sum rules we have defined here.  For example, we
could define
\begin{eqnarray}
\widetilde{S}_{CS} &\equiv& \lim_{Q^2 \rightarrow \infty} 
  \,\int_0^1 \,{dx\over 2x}\, \left[ F_2^{W^+ p}(x) -
  F_2^{W^- n}(x) \right]  \nonumber \\
  &=&  \int_0^1\,dx\, \left[ \delta\bar{u}(x)+ 
 \delta\bar{d}(x) \right]  = S_{CS}, \quad\quad {\rm and} 
  \nonumber \\ 
\widetilde{S}_A^{(3)} &\equiv& \lim_{Q^2 \rightarrow \infty} 
  \,\int_0^1 \,{dx\over 2x}\, \left[ F_2^{W^- p}(x) -
  F_2^{W^- n}(x) \right] = S_A + \widetilde{S}_{CS} \nonumber \\
  &=&  1 + \int_0^1\,dx\, \left[ \delta\bar{u}(x)+ 
 \delta\bar{d}(x) \right] 
\label{SCStil}
\end{eqnarray}

Experimental prospects for accurate measurements of any of these 
sum rules are poor.  All these sum rules require precision 
measurements for
both neutrinos and antineutrinos, over a wide range of
$x$.  Since the sum rules are very sensitive to the small
$x$ region, it is important to have precise data down 
to very small $x$.  Furthermore, all the sum rules require 
data on both hydrogen and deuterium targets.  
As the WA25 group \cite{WA25} measured cross sections from both
neutrinos and antineutrinos on protons and deuterium,  
they could in principle construct the charge symmetry
sum rule.  However, as we have seen (viz., Fig.\ \ref{fig52} of
Section 6.2), the errors in the Adler sum rule are of
the order of 20\%, so the charge symmetry sum rule would
be consistent with zero at the 20\% level.  We discussed
in the previous section the difficulties in obtaining
precise neutrino data on light targets, over a wide range
of $x$.  

We can compare our ``charge symmetry'' sum rule with the
one proposed by Ma \cite{Ma92}.  He defined the following
sum rule (Eq.\ [14] of Ref.\ \cite{Ma92})
\begin{eqnarray}
S_{CS}^{(Ma)} &\equiv& \,\int_0^1 \,{dx\over 2x}\, \left[ 
 F_2^{W^+ p}(x) + F_2^{W^- p}(x) - \left(
 F_2^{W^+ n}(x) + F_2^{W^- n}(x)\right) \right] \quad .   
\end{eqnarray} 
We see that if the charge symmetry sum rule can be written as
$S_{CS}^{(Ma)} = S_{CS} + \widetilde{S}_{CS}$; and since
we have shown that $S_{CS} = \widetilde{S}_{CS}$, all three 
of these sum rules give precisely the same information.  

\subsection{Flavor symmetry sum rule}

We can define a ``flavor symmetry'' sum rule by comparing
the $F_3$ structure functions from charged current neutrino
interactions on protons and neutrons, i.e.
\begin{eqnarray} 
S_{FS} &\equiv& \,\int_0^1 \,{dx\over 2x}\, \left[ 
 xF_3^{W^+ n}(x) - xF_3^{W^+ p}(x) \right]  \nonumber \\  
  &=& \,\int_0^1 \,dx [ -\left( \bar{u}^p(x) + \bar{c}(x) 
  - d^p(x) -s(x) \right) \nonumber \\
  &+&  d^n(x) + s(x) - \bar{u}^n(x) - \bar{c}(x) 
   ] \nonumber \\
  &=& \int_0^1\,dx\,\left[ d_{\rm v}^n(x) - d_{\rm v}^p(x) + 
  \bar{u}^p(x) - \bar{u}^n(x) - 
  \bar{d}^p(x) + \bar{d}^n(x) \right] \nonumber \\
  &=& \left( 1 + \int_0^1\,dx\,\left[ \bar{u}^p(x) - \bar{d}^p(x)  
  - \bar{u}^n(x) + \bar{d}^n(x) \right]\right) \left(1 - 
  {\alpha_s(Q^2)\over \pi} \right).
  \label{Sfs}
\end{eqnarray}
If flavor symmetry holds for the nucleon sea, i.e.\ if 
$\bar{u}^p(x) = \bar{d}^p(x)$ and 
$\bar{u}^n(x) = \bar{d}^n(x)$ (or the weak condition that
the integrals over $x$ of these distributions are equal), 
then $S_{FS}$ will be equal
to one.  So any deviation of this sum rule from one signifies 
flavor symmetry violation in the nucleon sea.  There is a 
QCD correction to this sum rule.  It is the same as for
the Gross-Llewellyn Smith sum rule (see Section 6.3).  
In Eq.\ \ref{Sfs} we 
include the lowest-order QCD correction.  

Experiments with muons suggest the magnitude of flavor symmetry
breaking which we might expect.  As we discussed in Sect.\ 3.2, 
theoretical calculations predict that sea quark CSV effects will
be very small.  In that case the antiquark contributions to 
Eq.\ \ref{Sfs} will be identical to those measured by the NMC group 
\cite{Ama91}.  For $Q^2 = 10$ GeV$^2$, we then expect the integral
over the antiquark distributions to give a 30\% effect.  However, there
will be significant experimental difficulties obtaining accurate data 
for the sum rule.  The factor $1/x$ in the integrand requires precise 
neutrino data on both protons and neutrons (i.e., deuterium), at very
small $x$.  Even small differences between the $F_3$ structure functions
on protons and deuterium become magnified when one integrates the
difference between them.  

Our flavor symmetry sum rule was obtained by combining $F_3$
structure functions for neutrinos on protons and neutrons.  We
could define an analogous function by utilizing antineutrino
structure functions on protons and neutrons,  
\begin{eqnarray}
\bar{S}_{FS} &\equiv& \,\int_0^1 \,{dx\over 2x}\, 
\left[ xF_3^{W^- p}(x) - xF_3^{W^- n}(x) \right]  \nonumber \\ 
  &=& \int_0^1\,dx\,\left[ u_{\rm v}^p(x) - u_{\rm v}^n(x) + 
  \bar{u}^p(x) - \bar{u}^n(x) - 
  \bar{d}^p(x) + \bar{d}^n(x) \right] \nonumber \\
  &=& \left( 1 + \int_0^1\,dx\,\left[ \bar{u}^p(x) - \bar{d}^p(x)  
  - \bar{u}^n(x) + \bar{d}^n(x) \right]\right) \left(1 - 
  {\alpha_s(Q^2)\over \pi} \right) \quad .
  \label{Sbarfs}
\end{eqnarray} 
Comparing Eq.\ \ref{Sbarfs} with Eq.\ \ref{Sfs} we see
that $\bar{S}_{FS} = S_{FS}$;
consequently both of these sum rules give
exactly the same information.  

Ma defines an additional sum 
rule, $S'$ (Eq.\ [23] of Ref.\ \cite{Ma92}), through
\begin{eqnarray}
S' &\equiv& \,\int_0^1 \,{dx\over x} \left[ xF_3^{W^+ p}(x) -
 xF_3^{W^- p}(x) - xF_3^{W^+ n}(x) + xF_3^{W^- n}(x) 
\right]  \nonumber \\
 &=& \left(-4 - 4\int_0^1\,dx\, \left[ \bar{u}^p(x) - \bar{d}^p(x) - 
 \bar{u}^n(x) + \bar{d}^n(x) \right]\right) \left(1 - 
  {\alpha_s(Q^2)\over \pi} \right) .  
\end{eqnarray} 
This is related to the two previously defined sum rules by 
$S' = - 2(S_{FS}+ \bar{S}_{FS})$, and gives the
same information as either of those.  It requires knowledge of
both neutrino and antineutrino parity violating 
structure functions on both the proton and neutron.

\section{Conclusions}
\mb{.5cm}

We have reviewed the validity of charge symmetry for parton distributions.  
We calculated CSV contributions for both valence and sea quarks.  
The ``majority'' valence quark distribution, and the sea quark
distribution, are predicted to obey charge symmetry to at least 1\% for
all values of $x$.  The ``minority'' valence quark distribution, however,
is predicted to show CSV effects of between 3-7\% at large Bjorken $x$.  
This prediction appears to be robust, as all calculations of quark
CSV predict effects of this magnitude, and reasonably model-independent
estimates confirm these results.  

This violation of charge symmetry is large.  If this turns out to
be correct, we should probably re-evaluate all phenomenological
parton distributions, introduce some explicit charge symmetry
violation and re-fit existing cross sections.  Towards this end, 
we have redefined a number of observables, using a formalism 
which does not assume explicit parton charge symmetry.   One 
difficulty in searching for CSV effects at present is that all
phenomenological parton distributions assume charge symmetry at
the outset, so any existing CSV effects have been absorbed into
the current parton distributions.  This makes it difficult to
search for experimental violation of charge symmetry.  Of course,
if CSV effects were extremely large, we would already have seen 
this in existing experimental data. 

If charge symmetry is indeed violated at the predicted level for the minority
valence quarks, then one should be able to measure such effects.  
First, we reviewed the status of current tests of parton charge
symmetry.  There are essentially two such tests.  The first involves
the comparison of charge-current cross sections induced by neutrinos
to those from antineutrinos.  This comparison can detect CSV in valence
quark distributions; the test is only valid at large $x$.  Existing data 
gives only weak upper limits on valence quark CSV.  The second test of 
parton CSV is the so-called 
``charge ratio,'' the ratio of $F_2$ structure functions in DIS 
processes induced by leptons, to the $F_2$ structure functions 
in charge-changing weak processes from neutrino beams (both of these
need to be measured on isoscalar targets).   Despite the fact that
this comparison requires a large number of corrections, it appears that 
most of these corrections are under control.  

Recent data, muon induced DIS on deuterium from the NMC group 
\cite{Ama91}, and neutrino reactions on iron from the CCFR group
at FNAL \cite{Sel97}, allow us to place rather tight constraints
on parton CSV amplitudes.  Such relations should hold at all values
of $x$.  These 
experiments allow us to set limits of roughly 10\% on
charge symmetry violating contributions to parton
distributions, in the region $0.2 \le x \le 0.4$.  However, in the 
region $x < 0.2$, there appears to be a discrepancy between the 
electromagnetic and weak $F_2$ structure functions.  

In this review, we have suggested several
experiments specifically designed either to detect CSV in
parton distributions, or to set more stringent
upper limits on parton CSV.  Probably the most sensitive test
would be a comparison of the $F_2$ structure functions for 
charge-changing weak processes induced by electrons and positrons on
an isoscalar target.  At sufficiently high energies (such as are 
accessible at HERA), comparison of $e^+ -D$ and $e^- -D$ processes 
should be a powerful and relatively clean test of parton CSV.  Other
potentially useful reactions which are quite sensitive to parton CSV
are Drell-Yan processes for
charged pions on isoscalar nuclear targets, or semi-inclusive
leptoproduction of pions on isoscalar targets.  We made estimates
of the magnitude of CSV effects expected in these reactions.   

We also discussed the contributions of CSV effects in various 
sum rules.  We showed that CSV contributions to the
nucleon sea have no effect on the Gross-Llewellyn Smith 
sum rule, but in principle they affect both the Adler
and Gottfried sum rules.  In fact, the existing experimental
``test'' of the Adler sum rule contains a 
contribution from CSV in sea quark distributions, and can be used to
place an upper limit on parton sea quark CSV.  
We introduced two new sum rules, a ``charge symmetry'' 
sum rule which is zero if charge symmetry is exact, and
a ``flavor symmetry'' sum rule which is one if flavor
symmetry is exact in the nucleon sea.  Both sum rules
require measuring the structure functions for neutrinos
and/or antineutrinos on isoscalar targets, with particular attention
to the small-$x$ region.  Limits on CSV from present experiments are 
in the neighborhood of 10\% of the average parton distributions.  

\vspace{2.5cm}
{\bf Acknowledgments}
\mb{.5cm}

This review includes research supported in part by the US National
Science Foundation under research contract NSF-PHY-9722076, and 
by the Australian Research Council.  One author [JTL] would 
also like to thank the Special Research Centre for the Subatomic
Structure of Matter for its hospitality during a period when 
this review was written.  The authors acknowledge several discussions  
with A. Pang regarding sections of this review.  In addition, the authors 
would like to acknowledge discussions with and contributions by C. Benesh, 
C. Boros, S. Braendler, G.T. 
Garvey, R.L. Jaffe, G.Q. Liu, W. Melnitchouk and E.N. Rodionov.  
One author [JTL] acknowledges several discussions with S. Kumano, 
and wishes to thank C. Keslin and D. Murdock for assistance with the 
figures for this review.

\end{document}